\documentclass[manuscript]{aastex}
\usepackage{amsmath}
\usepackage{xspace}
\usepackage{graphicx}
\usepackage{epstopdf}
\usepackage{placeins}

\newcommand\hh{H$_2$}

\newcommand\vol{cm$^{-3}$}

\shorttitle{UV-heated outflow cavity walls}
\shortauthors{Lee et al.}

\begin{document}
\title{The warm CO gas along the UV-heated outflow cavity walls: a possible interpretation for the \textit{Herschel}/PACS CO spectra of embedded YSOs}

%\author{Seokho Lee \begin{CJK}{UTF8}{mj} (이석호)\end{CJK}}
\author{Seokho Lee}
\affil{School of Space Research, Kyung Hee University, Yongin-shi, Kyungki-do 449-701, Korea}

%\author{Jeong-Eun Lee \begin{CJK}{UTF8}{mj} (이정은) \end{CJK}}
\author{Jeong-Eun Lee}
\affil{School of Space Research, Kyung Hee University, Yongin-shi, Kyungki-do 449-701, Korea; jeongeun.lee@khu.ac.kr}

\and 
\author{Edwin A. Bergin}
\affil{Department of Astronomy, University of Michigan, 830 Dennison Building, 500 Church Street, Ann Arbor, MI 48109}
%\end{CJK}

\begin{abstract}

A fraction of the mid-$J$ ($J$= 14--13 to $J$= 24--23) CO emission detected by the \textit{Herschel}/PACS observations of embedded young stellar objects (YSOs) has been attributed to the UV-heated outflow cavity walls. We have applied our newly developed self-consistent models of Photon-Dominated Region (PDR) and non-local thermal equilibrium line Radiative transfer In general Grid (RIG) to the \textit{Herschel} FIR observations of 27 low mass YSOs and one intermediate mass YSO, NGC7129-FIRS2. When the contribution of the hot component (traced by transitions of $J>~24$) is removed, the rotational temperature of the warm component is nearly constant with $\sim250$~K. This can be reproduced by the outflow cavity wall ($n \geq 10^6\, \mathrm{cm}^{-3}$, $\log G_{0}/n \geq-4.5$, $\mathrm{log}~G_0\ge 3$, $T_{\rm gas} \ge 300~$K, and X(CO)$ \ge 10^{-5}$) heated by  a UV radiation field with a black body temperature of 15,000~K or 10,000~K. However, a shock model combined with an internal PDR will be required to determine the quantitative contribution of a PDR relative to a shock to the mid-$J$ CO emission.

\end{abstract}
%
%
%  INTRODUCTION
%
%
\section{Introduction }\label{sec:intro}
 Embedded young stellar objects (YSOs) are associated with energetic phenomena: jets, outflows, and high energy photons emitted by accretion shocks on the surface of protostars and disks. 
 These phenomena determine the physical conditions of the surrounding material, in particular, in close proximity to  the central object.
 However, it is difficult to directly detect emission from the warm/hot gas and dust closest to the forming star because of the thick enshrouding envelope.

In this regard, far infrared (FIR) spectroscopy can be a powerful tool for the study of embedded YSOs because the energetic photons produced by accretion are absorbed and re-emitted in this wavelength regime.  
 FIR spectroscopic observations of twenty-eight low-mass embedded protostars \citep[e.g.,][]{Giannini2001,Nisini2002,van Dishoeck2004} were carried out for the first time with the Long Wavelength Spectrometer \citep[LWS;][]{Clegg1996} aboard the Infrared Space Observatory (ISO).
 These observations discovered widespread CO emission arising from a  rotational states from  $J$=~14 to $J$=~29 \citep[for NGC1333 IRAS 4,][]{Giannini2001,Maret2002}.
 A Large Velocity Gradient (LVG) analysis of these observations suggests that the FIR CO emission is radiated from the gas with the temperature of  a few hundreds to $\sim$1000~K and a density of $10^5 \sim 10^6$~cm$^{-3}$ \citep[e.g., ][]{Giannini2001,Nisini2002,van Dishoeck2004}. 
A non-dissociative shock is the leading candidate for the heating mechanism producing the high-$J$ CO emission \citep[e.g.,][]{Nisini2002}; however \citet{Ceccarelli2002} proposed that super-heated a disk surface layer that is exposed to ultraviolet photons  could account for the FIR CO emission in Elias 29.

 The \textit{Herschel} Space Observatory \citep{Pilbratt2010} provides higher spatial resolution and sensitivity when compared to ISO.
  Furthermore the Photodetector Array Camera and Spectrometer \citep[PACS, ][]{Poglitsch2010} covers the CO rotational lines from $J$=~14--13 to $J$=~49--48 \citep{Herczeg2012} and provided a wealth of new observations of energetic gas in protostars. 
A \textit{Herschel} open time key program, ``\textit{Herschel} Orion Protostar Survey (HOPS)" observed 22 protostars in Orion.  \citet{Manoj2013} noted that these sources span two orders of magnitude in bolometric luminosity ($0.2~\mathrm{L}_\odot\leq~L_{\rm  bol}\leq~28~\mathrm{L}_\odot$), while their CO rotation diagrams show that the CO emission can be characterized by two thermal components: warm gas with the rotational temperature of $T_{\rm rot}$ $\sim350$~K and hot gas with $T_{\rm rot} \,\sim 700-900$~K. The rotational temperature of $\sim350$~K appears to be universal in the mid-$J$ range ($14 \leq J\leq 24$) and independent of the bolometric luminosity. Additional observations of the CO rotational emission were obtained as part of the \textit{Herschel} key program, ``Water in star forming regions with \textit{Herschel}" (WISH) observed 18 embedded protostars \citep{Karska2013}, and ``Dust, Ice, and Gas In Time" (DIGIT) observed 30 sources \citep{Green2013}.  These sources also have properties similar to those observed by the HOPS program, i.e., all programs found the universal 350~K component in the mid-$J$ CO ladder.

 An LVG analysis indicates that there are two possible explanations for the observed CO emission. One is sub-thermally excited gas with a high temperature ($>~1000~\rm{K}$) and  low density ($< 10^{6}$~cm$^{-3}$) \citep{Neufeld2012,Manoj2013}. In this case, one component can fit the CO emission over the entire PACS range. The other is thermally excited gas with high density ($> 10^6$~cm$^{-3}$) and both warm  ($\sim300~\rm{K}$) and hot ($700 - 800$~K) temperatures, which are shown in the rotational diagram.

 \citet{van Kempen2010a} and \citet{Visser2012} explored the origin of the mid-$J$ CO emission as arising from the UV heated gas along the outflow cavity walls (Photon Dominated Region; PDR) and  small-scale C-type shocks inside the walls. Furthermore, \citet{Visser2012} suggested that the PDR contributes more to the FIR CO emission as the protostar evolves. However, there is caveats in their model; they combined an \textit{approximated} two-dimensional (2D) PDR model and one-dimensional (1D) shock model, and the UV spectrum for gas energetics and chemistry was different in the PDR model.

 \citet{Manoj2013} argued that the PDR is a minor contributor to the mid-$J$ CO emission based on three points. First, it is difficult for the PDR process to produce a similar gas temperature regardless of the bolometric luminosity, $L_{\rm bol}$. The gas temperature in the PDR is roughly proportional to $L_{\rm UV}/n$, where $L_{\rm UV}$ is the UV luminosity (considered to be proportional to $L_{\rm bol}$) and $n$ is the gas density. The density at a specific radius is not necessarily correlated with $L_{\rm bol}$. As a result, there is no reason to have a similar $L_{\rm UV}/n$ over a large range of $L_{\rm bol}$. Second, because of the dilution of UV photons, the PDR cannot produce consistent gas conditions over the large range of radius as required to match the CO mid-$J$ emission. Finally, resolved spectra of lower $J$ ($\le$~10) CO emission in NGC1333 \citep{Y&inodot;ld&inodot;z2010},  shows that the contribution of a ``broad" spectral component is consistent with shocked gas  and dominates $J$=~10--9 emission, implying that the PDR is a  minor contribution to the CO emission in PACS range.
In addition, \citet{Karska2013} also argued that the UV heated gas is a minor component to the mid-$J$ CO emission because of a strong correlation between the fluxes of CO $J$=~14--13 and $J$=~24--23  and the flux of H$_2$O $2_{12}-1_{01}$, which traces shocked gas. 
  
 However, an internal PDR must exist some level in the protostellar  stage.  Furthermore it may not be negligible because FUV observations toward classical T Tauri stars find that these stars emit the UV photons at a few percent of the accretion luminosity \citep[e.g., ][]{Herczeg2002,Yang2012}. 
 These UV photons can affect the physical and chemical properties of exposed gas within the outflow cavity and along the walls \citep[][hereafter Paper I]{van Kempen2010a,Visser2012,Lee2014b}.  
In this regard, \citet{Y&inodot;ld&inodot;z2012} used spectrally resolved observations of $^{13}$CO~$J$=6--5 finding that the narrow emission lines ($\Delta v <\, 2\, {\rm km \, s}^{-1}$) towards NGC1333-IRAS4A are consistent with emission from
UV heated outflow cavity walls, which encapsulate the broad outflow lines ($\Delta v >\, 10\, {\rm km \, s}^{-1}$).   They find that the mass of the UV heated gas is at least comparable to the mass of  outflowing gas.
 In addition, UV photons produced from accreting protostars are required to explain the emission from ionized hydrides (CH$^+$ and OH$^+$) that are detected with \textit{Herschel}/HIFI \citep{de Graauw2010} and are inferred to emit from within 100 AU of the young star \citep{Kristensen2013}. Therefore, a quantitative test with a self-consistent PDR model is require to study the importance of PDR for the mid-$J$  CO transitions in the embedded protostellar objects.
 
 We developed a 2D PDR model that self-consistently calculates the gas energetics and chemistry for a given UV spectral type (Paper I). This model was applied to HH46, assuming the Draine interstellar radiation field and a $T_{\rm eff} = 1.5\times 10^4\, \mathrm{K}$ black body radiation field (hereafter BB1.5).  We find that the PDR reproduces the observed mid-$J$ CO emission, while the $T_{\rm eff} = 10^4\, \mathrm{K}$ black body radiation (hereafter BB1.0) results in a lower rotational temperature in agreement with \citet{Visser2012}. According to  this model, the mid-$J$ CO emission is radiated from the thermally excited dense CO gas ($n \geq 10^6\, \mathrm{cm}^{-3}$) with a gas temperature higher than $\sim 300~{\rm K}$ and the CO abundance above $10^{-5}$. 

 In this paper, we apply our self-consistent PDR model to a large number of selected embedded protostars and  test whether the UV heated cavity walls can reproduce the universal rotational temperature and fluxes observed in the mid-$J$ CO ladder.
 We present properties of our sources in Section~\ref{sec:sources}, and the PDR model and adopted physical parameters are described in Section~\ref{sec:model}.
 We present our modeling results in Section~\ref{sec:results}  and discuss the effect of physical parameters in Section~\ref{sec:discussions}. Finally, we summarize our conclusions in Section~\ref{sec:summary}.
 
%
% SOURCE
%
\section{Sources}\label{sec:sources}
 In order to test whether the universal rotational temperature can be produced by a PDR along the outflow cavity walls, we have applied our PDR model to the sources that have pre-existing determinations of  the density structure in the literature \citep{J&oslash;rgensen2002, Kristensen2012}. In addition, one intermediate mass embedded protostar, NGC7129-FIRS2 \citep{Crimier2010,Fich2010} has been modeled as an example of a high luminosity source.
   
\begin{figure*}
\includegraphics[width=1.0 \textwidth]{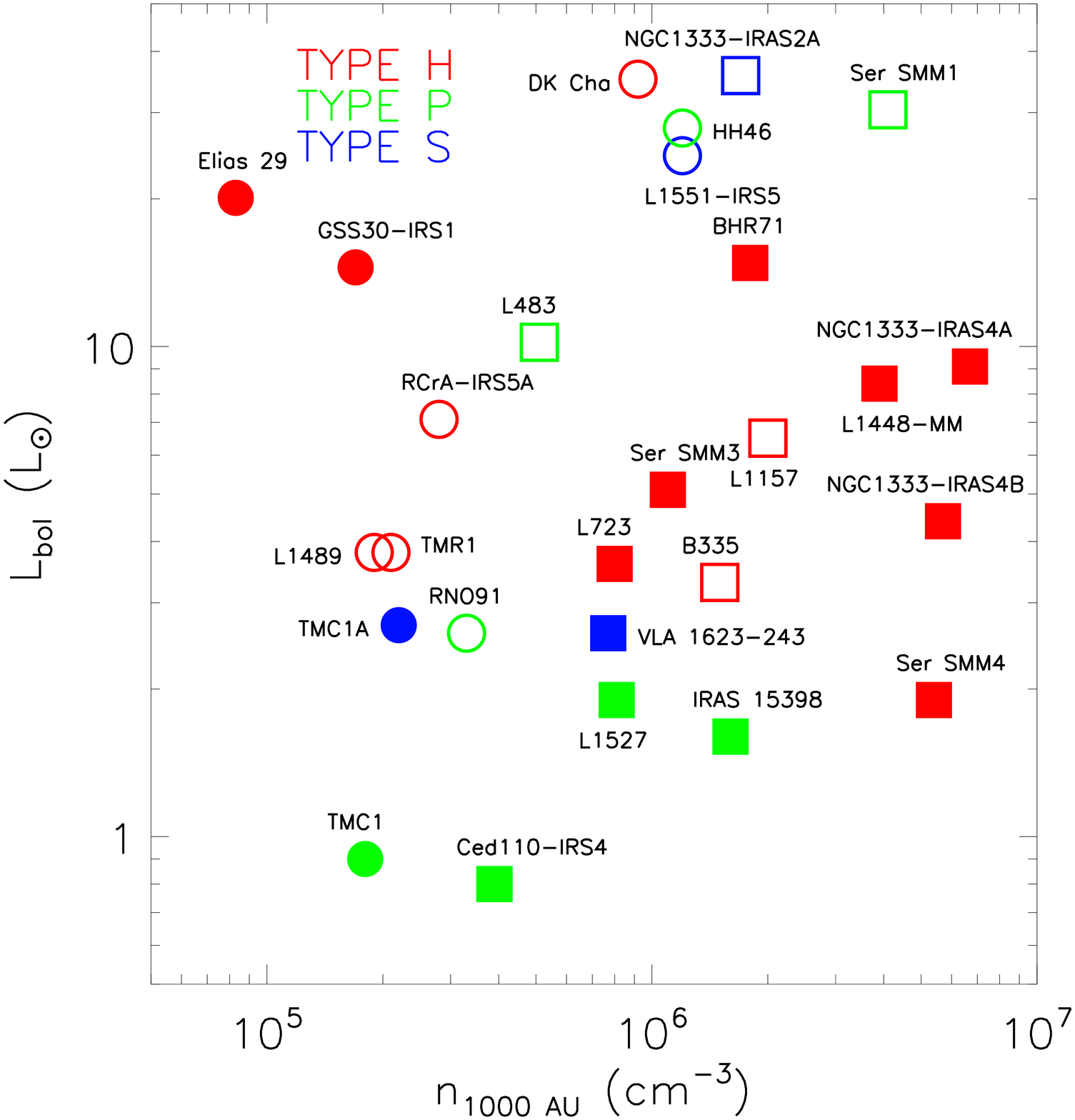}
\caption{The molecular hydrogen number density at 1000~AU ($n_{1000 \rm AU}$) and the bolometric luminosity ($L_{\rm bol}$) of the sources. Class I sources (circle) are located upper left of Class 0 sources (square). NGC7129-FIRS2 is excluded in this plot. Results in Table \ref{tb:result} are also plotted. Open (filled) symbols represent the compact (extended) sources. The color of red, green, and blue represents the source type of ``H", ``P", and ``S", respectively (see text).
}\label{fig:ep_source_type}
\end{figure*}

 The selected sources (0.8~$L_\odot\le L_{\rm bol} \le 500~L_\odot$) are listed in Table \ref{tb:source} and are plotted in the domain of bolometric luminosity versus the density at 1000~AU~($n_{1000\rm{AU}}$) (Figure \ref{fig:ep_source_type}).
 Class I sources (shown as circles) generally have lower values of $n_{\rm 1000 AU}$ than Class 0 sources (squares), as shown in Figure~\ref{fig:ep_source_type}.
 We classify sources as ``compact" and ``extended" depending on the distribution of the CO $J$=14--13 or CO $J$=16--15 emission (in Figure \ref{fig:ep_source_type}; ``compact" and ``extended" sources are marked with open and filled symbols, respectively) following the definition of \citet{Karska2013}.
    
 The observed and synthesized CO fluxes are represented as the total number of CO molecules emitting in a given $J$ level as follows \citep{Karska2013,Green2013},
\begin{equation}
\label{eq:NJ}
\mathcal{N}_{\rm OBS}(J)=\frac{4\pi D^{2}F_{J}}{h\nu_J A_J},
\end{equation}
where $F_{J}$ and $\nu_J$ denote the line flux and the frequency of the CO rotational transition from $J$ to $J-$1, $D$ is the distance to the source, $A_J$ is the Einstein coefficient, and $h$ is Planck's constant.

\begin{figure*}
\includegraphics[width=0.30 \textwidth]{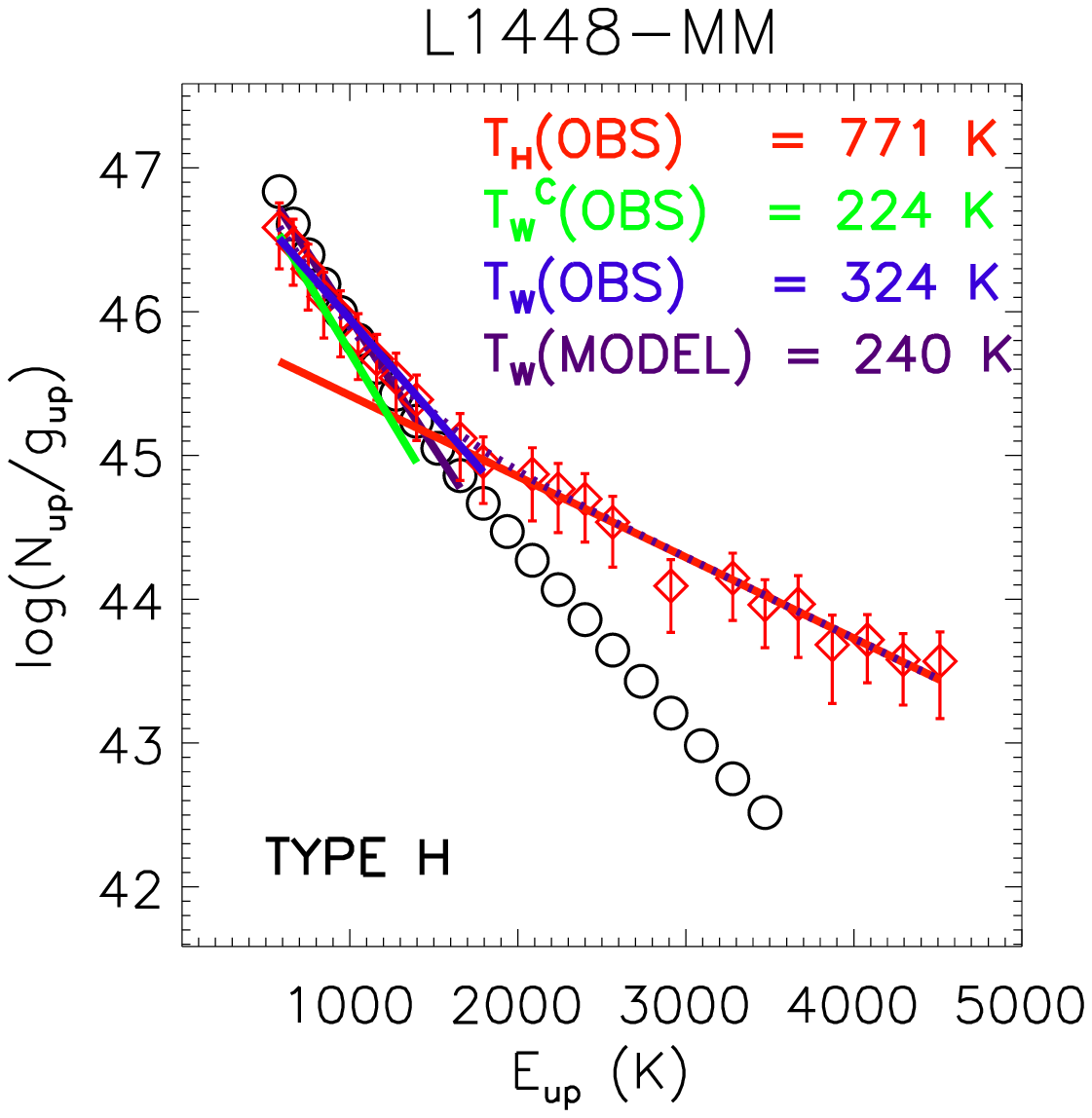}
\includegraphics[width=0.30 \textwidth]{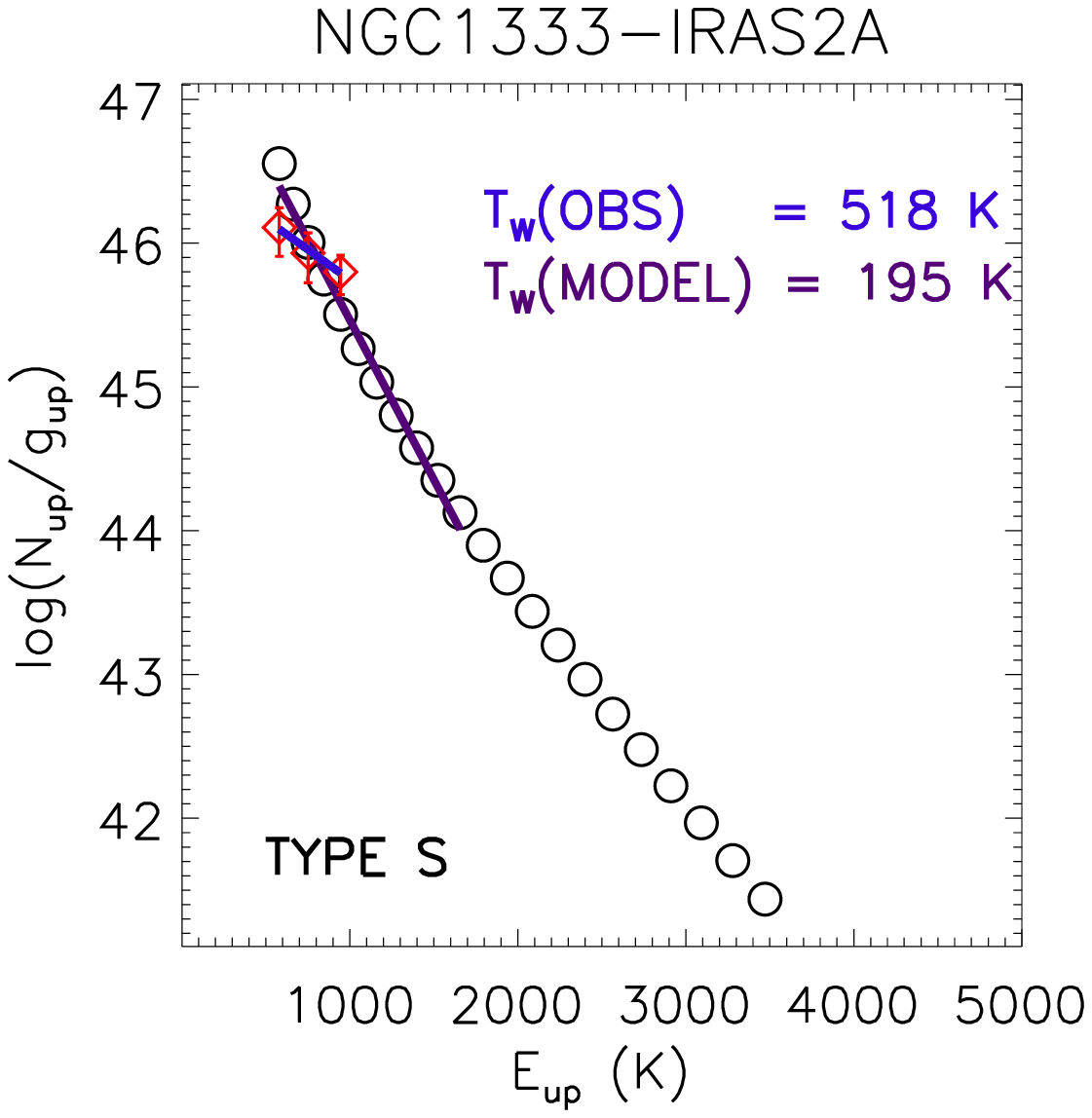}
\includegraphics[width=0.30 \textwidth]{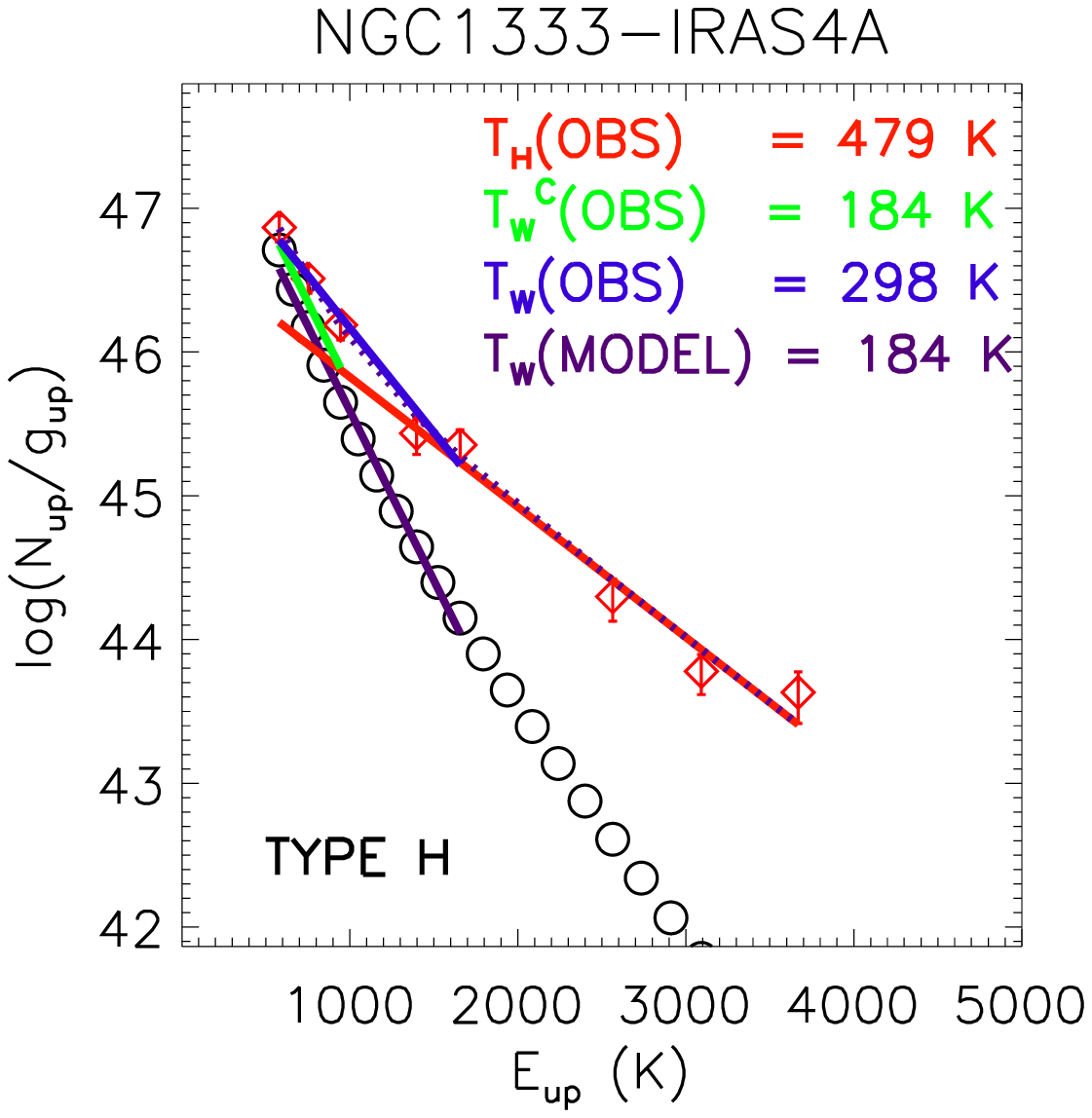}

\includegraphics[width=0.30 \textwidth]{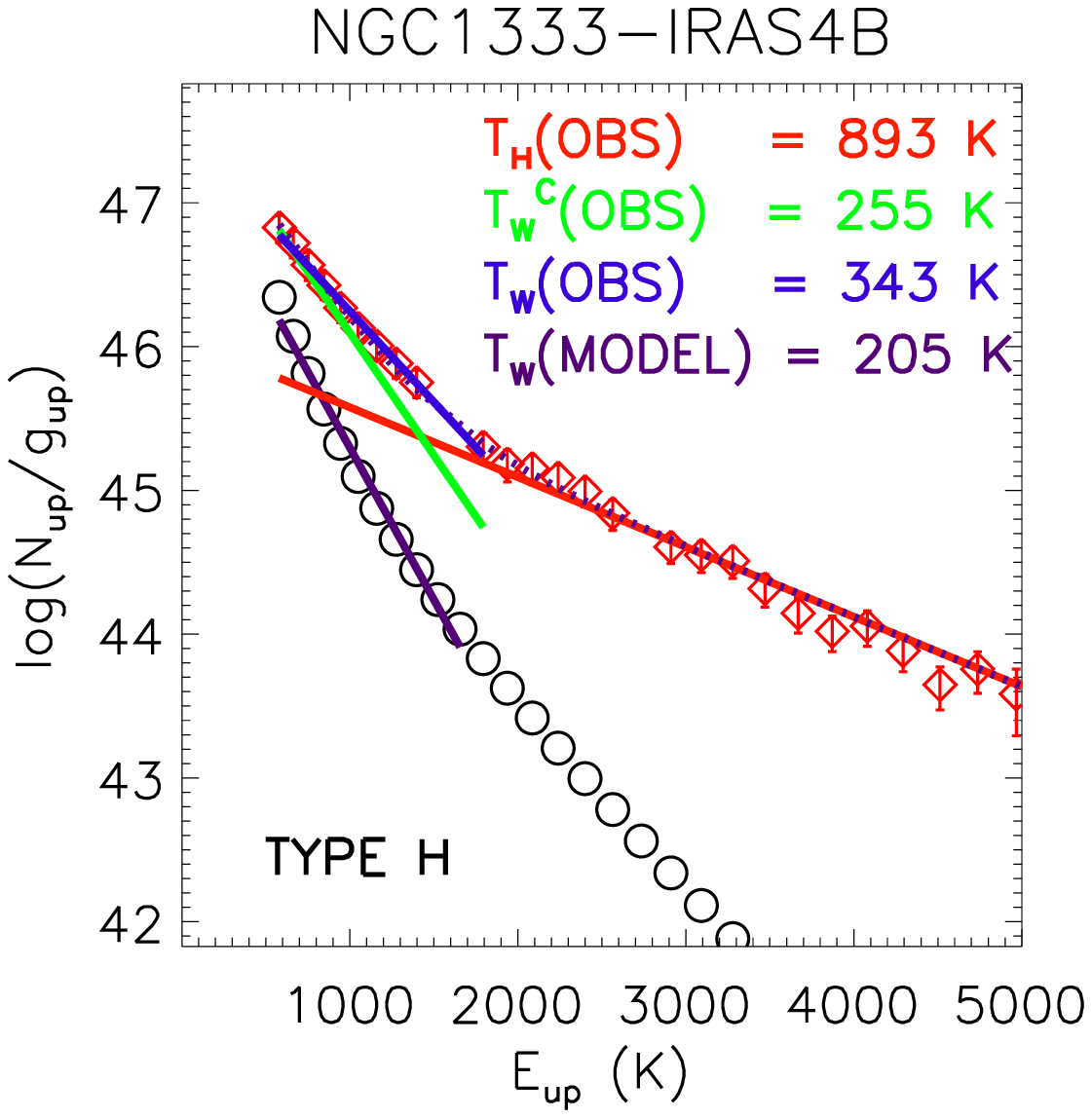}
\includegraphics[width=0.30 \textwidth]{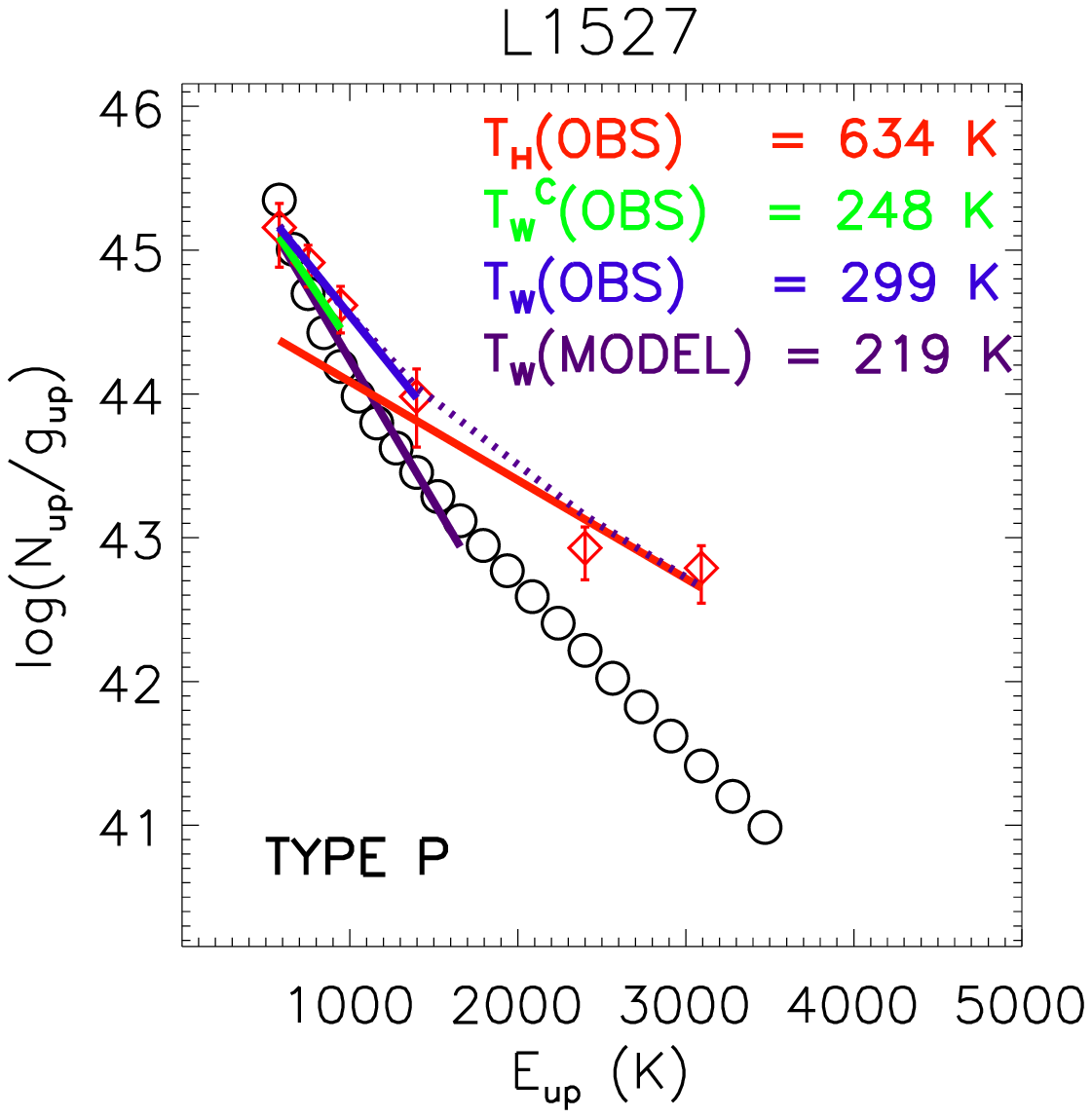}
\includegraphics[width=0.30 \textwidth]{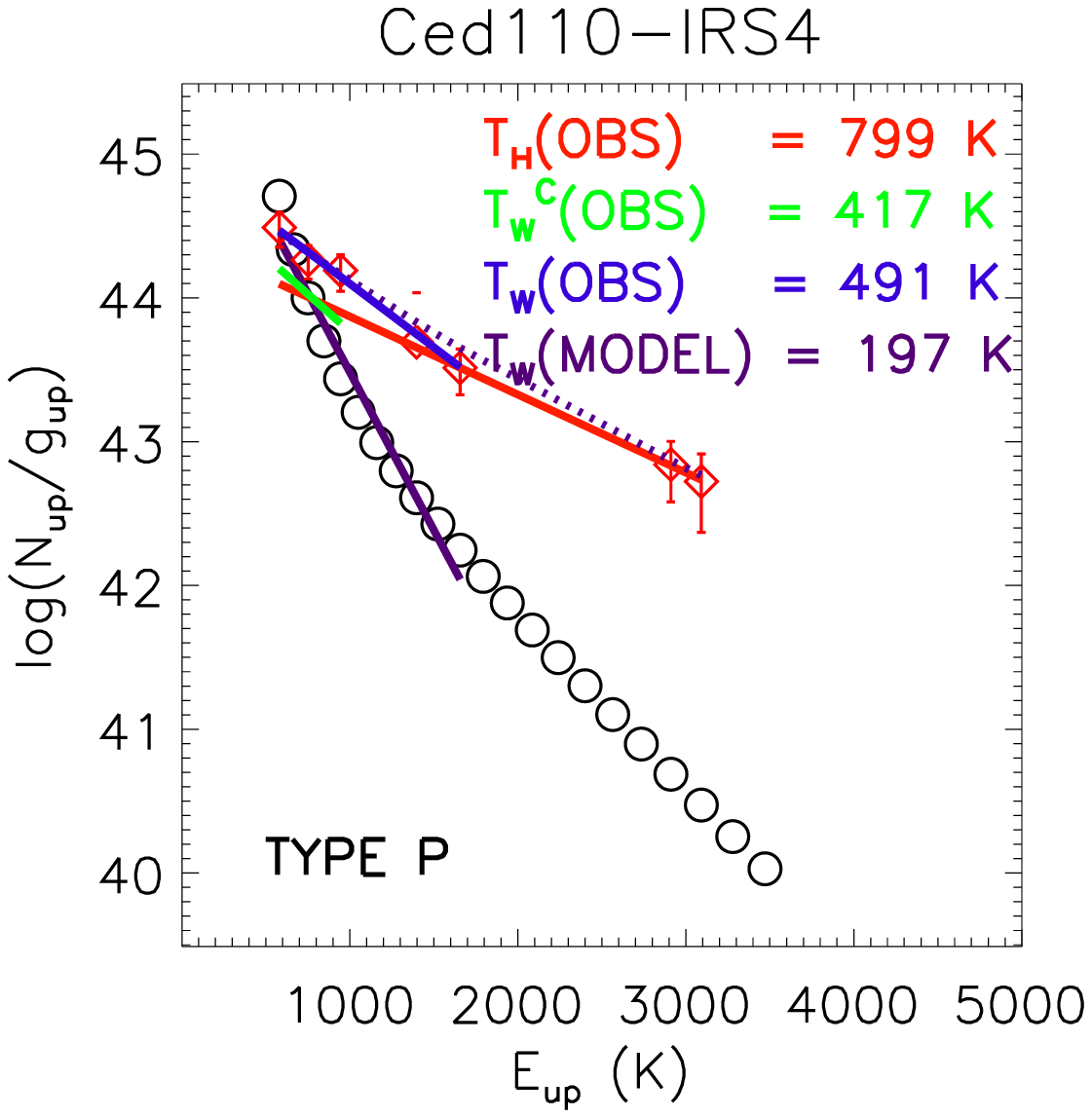}
\caption{ The CO rotational diagrams for L1448-MM, NGC1333-IRAS2A, NGC1333-IRAS4A, NGC1333-IRAS4B, L1527, and Ced110-IRS4, in units of total number of detected CO molecules 
(see Eq.~\ref{eq:NJ}) divided by degeneracy g. 
The  open red diamonds indicate the values derived from the \textit{Herschel}/PACS observations.
The red (``Hot" component) and blue lines (``Warm" component) are linear fits to the observed fluxes of the high-$J$ (E$_{\rm up}$~$>$~1700~K) and  mid-$J$ (550~K $\leq$ E$_{\rm up}$ $\leq$ 1700~K) transitions, respectively. The green lines are fitted to the mid-$J$ fluxes after subtracting the contribution of the ``Hot" component from the total fluxes. Dotted lines represent the sum of the red and green lines.
The open black circles represent the best-fit model to the corrected mid-$J$ CO fluxes, and 
 the purple line represents the linear-fit of the best-fit model fluxes. The rotational temperature $T_{\rm rot}$ derived from  each color line and the source type (see text) are presented  in  the upper right of the box.}\label{fig:rd1}
\end{figure*}

\begin{figure*}
\includegraphics[width=0.30 \textwidth]{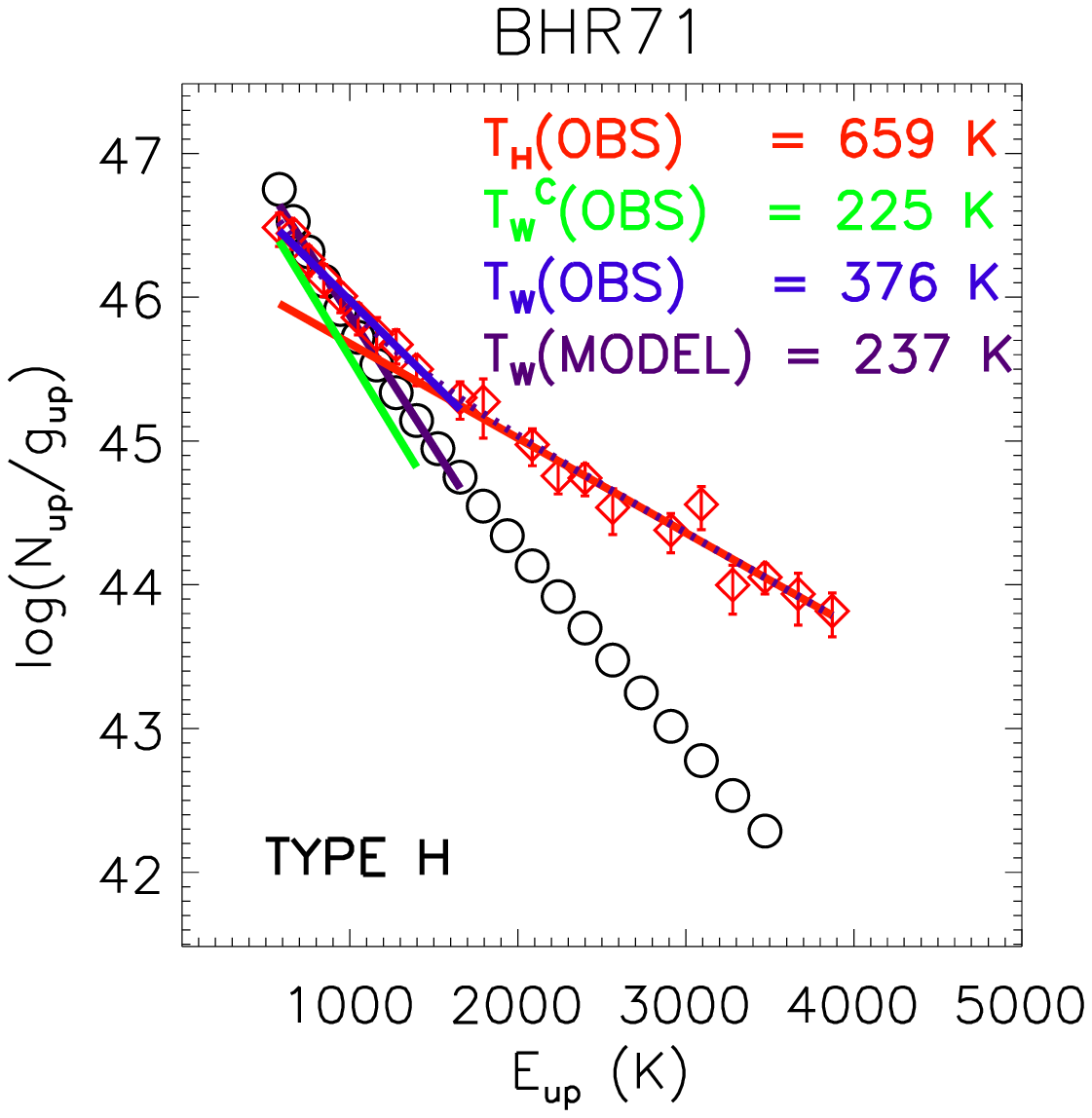}
\includegraphics[width=0.30 \textwidth]{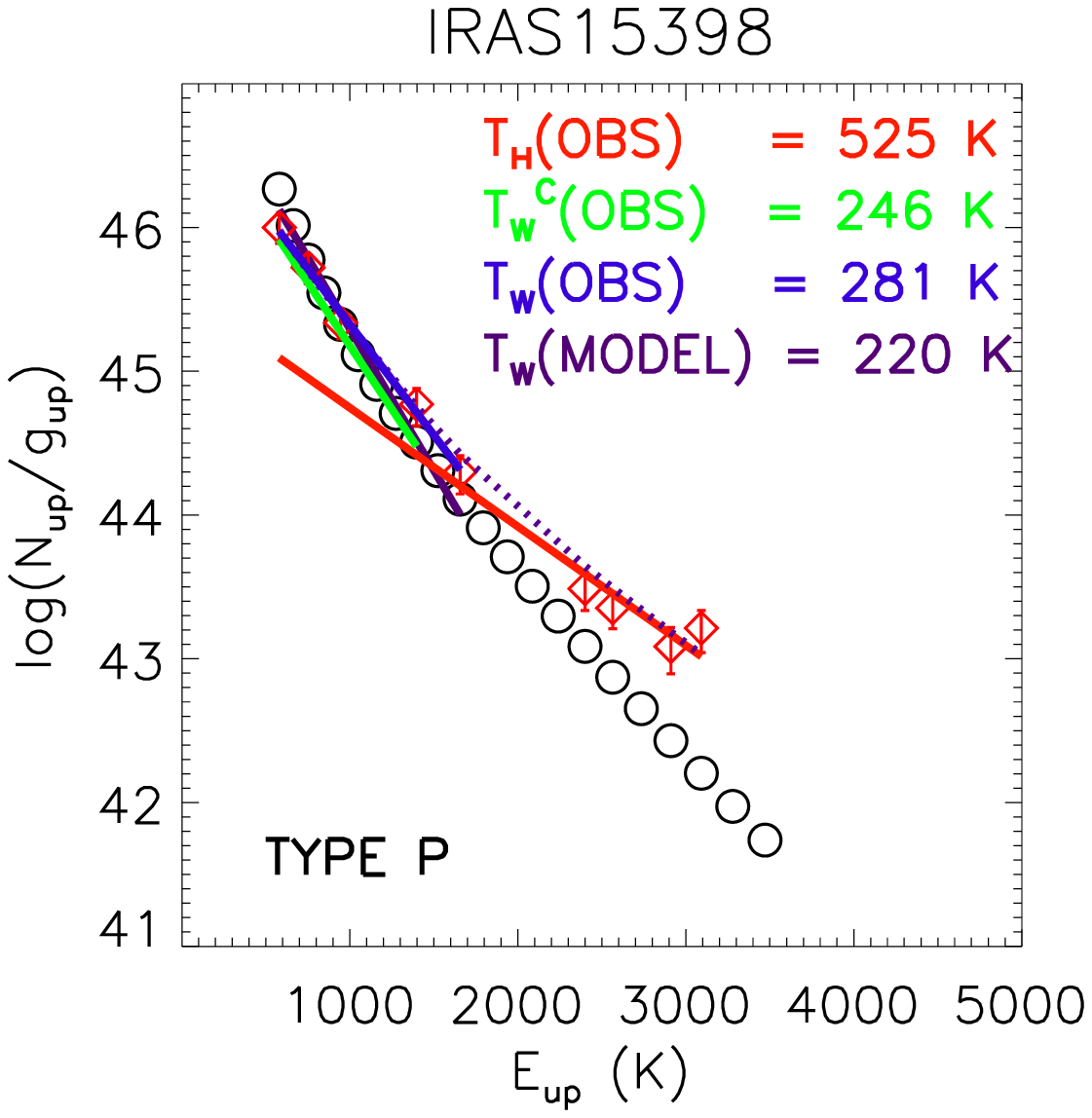}
\includegraphics[width=0.30 \textwidth]{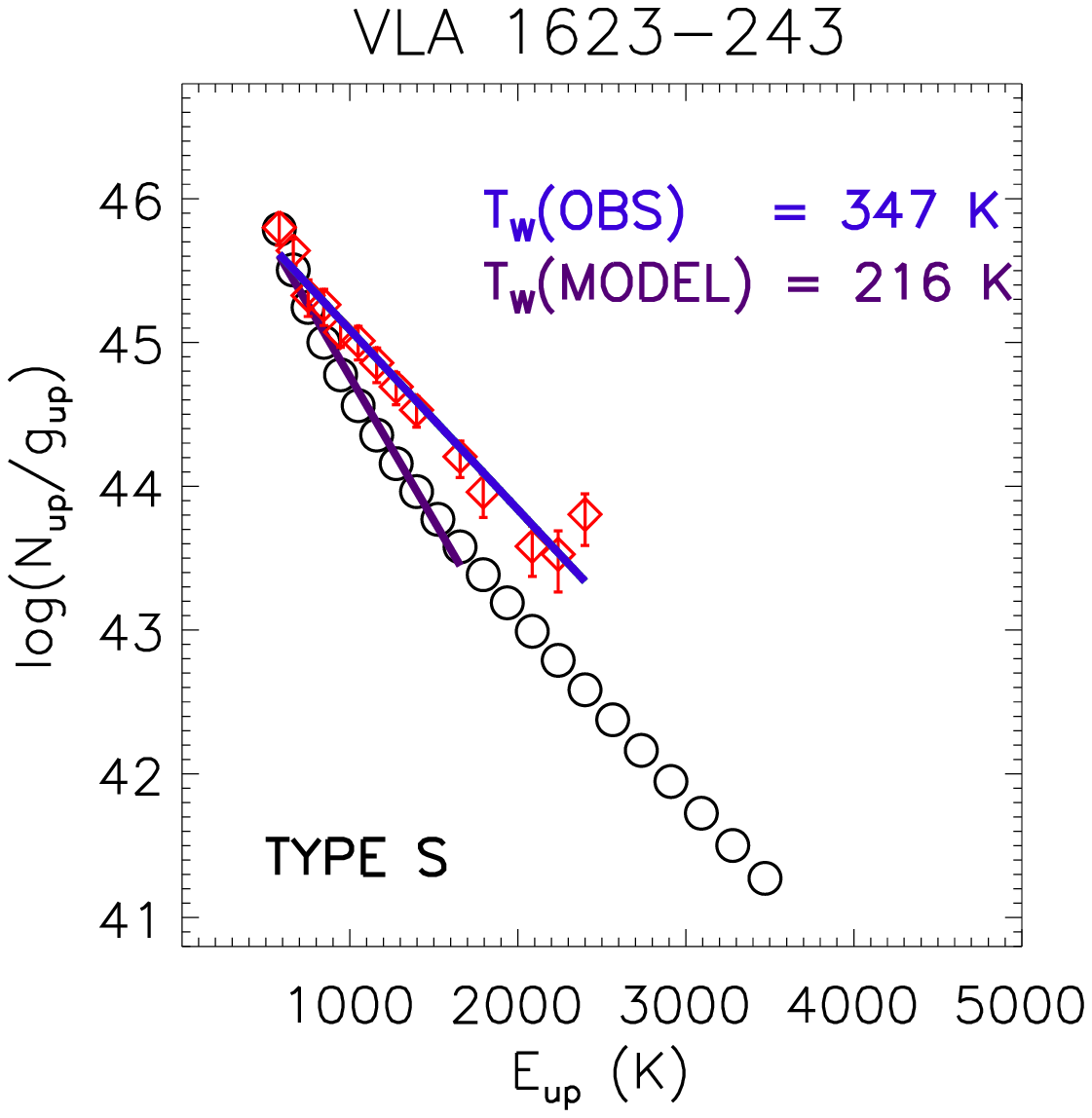}

\includegraphics[width=0.30 \textwidth]{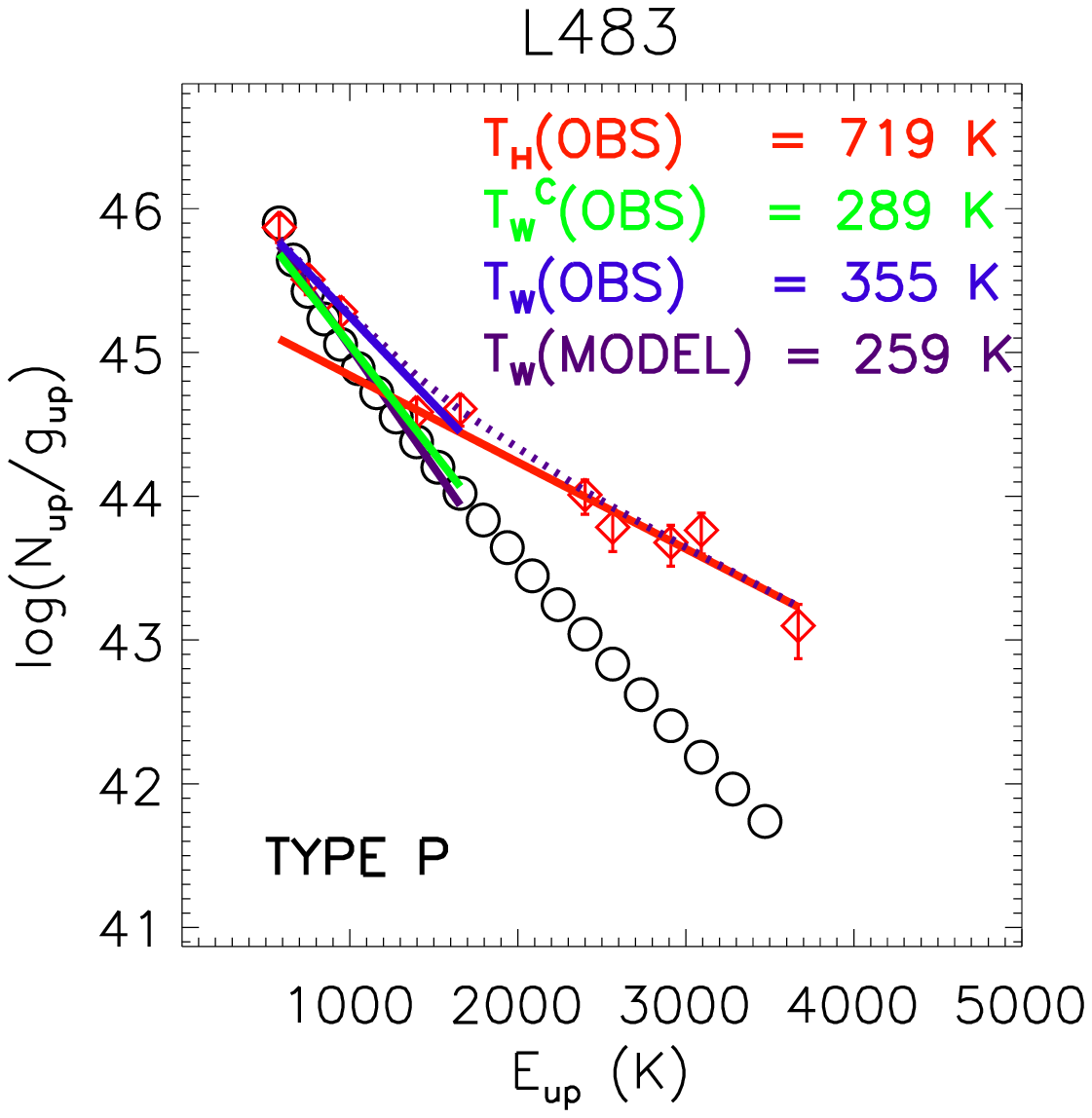}
\includegraphics[width=0.30 \textwidth]{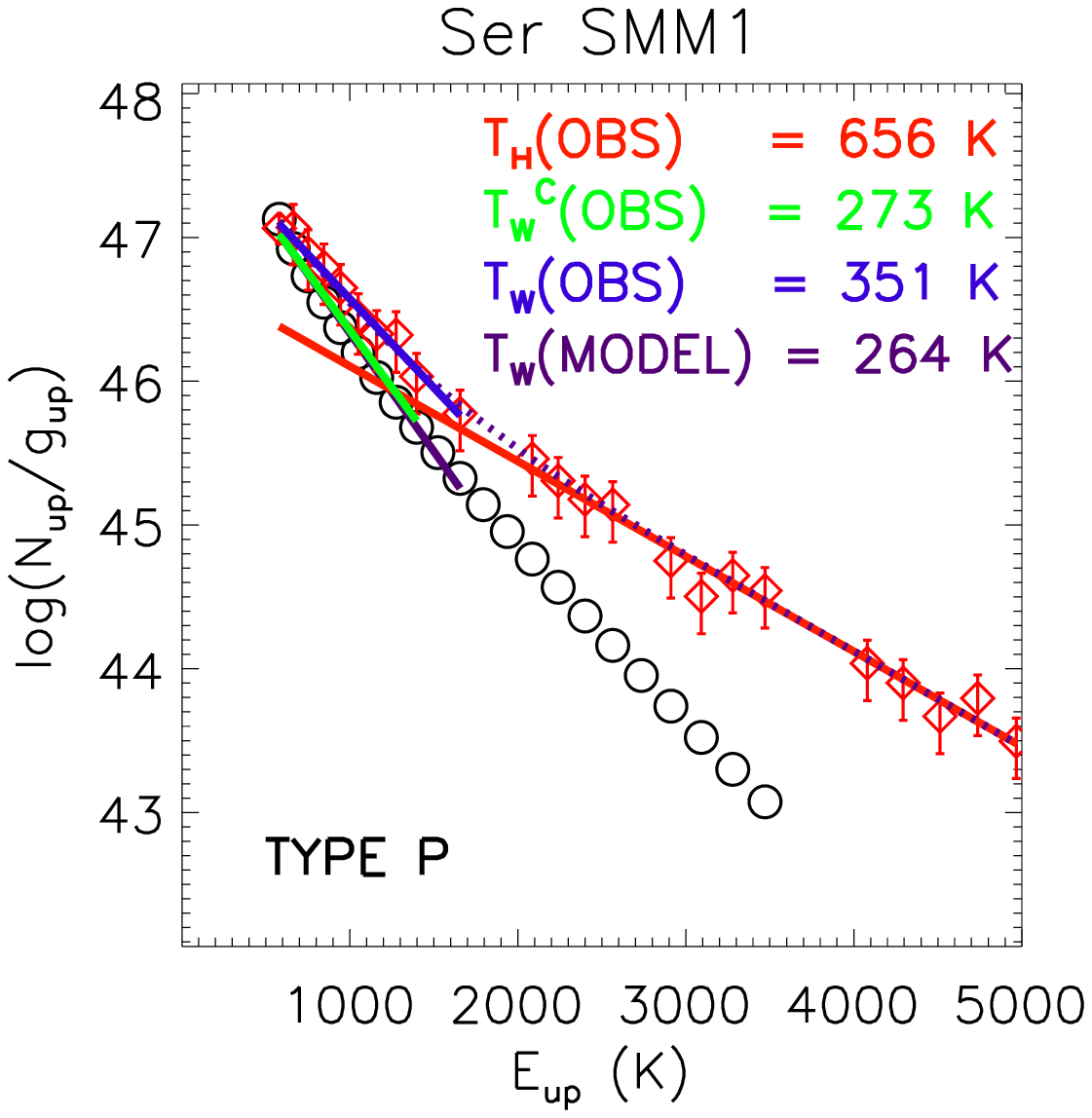}
\includegraphics[width=0.30 \textwidth]{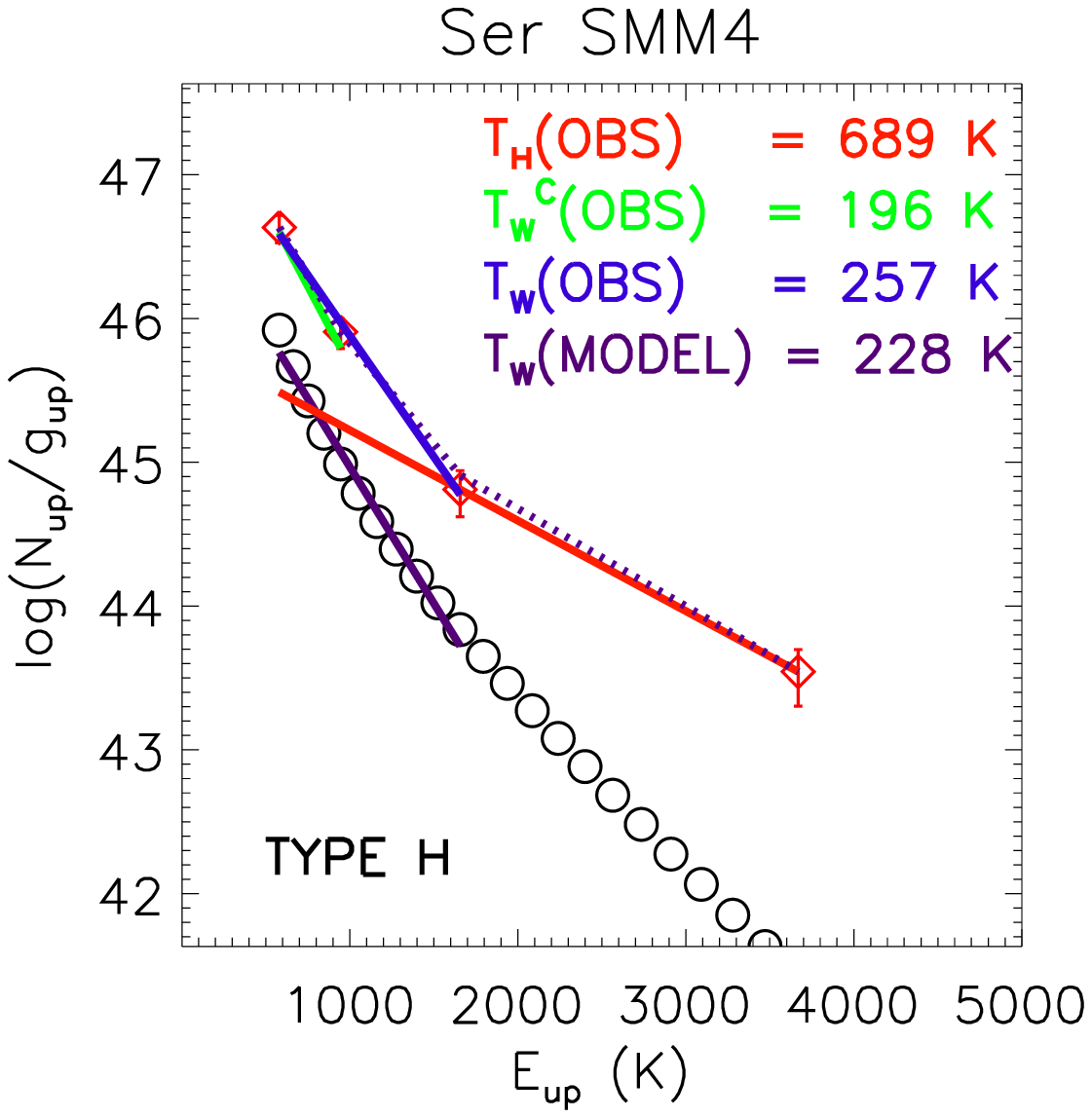}

\includegraphics[width=0.30 \textwidth]{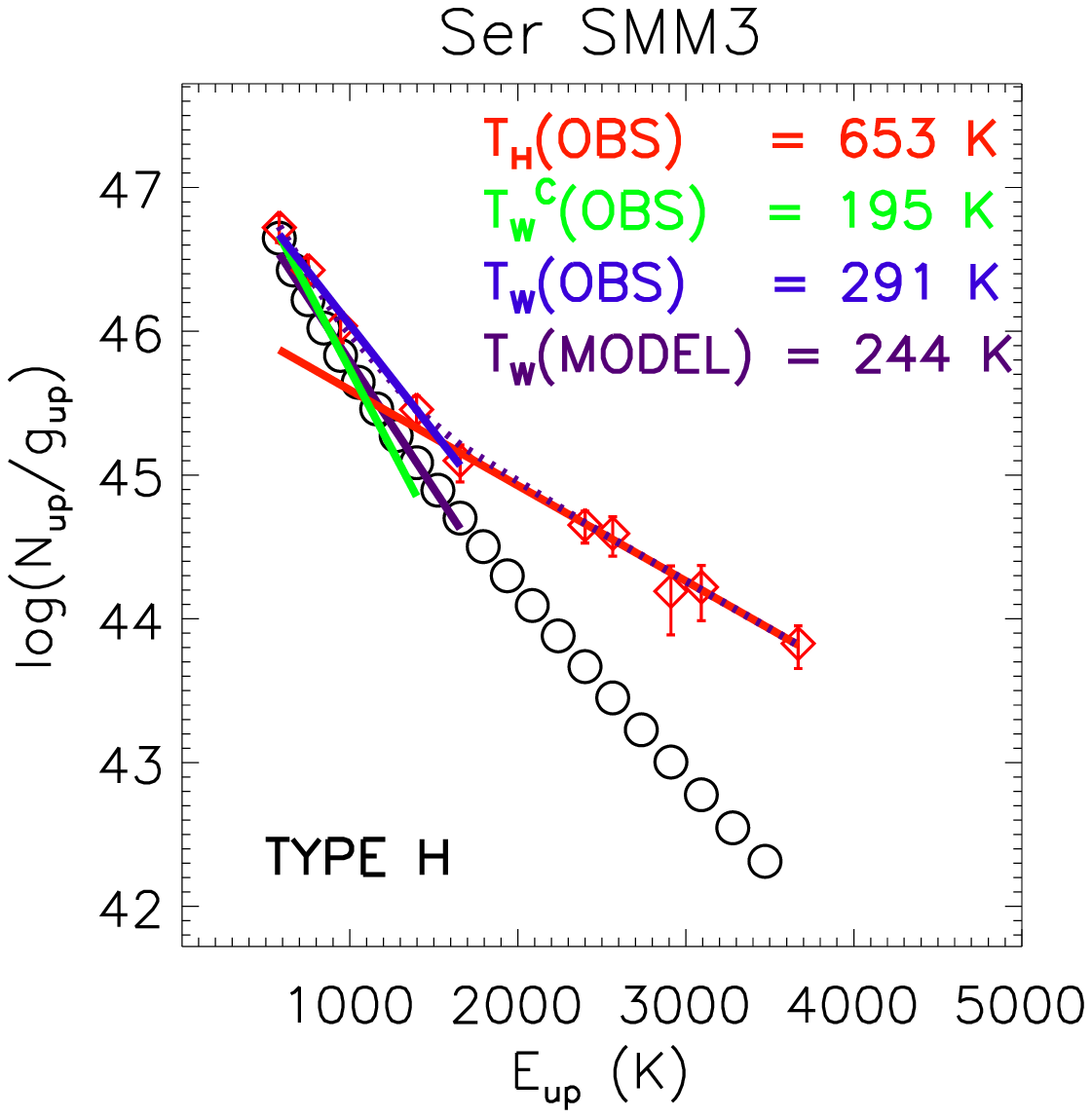}
\includegraphics[width=0.30 \textwidth]{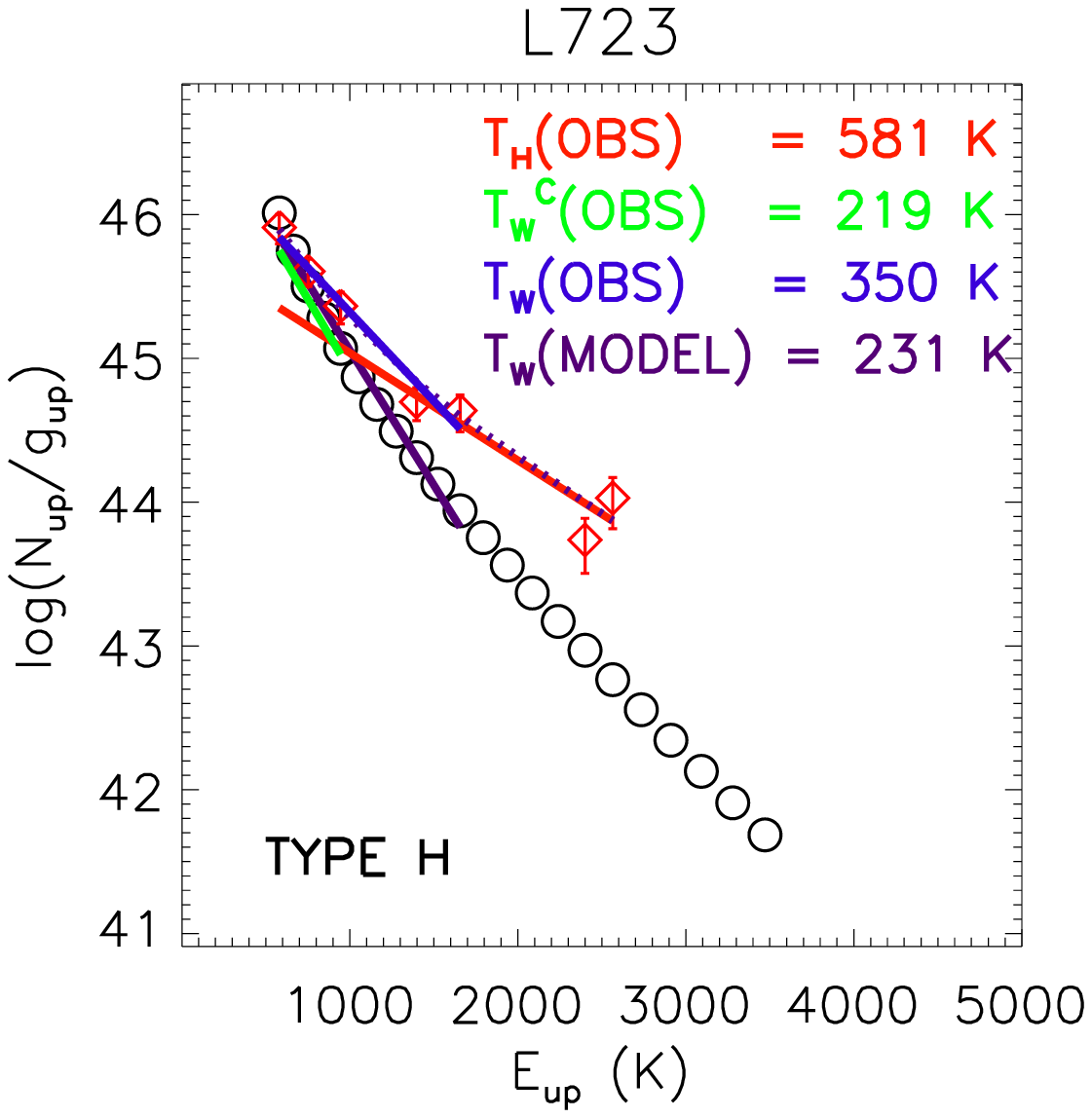}
\includegraphics[width=0.30 \textwidth]{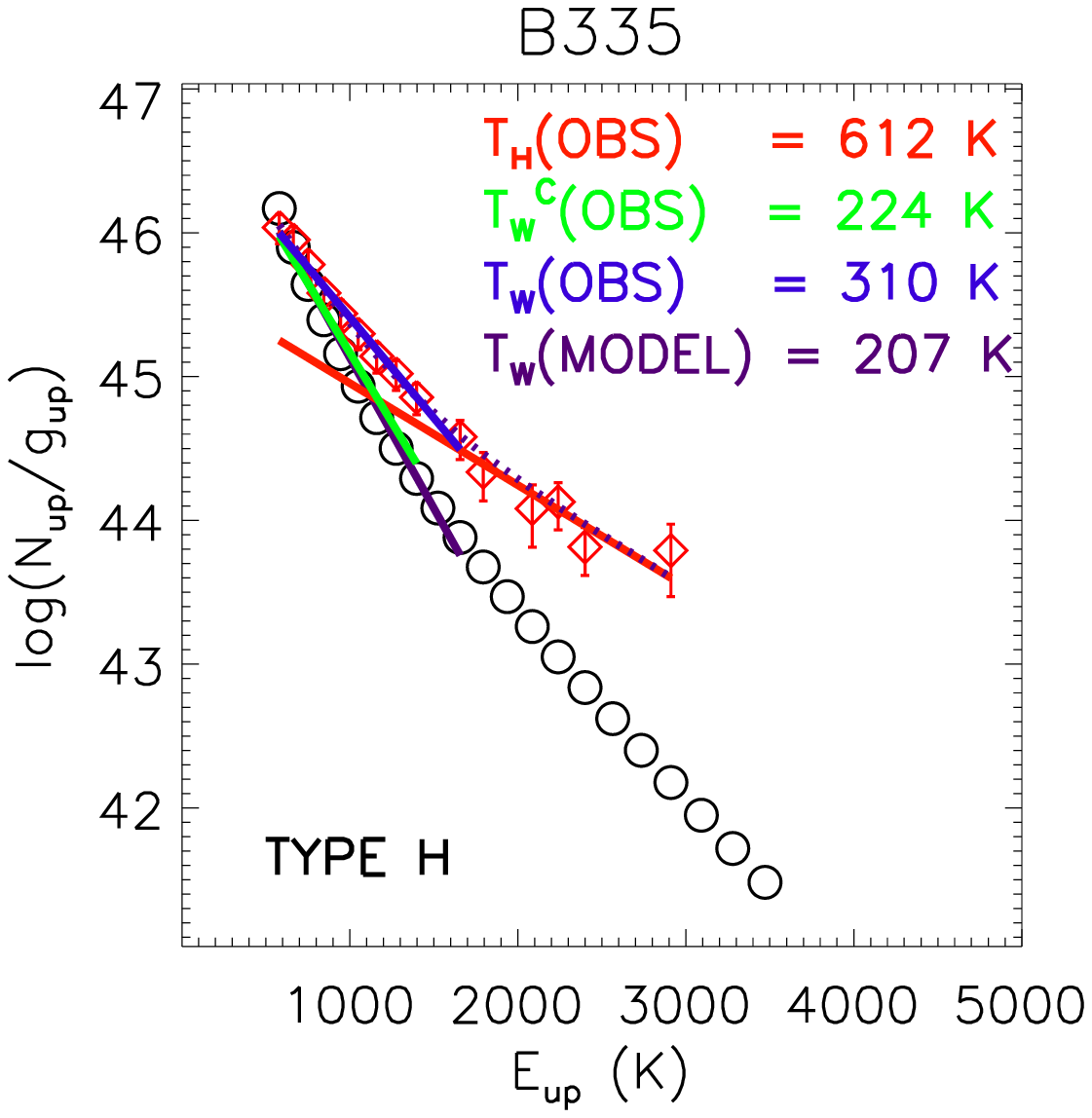}

\caption{The same as Figure~\ref{fig:rd1} except for  BHR71, IRAS 15398, VLA 1623-243, L483, Ser SMM1, Ser SMM4, Ser SMM3, L723, and B335. }\label{fig:rd2}
\end{figure*}

\begin{figure*}
\includegraphics[width=0.30 \textwidth]{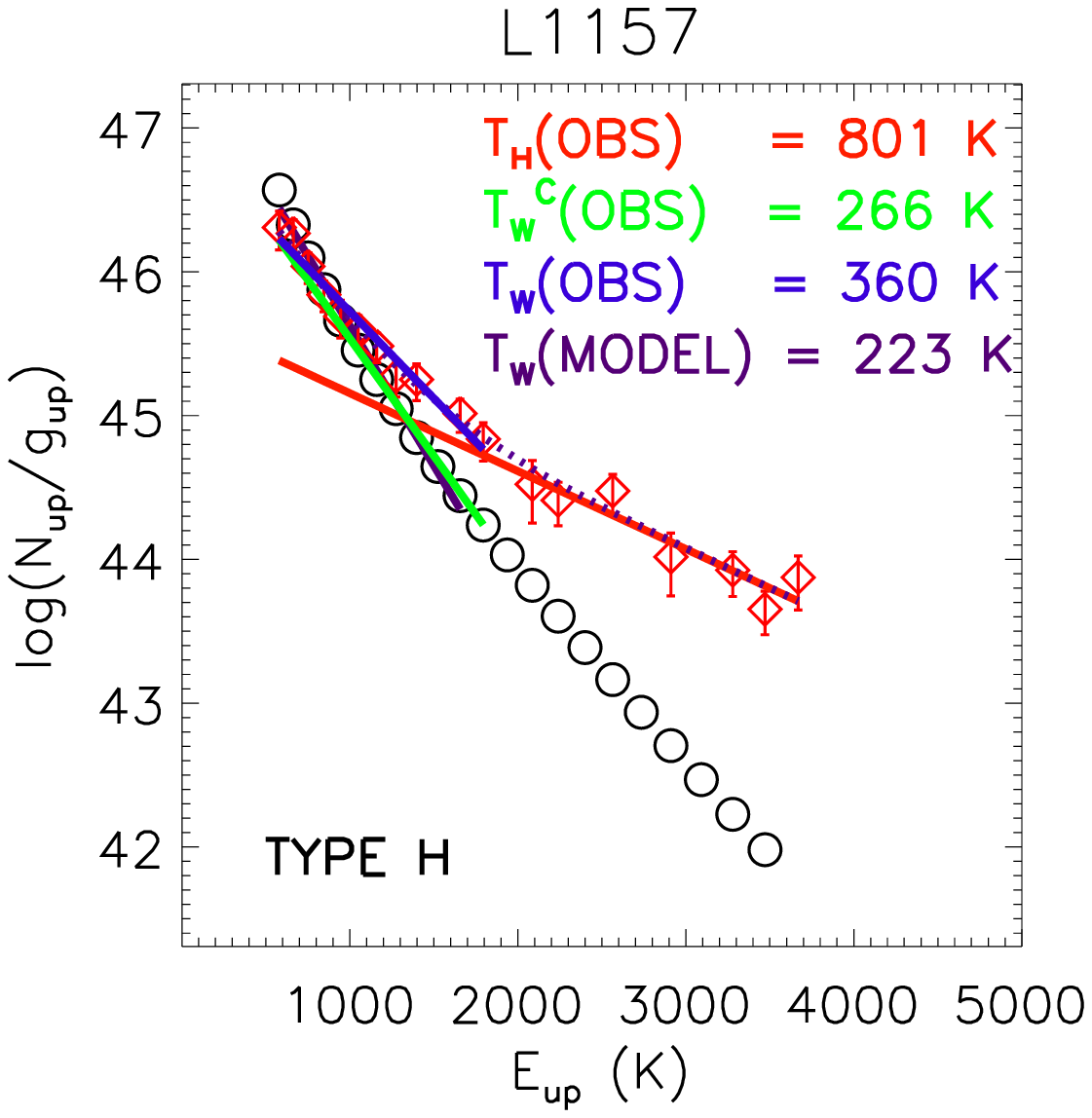}
\includegraphics[width=0.30 \textwidth]{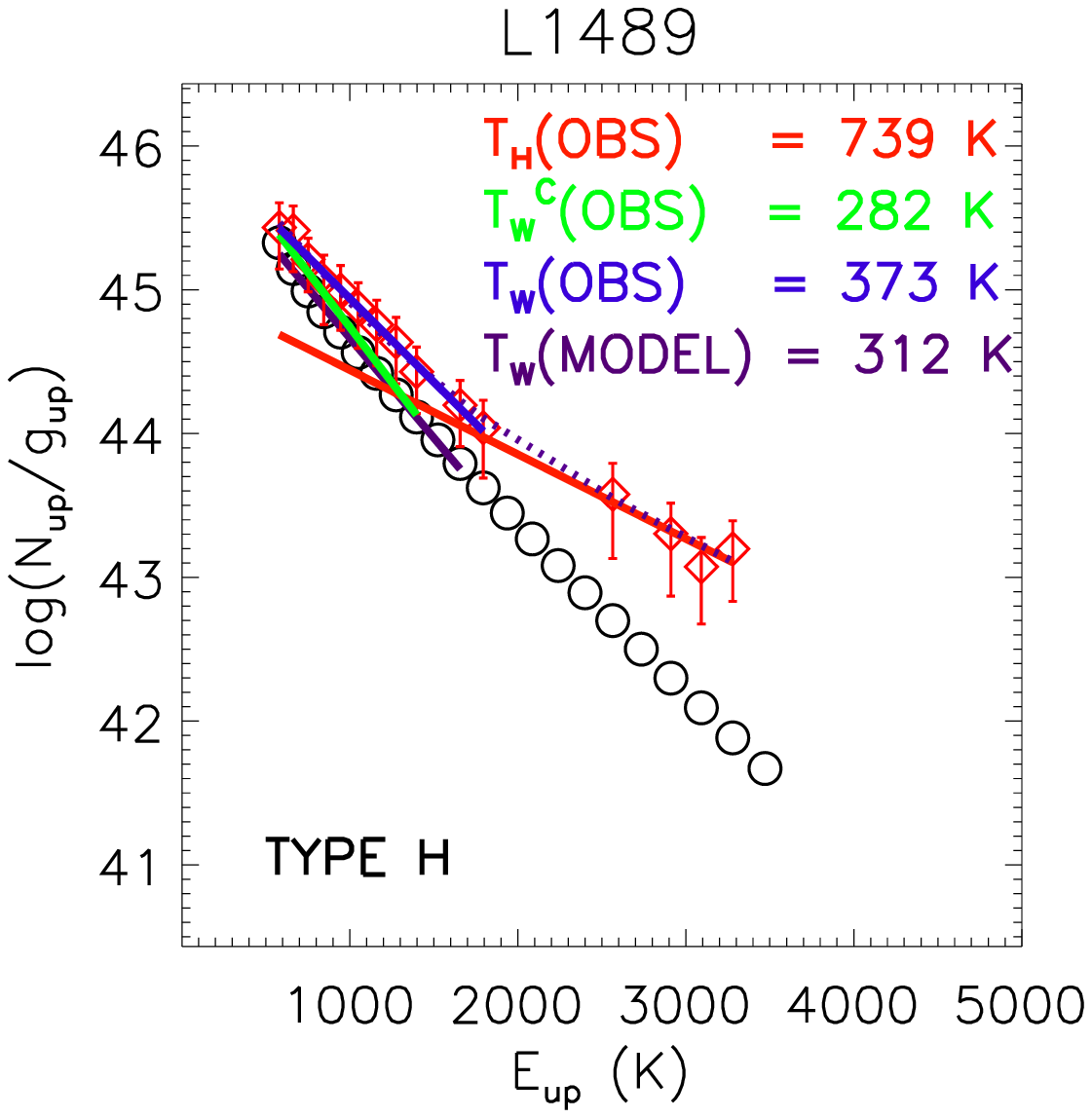}
\includegraphics[width=0.30 \textwidth]{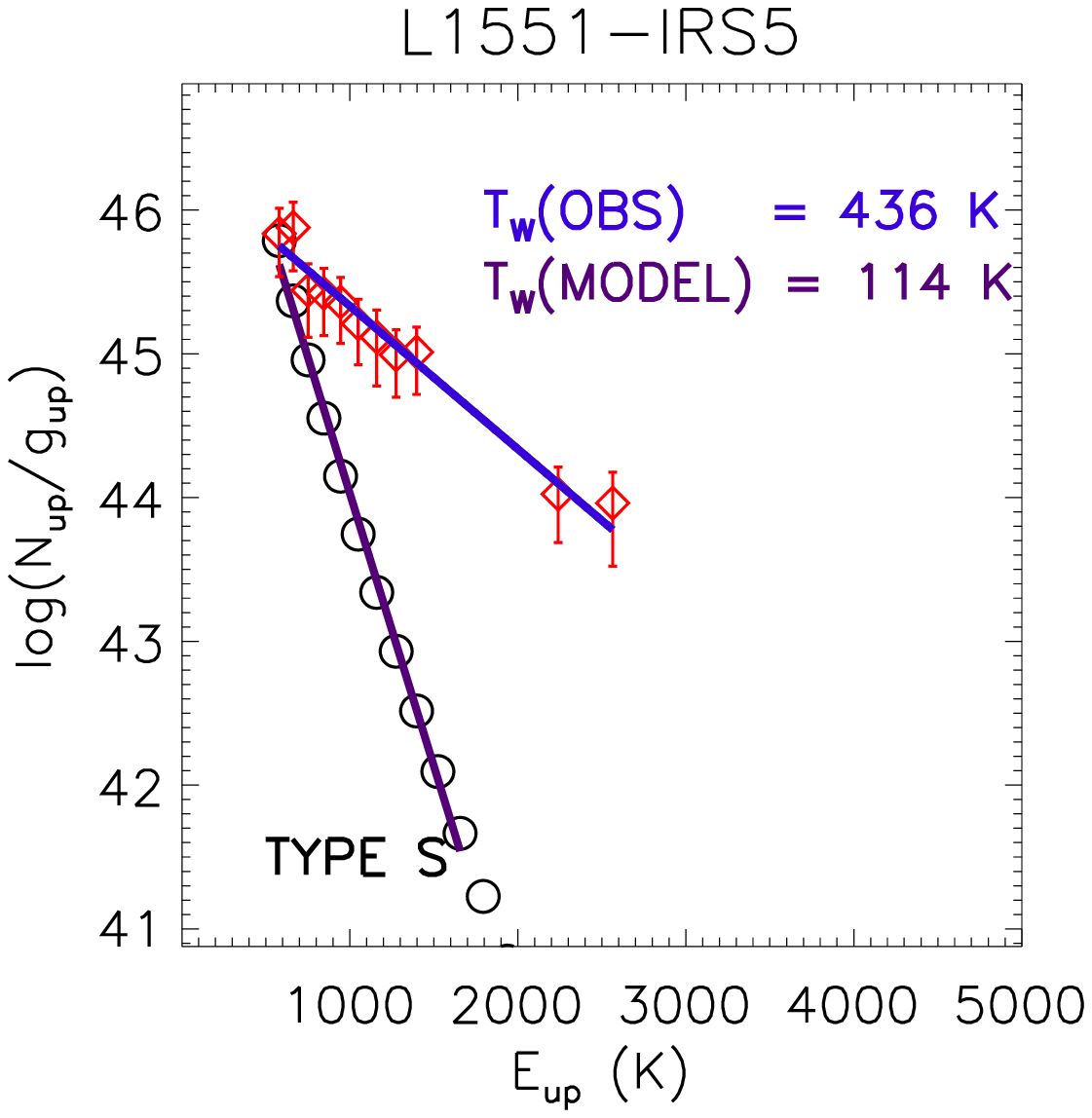}

\includegraphics[width=0.30 \textwidth]{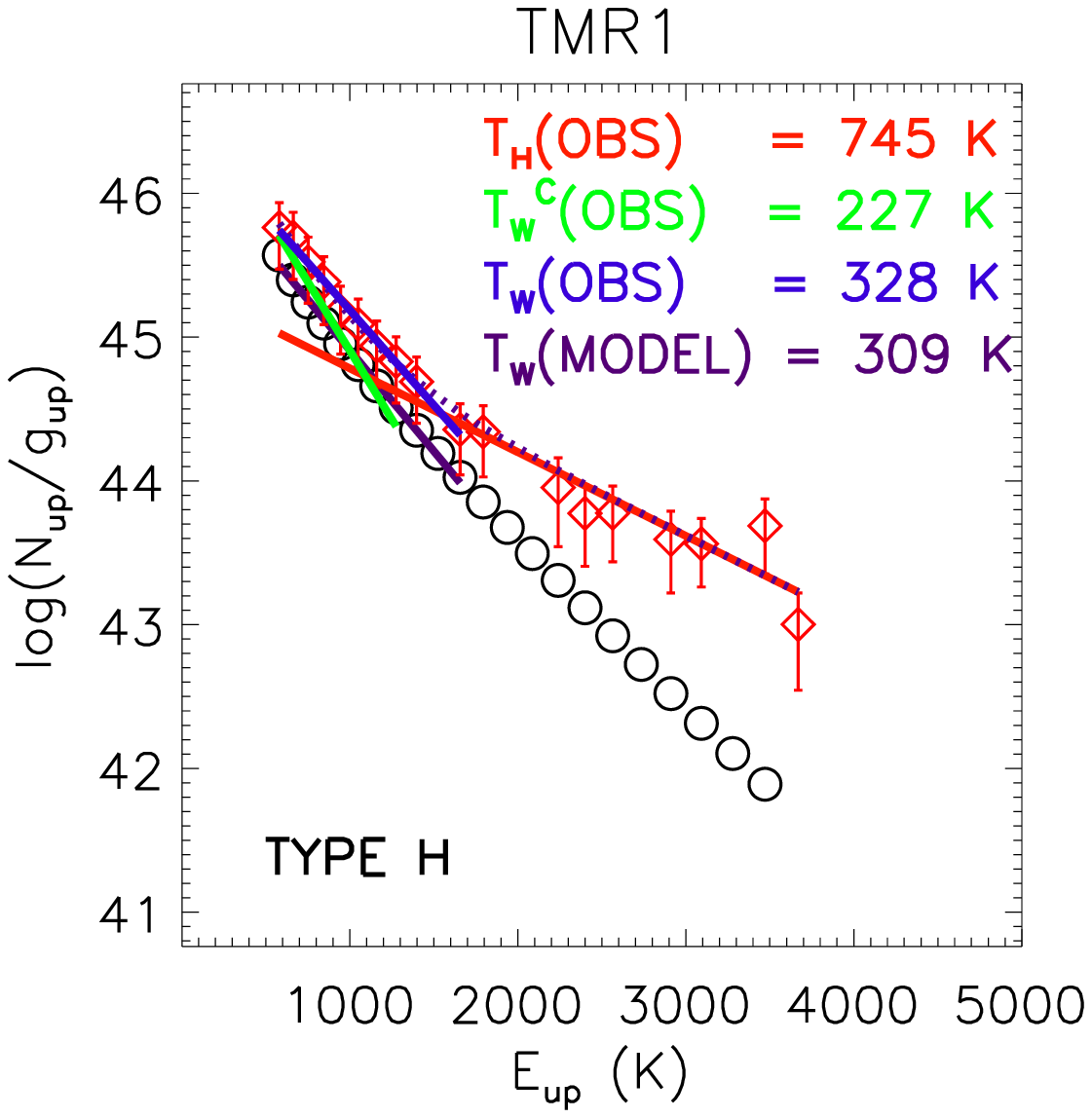}
\includegraphics[width=0.30 \textwidth]{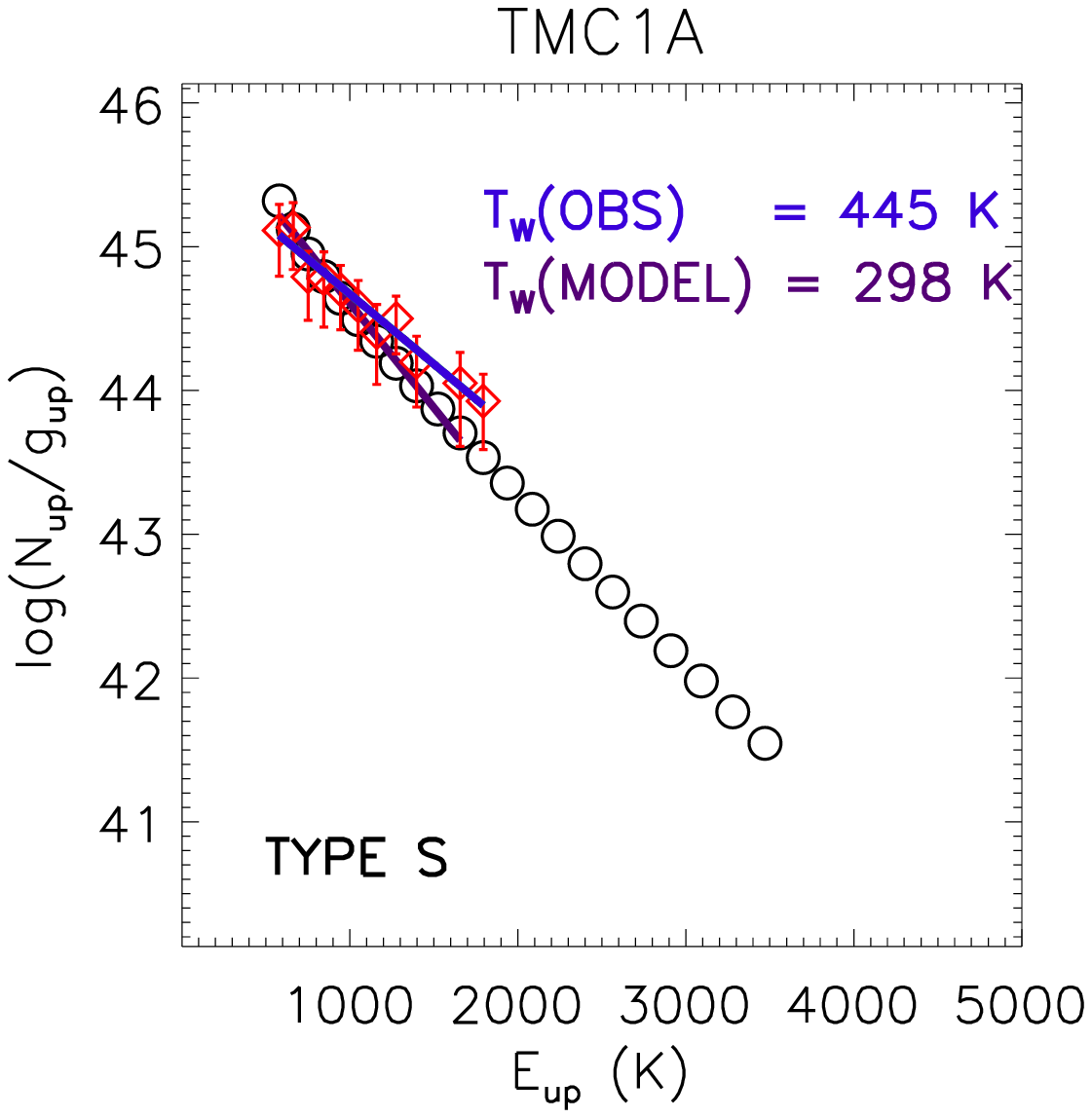}
\includegraphics[width=0.30 \textwidth]{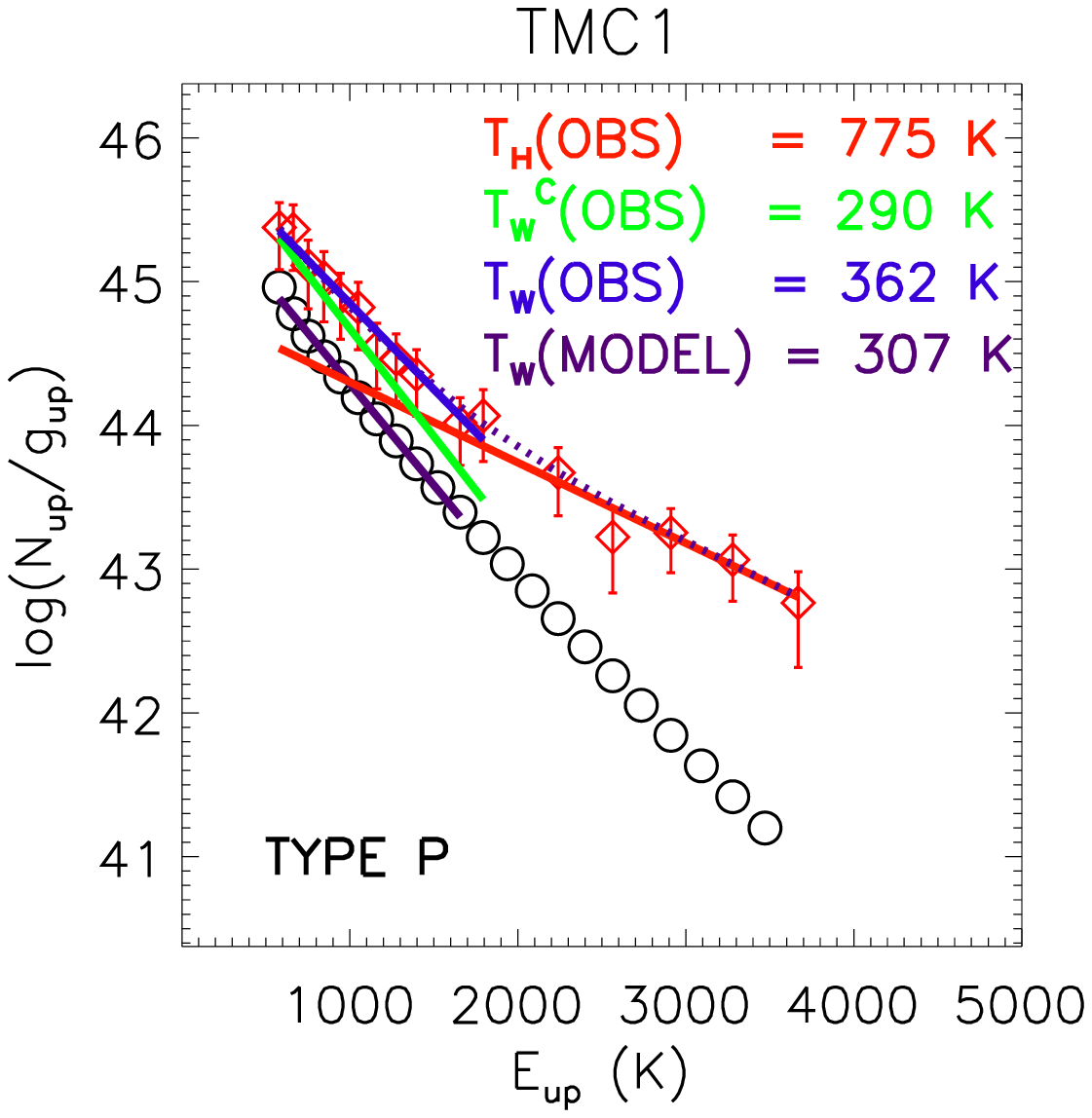}

\includegraphics[width=0.30 \textwidth]{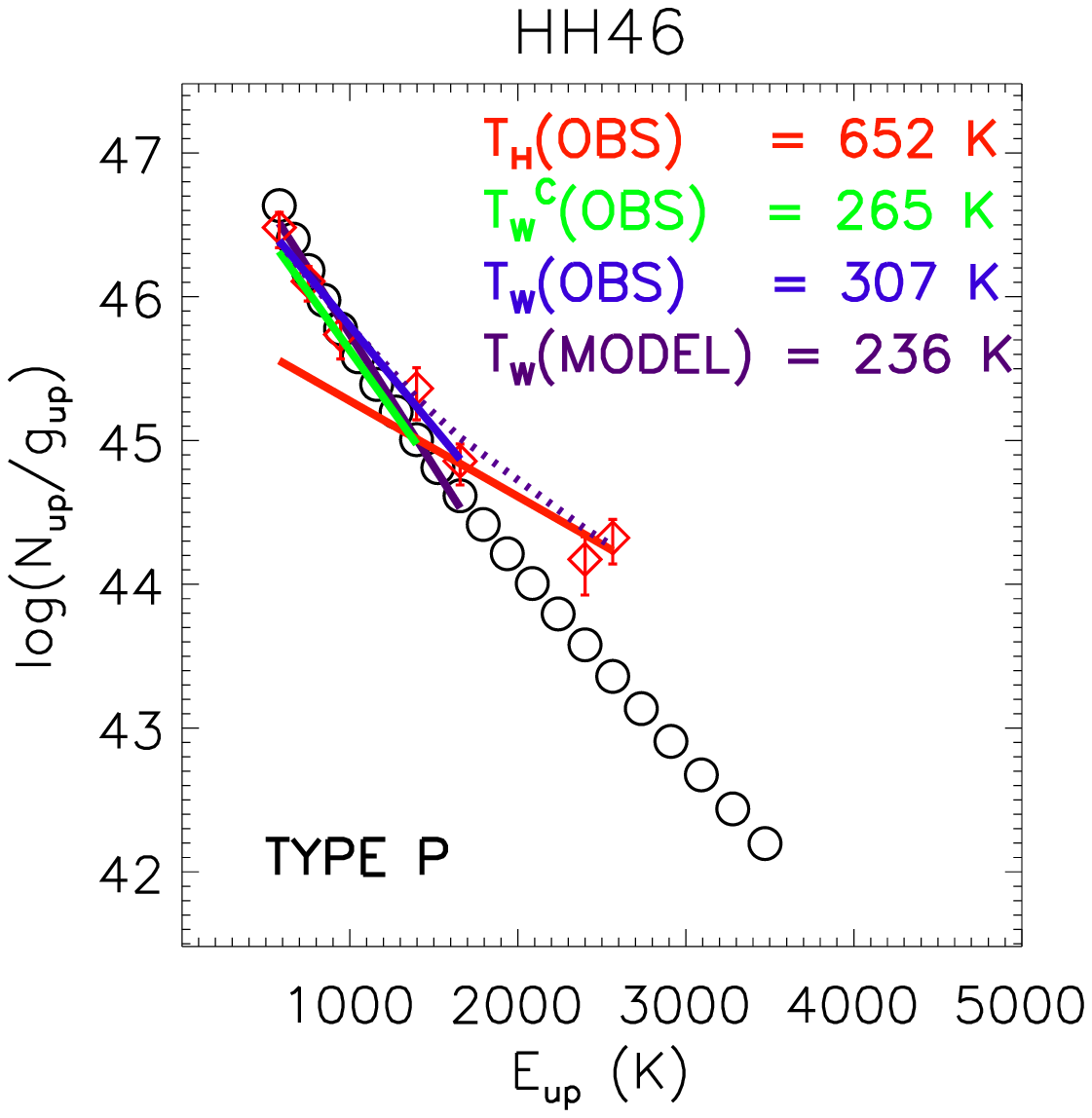}
\includegraphics[width=0.30 \textwidth]{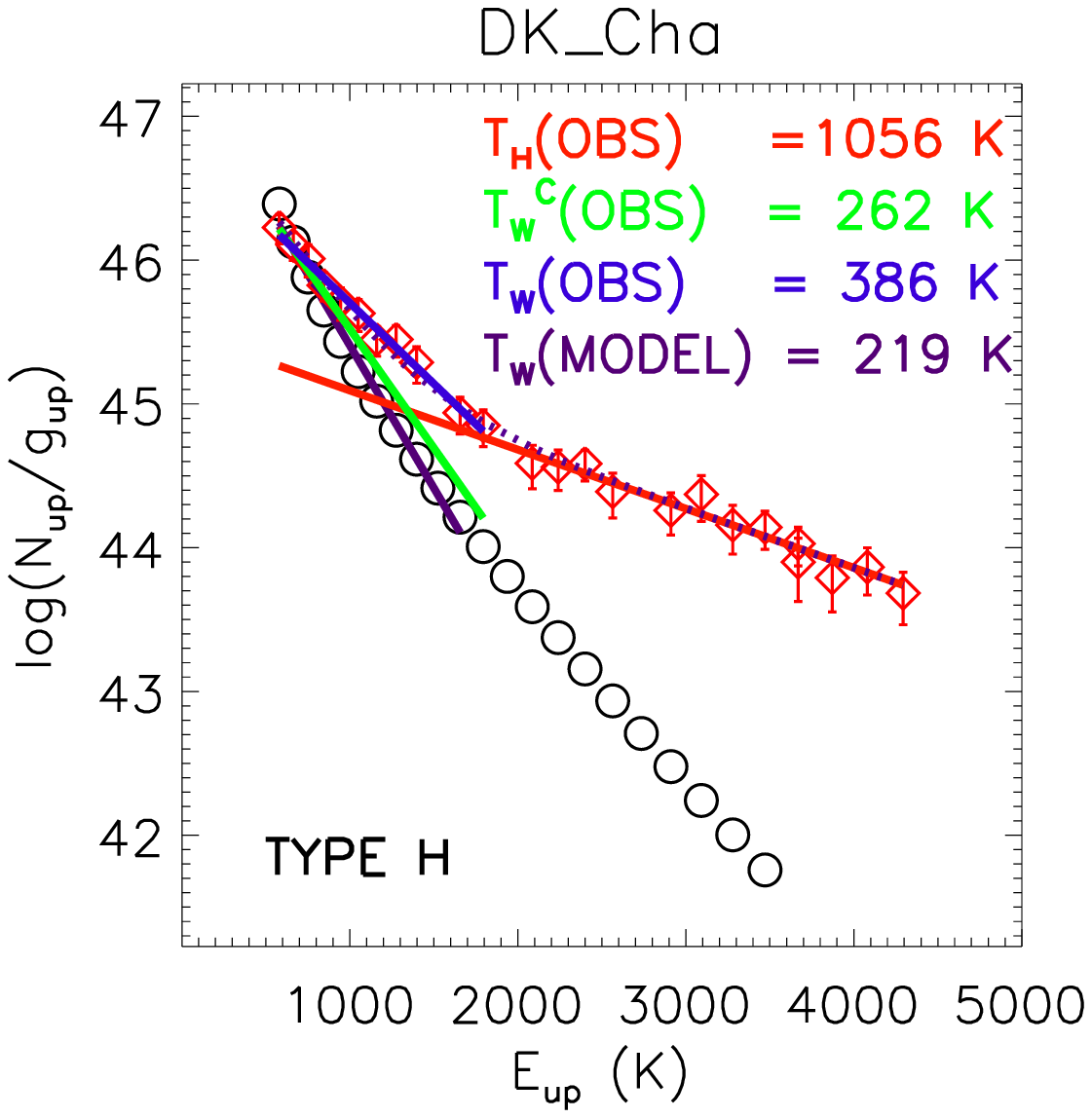}
\includegraphics[width=0.30 \textwidth]{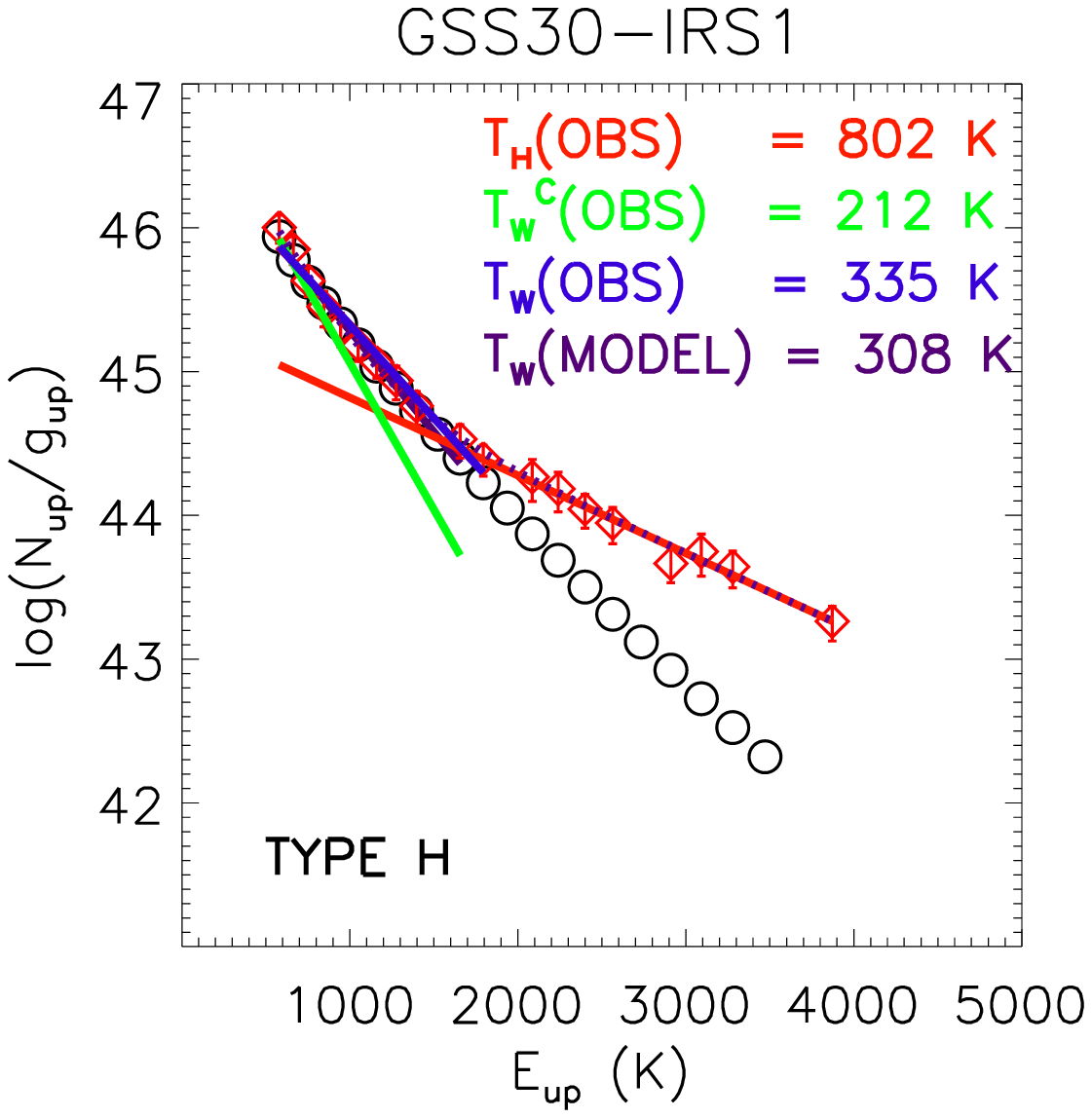}

\caption{The same as Figure~\ref{fig:rd1} except for L1157, L1489, L1551-IRS5, TMR1, TMC1A, TMC1, HH46, DK Cha, and GSS30-IRS1}\label{fig:rd3}
\end{figure*}

\begin{figure*}
\includegraphics[width=0.30 \textwidth]{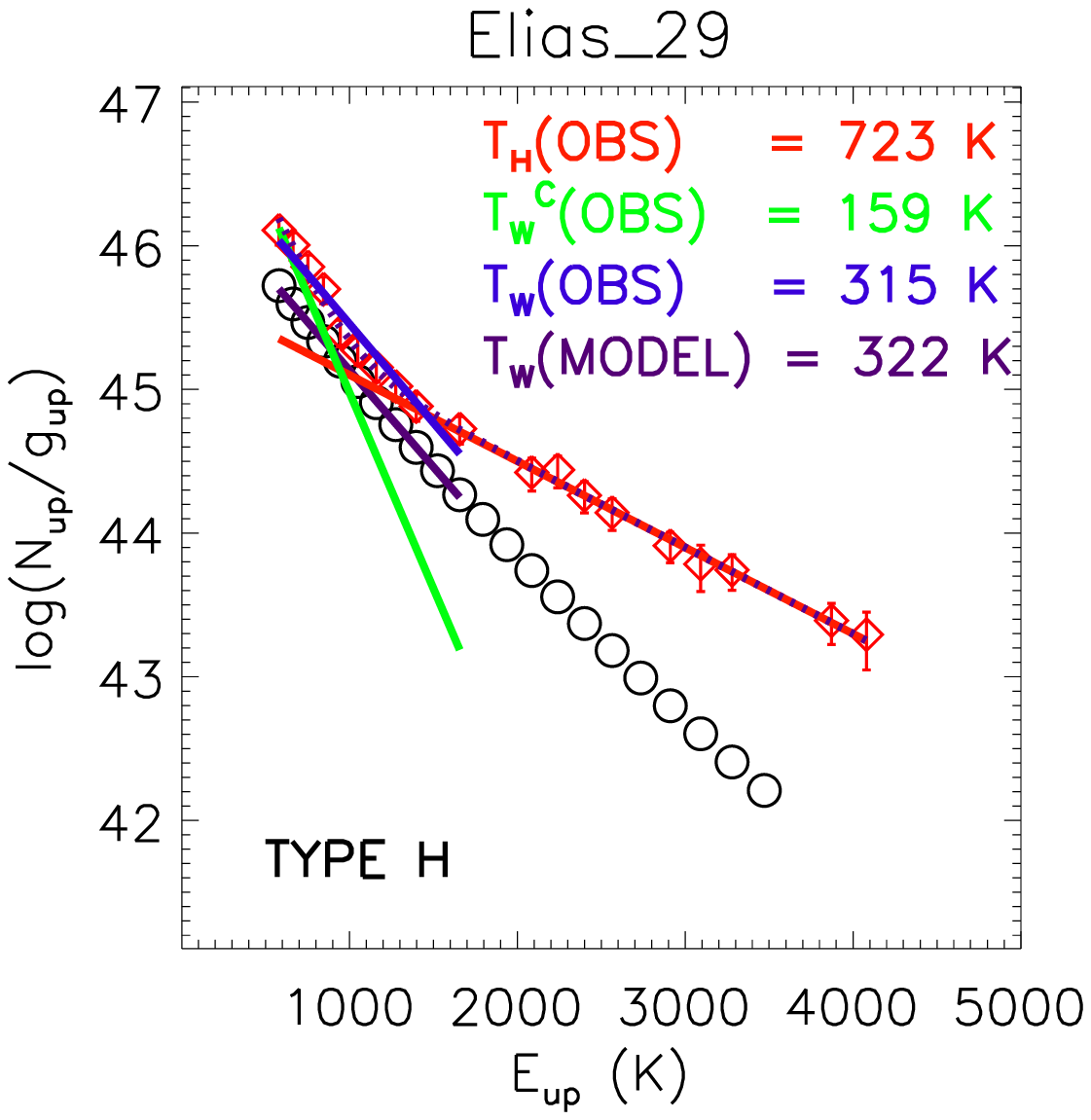}
\includegraphics[width=0.30 \textwidth]{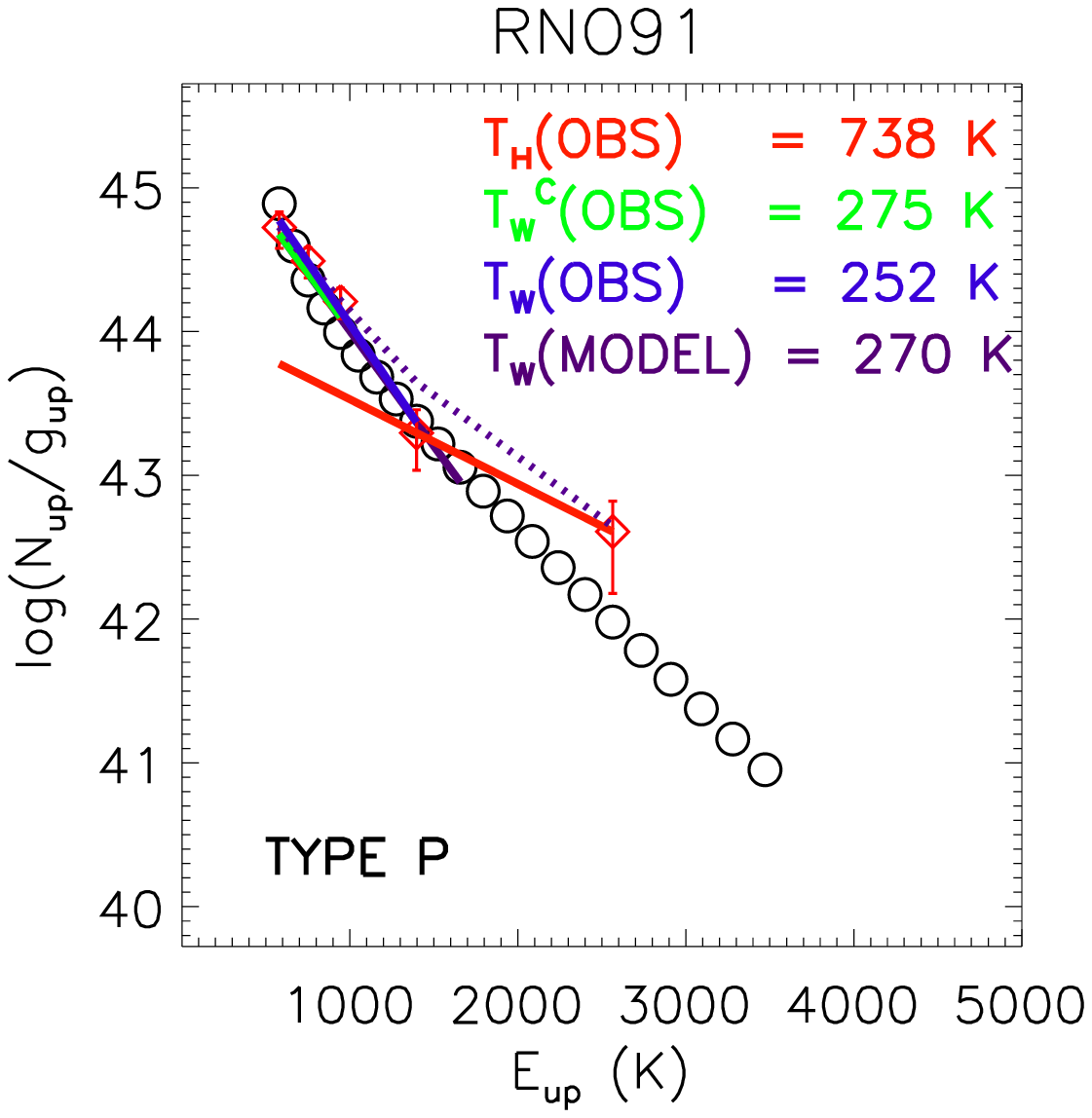}

\includegraphics[width=0.30 \textwidth]{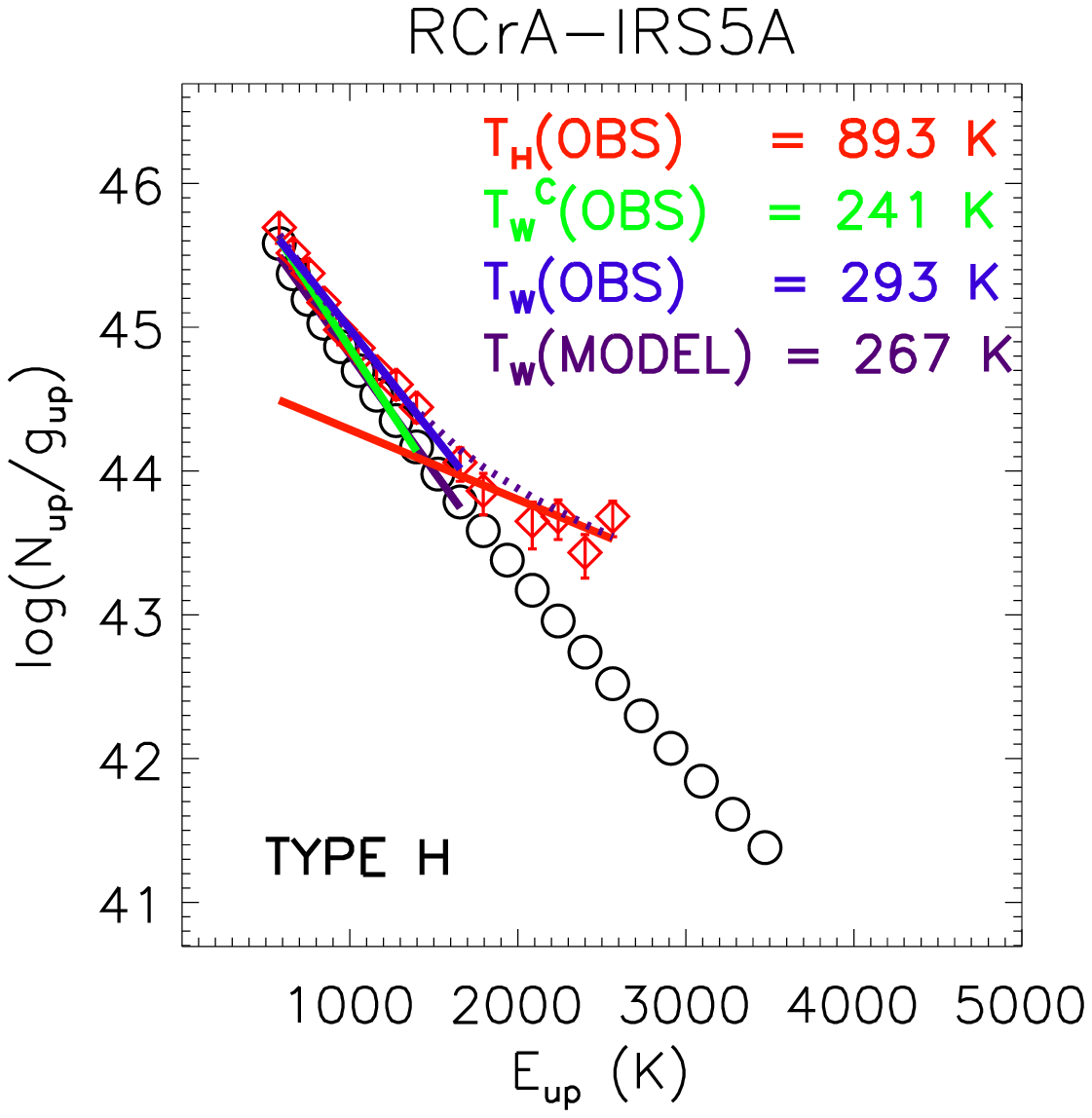}
\includegraphics[width=0.30 \textwidth]{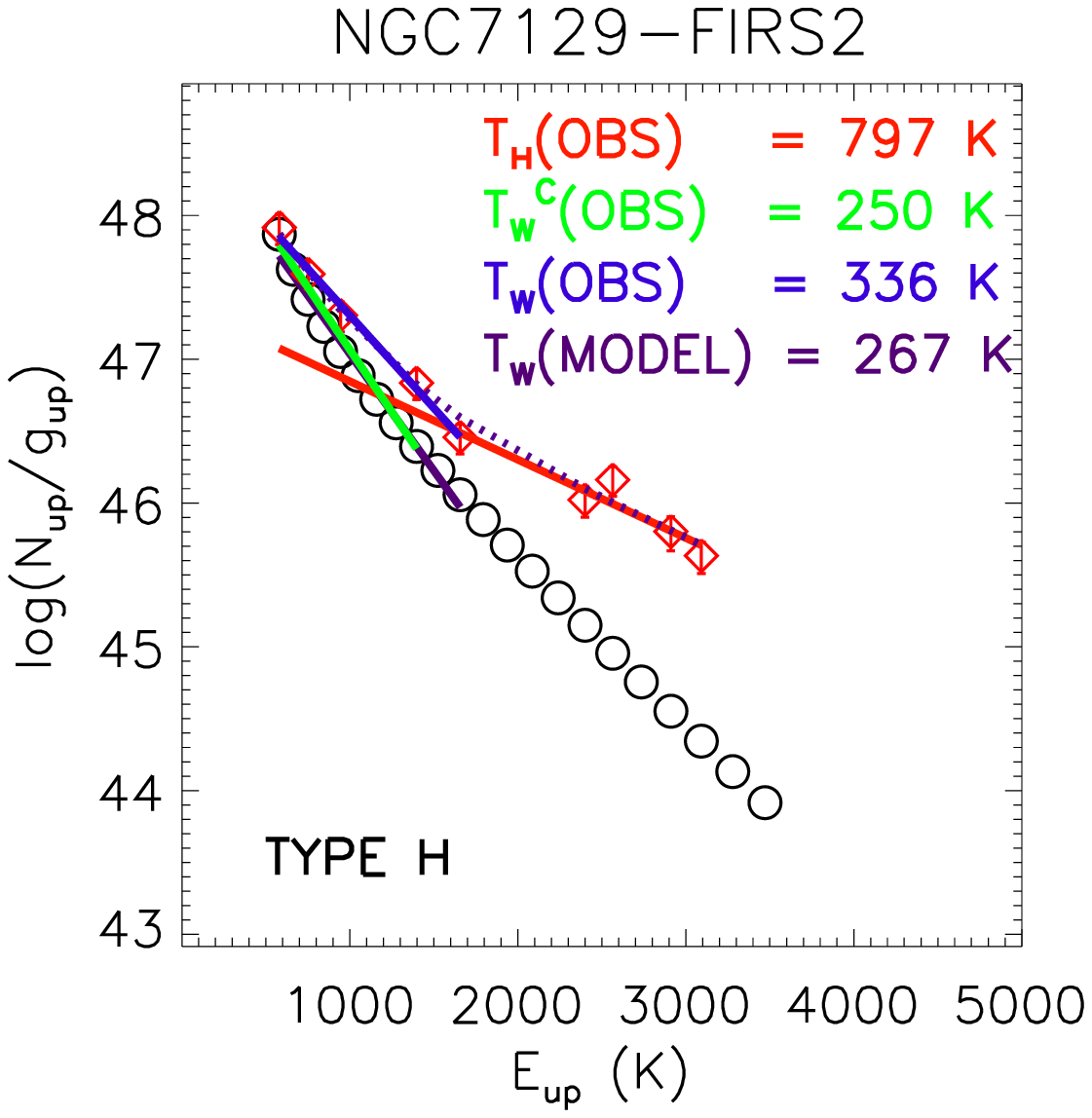}
\caption{The same as Figure \ref{fig:rd1} except for Elias 29, RNO91, RCrA-IRS5, and NGC7129-FIRS2}\label{fig:rd4}
\end{figure*}

 Rotational diagrams for our sources are plotted in Figures~\ref{fig:rd1}--\ref{fig:rd4}.
The high-$J$ ($J>$~24) CO transitions, which produce a high rotational temperature of $\sim$700-900~K, were detected toward most of our sources except NGC1333-IRAS2A and TMC1A. This hot component is generally interpreted as emitting from shocked gas as opposed to the UV heated cavity walls \citep[][Paper I]{Visser2012}. To determine the potential emission from the PDR  we must first remove the contribution of the hot component from the mid-$J$ CO emission.
 
 We calculate two rotational temperatures from the observed mid-$J$ CO lines. The observed rotational temperature of the Warm component, $T_{\rm W}(\mathrm{OBS})$ is linear-fitted from the total observed fluxes, while the Corrected rotational temperature of the Warm component, $T_{\rm W}^{\mathrm{C}}(\mathrm{OBS})$ is derived after subtracting the contribution by the hot component from the total mid-$J$ CO fluxes. For this subtraction, we calculate the mid-$J$ fluxes emerging from the hot component using the rotational temperature  of the Hot component derived from the observed high-$J$ CO fluxes at $J >$ 24 ($T_{\rm H}(\mathrm{OBS})$). 
 $T_{\rm H}(\mathrm{OBS})$, $T_{\rm W}^{\mathrm{C}}(\mathrm{OBS})$, and $T_{\rm W}(\mathrm{OBS})$ for each source are listed in Table~\ref{tb:result} and  plotted as red, green, and blue color lines, respectively, in Figures~\ref{fig:rd1}--\ref{fig:rd4}.

 We classify sources as TYPE H, P, and S. 
 TYPE H sources, which represent half of our sample, are contaminated significantly by the ``HOT" component, so $T_{\rm W}^{\mathrm{C}}(\mathrm{OBS}) < T_{\rm W}(\mathrm{OBS})- 3 \sigma_{\rm W}(\mathrm{OBS})$ where $\sigma_{\rm W}(\mathrm{OBS})$ is a linear fit error of $T_{\rm W}(\mathrm{OBS})$. 
 TYPE P (``PURE") sources, which cover a quarter of our sources, are not contaminated by the hot component, so $T_{\rm W}^{\mathrm{C}}(\mathrm{OBS}) > T_{\rm W}(\mathrm{OBS})- 3 \sigma_{\rm W}(\mathrm{OBS})$.
 Finally, TYPE~S are  best fit with  a ``SINGLE" temperature regardless whether it is hot or warm.
 If the UV heated outflow cavity wall reproduces $T_{\rm W}^{\mathrm{C}}(\mathrm{OBS})$ as well as the corrected fluxes for the TYPE H sources, the hot component is important to produce the universal $T_{\rm rot}$ of 350~K. 
 However, for the TYPE~P sources, the UV heated outflow cavity walls can be tested directly for the universal 350~K rotational temperature. 
 The type of each source is described in Table~\ref{tb:result} and inside each panel in Figs.~\ref{fig:rd1}--\ref{fig:rd4}. 
 This classification is not correlated with $L_{\rm bol}$, 
$n_{\rm 1000 AU}$, or evolutionary stage as shown in Figure~\ref{fig:ep_source_type}.

\begin{figure*}
\includegraphics[width=0.50 \textwidth]{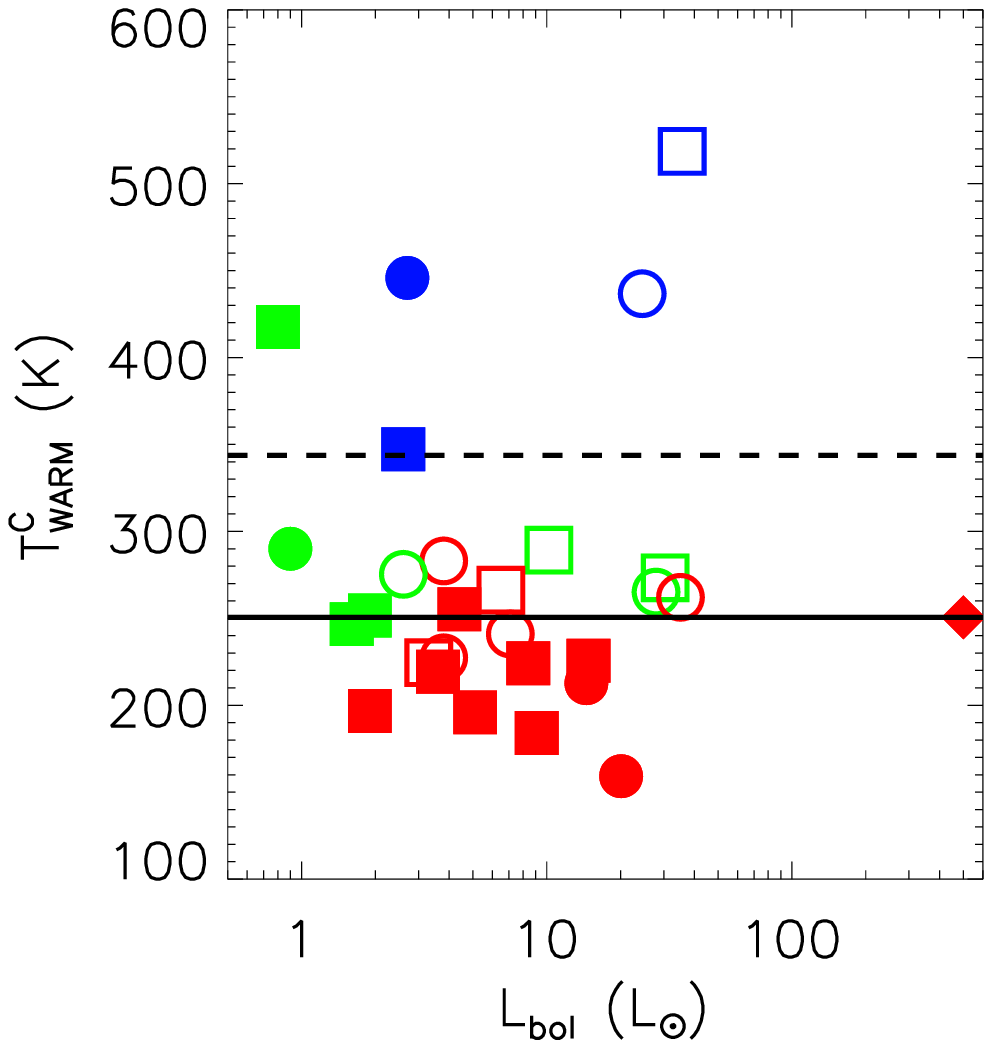}
\includegraphics[width=0.50 \textwidth]{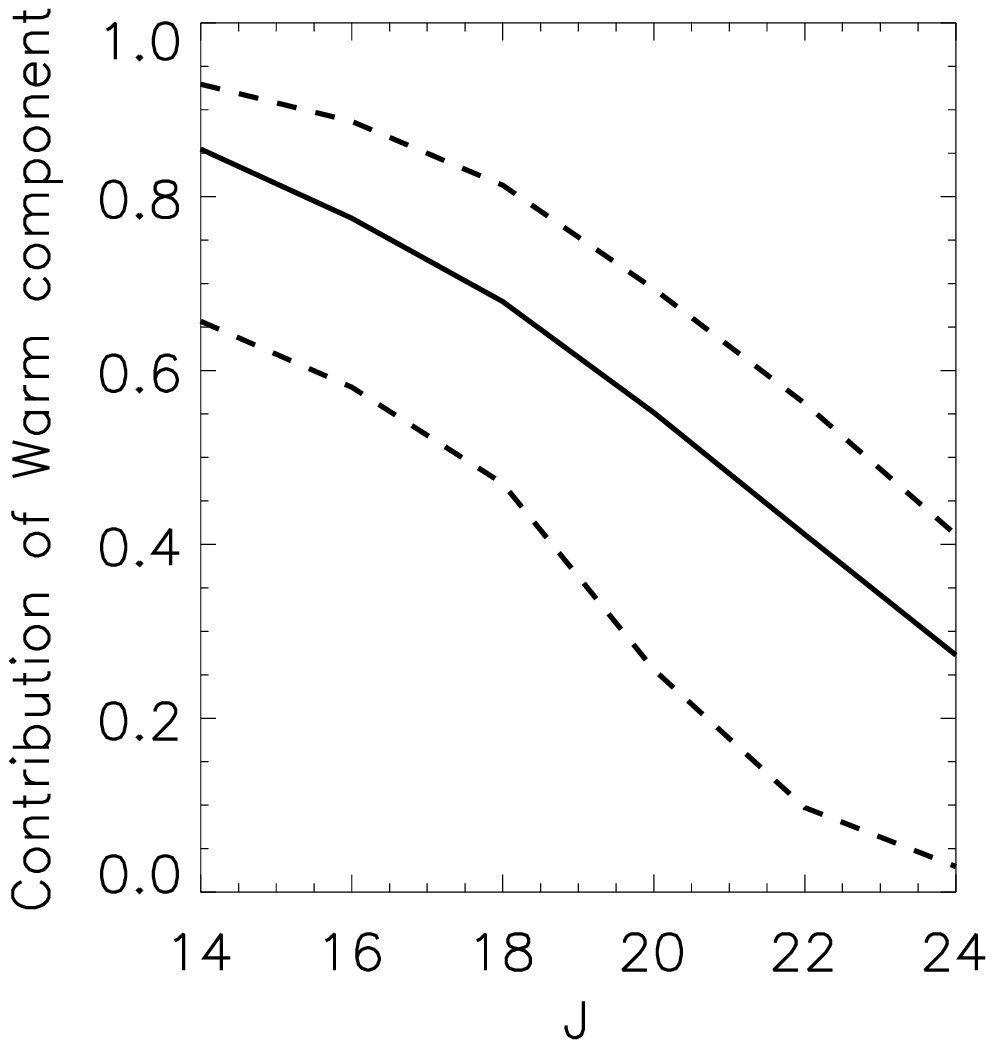}
\caption{Left: the bolometric luminosity ($L_{\rm bol}$) and the rotational temperature $T_{\rm W}^{\rm C} ({\rm OBS})$ of the sources. Right: the contribution of warm component to the total flux of the level J. The symbol is the same as in Figure~\ref{fig:ep_source_type} except that NGC7129 FIRS2 (diamond) is included. In the left panel, solid and dotted lines indicates the median of $T_{\rm W}^{\rm C} ({\rm OBS})$ (when excluded TYPE S sources as presented by the blue color) and $T_{\rm W}({\rm OBS})$, respectively. In the right panel, the solid and dotted lines represent median, maximum, and minimum values, respectively. 
}\label{fig:ep_source}
\end{figure*}

\begin{figure*}
\includegraphics[width=0.50 \textwidth]{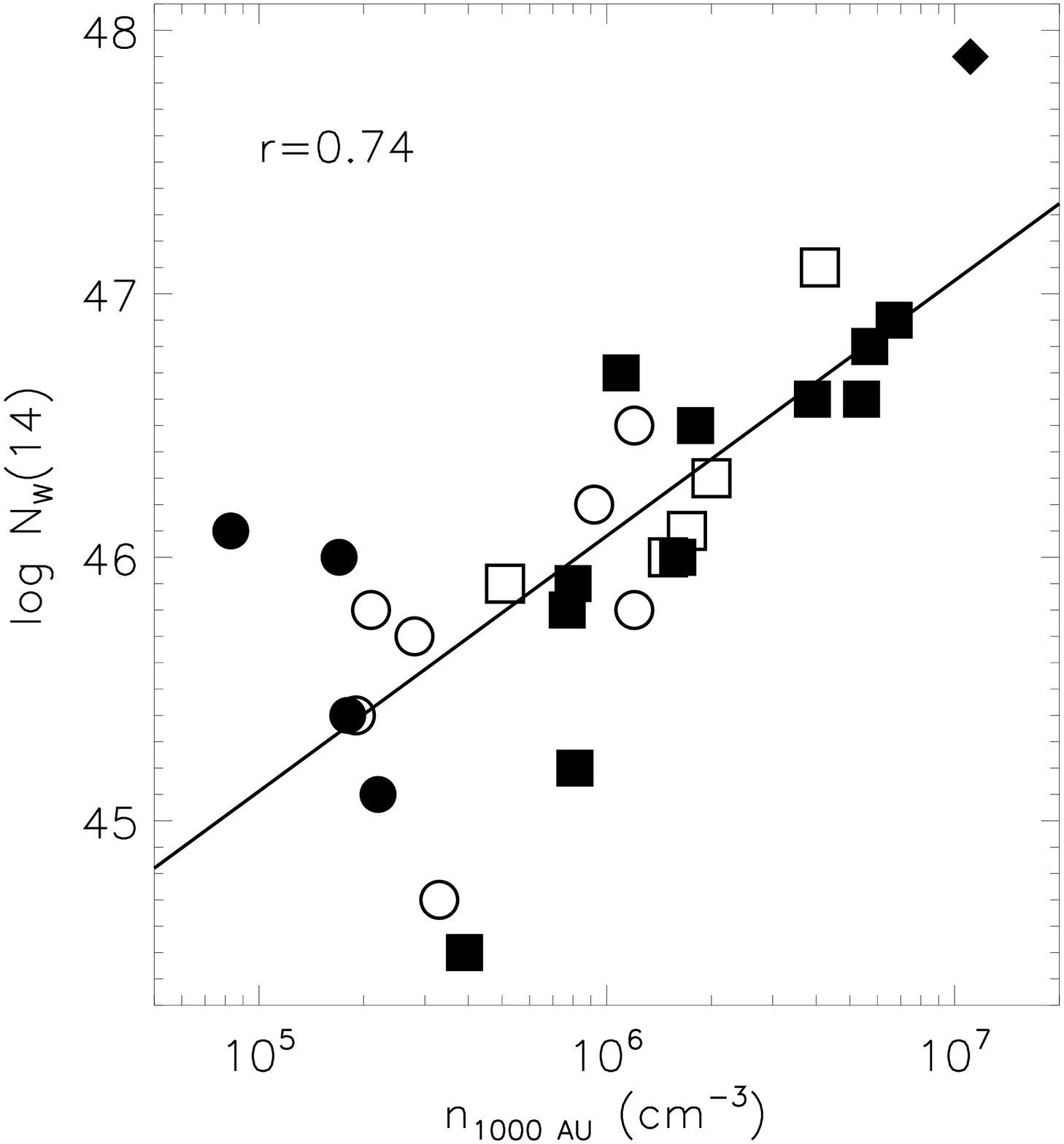}
\includegraphics[width=0.50 \textwidth]{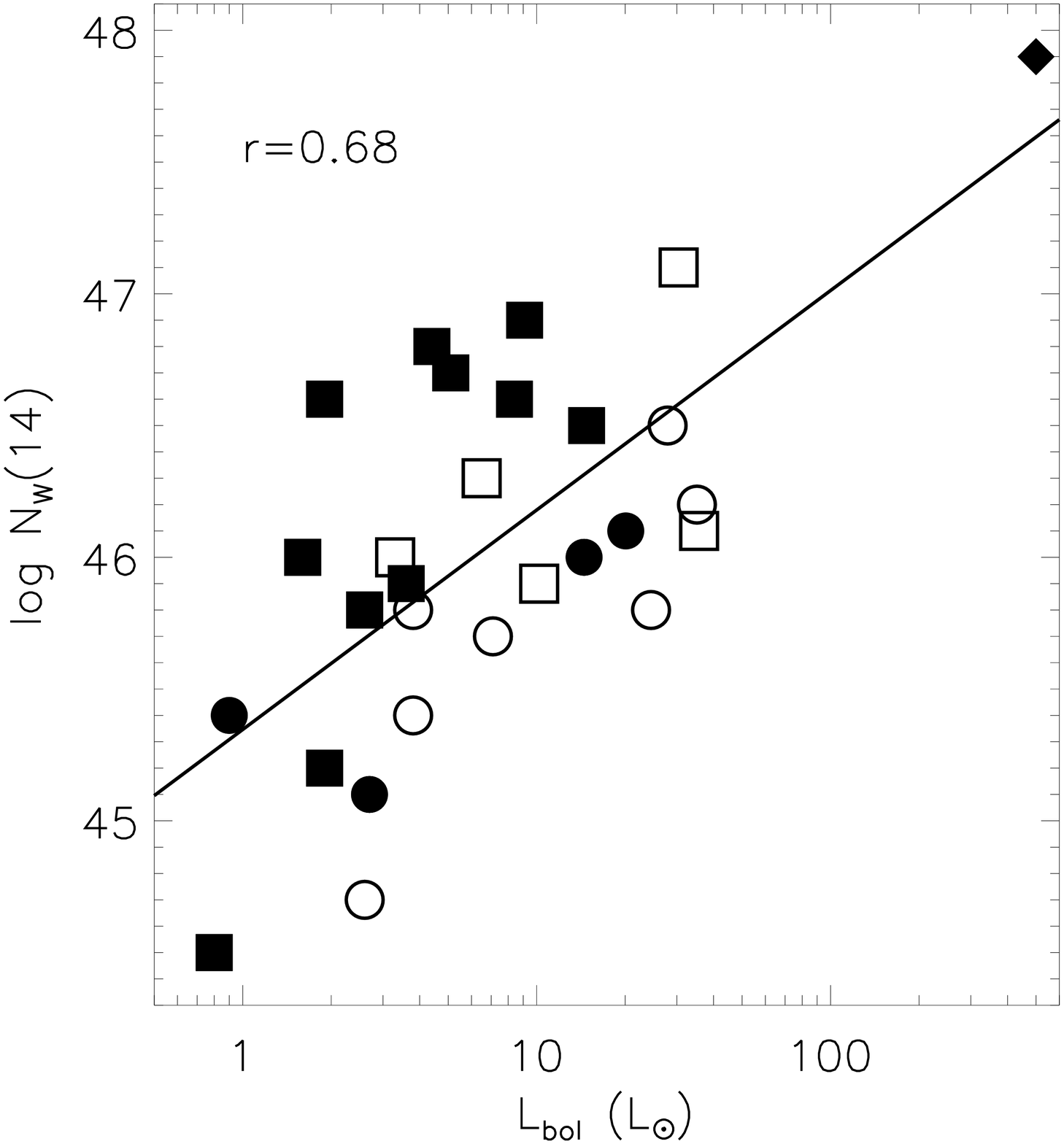}
\caption{Correlation of the corrected CO number in $J$=~14 with density at 1000~AU (left) and the bolometric luminosity (right). The Pearson correlation coefficient $r$ is presented on the panel, which is larger than $r$=~0.55, the value in the confidence level of 3 sigma, indicative of a tight correlation between the parameters. The symbol is the same as in Figure~\ref{fig:ep_source}.  
}\label{fig:ep_source2}
\end{figure*}

 The mid-$J$ CO fluxes, corrected for the hot component, still exhibit a nearly constant rotational temperature with $T_{\rm W}^{\mathrm{C}}(\mathrm{OBS})\sim$250~K, as marked with the solid line in Figure~\ref{fig:ep_source} (left panel). The right panel of Figure~\ref{fig:ep_source} shows the relative contribution of the warm component to the total mid-$J$ CO fluxes; more than half of the  flux for a given mid-$J$ transition are emitted by the warm gas component at $J\le$21. The CO number in $J$=~14 from only the warm component $N_{\rm W}(14)$ is correlated with $n_{\rm 1000AU}$ and $L_{\rm bol}$ (Figure~\ref{fig:ep_source2}). 
%density at 1000~AU and bolometric luminosity (Figure~\ref{fig:ep_source2}). 
These correlations have been shown in \citet{Karska2013}, when $N_{\rm W}(14)$  was calculated from the total flux of $J$=14--13 emitted by both warm and hot components.
%%%%%%%%%%%%%%%%%%%%%
%
%  MODEL
%
%%%%%%%%%%%%%%%%%%%%%%%%

\section{Model}\label{sec:model}

\subsection{Density distribution}\label{sec:density}
 We assume that the density in the envelope has a power law distribution of a spherically symmetric sphere, excluding the outflow cavity. 
 For our study, the envelope density structure of each source is determined from previous
 efforts in the literature  \citep{J&oslash;rgensen2002,Kristensen2012}, using the 1D radiative transfer program DUSTY \citep{Ivezic1997}. 
 Within this framework the outflow cavity is carved out using the function given below in the Cartesian coordinate system \citep{Bruderer2009a},
\begin{eqnarray}\label{eq:bound}
  z & = & \delta_0 \times (x^2+y^2) \\ \nonumber
   & = & \left( \frac{1}{10^4\,{\rm AU} \tan^2(\alpha/2)}\right) \times (x^2 + y^2)
 \end{eqnarray}
 where $z$ is the outflow axis and $\alpha$ is the full opening angle at $z$~=~$10^4$~AU. For the density inside the outflow cavity, we adopt the density of shocked gas, $n$~=~6.3$\times 10^3$ cm$^{-3}$ \citep{Neufeld2009}, which should be the upper limit for the outflow cavity.  

\begin{figure*}
\includegraphics[width=1.0\textwidth]{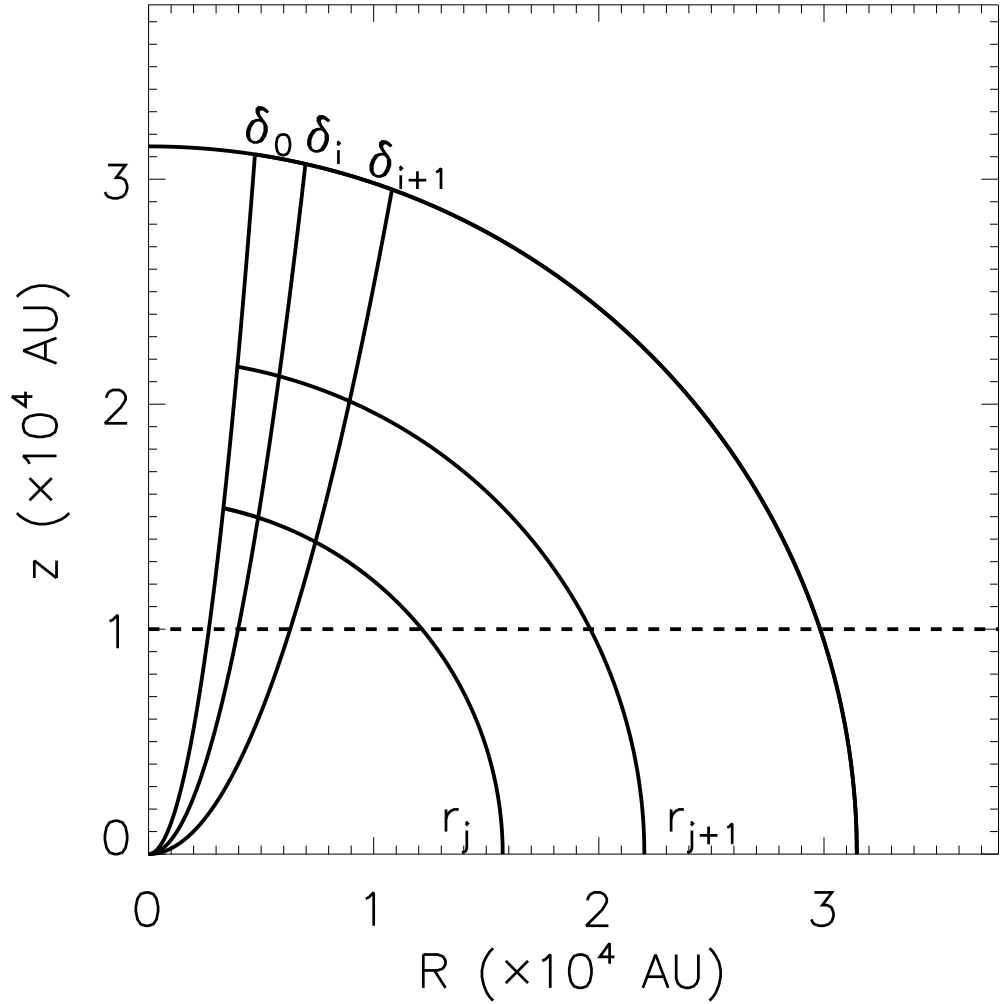}
\caption{Illustration of ($r$, $\delta$) coordinates. The $\delta_0$ indicates the boundary between the outflow cavity and the envelope described with the equation~\ref{eq:bound}. The general $\delta_{\rm i}$ is described with the same equation except for a larger opening angle at z= 10000~AU.
  }\label{fig:grid_cartoon}
\end{figure*}

 We introduce a new coordinate axis $\delta \equiv z/(x^2 + y^2)$ instead of $\theta$ in the spherical coordinate system ($r$, $\theta$) as shown in Figure~\ref{fig:grid_cartoon}.
 While the $\theta$ coordinate describes a circular conical surface, the $\delta$ coordinate provides a circular paraboloid. 
 Both PDR and non-local thermal equilibrium line radiative transfer models explore scales ranging from $\sim10$~AU to $\sim10^4$~AU, resolving the very narrow regions near the outflow cavity wall surface where the warm CO gas exists. 
 As the boundary between the outflow cavity and the envelope ($\delta_0$ in Eq.~\ref{eq:bound}) is a point of the $\delta$ coordinate,  the ($r$, $\delta$) coordinates can simply describe the density profile of thin layers near the surface (see Figure 16 in Paper I). 
 Therefore, we use the ($r$, $\delta$) coordinates through all the procedures except RADMC-3D\footnote{http://www.ita.uni-heidelberg.de/$^\sim$dullemond/software/radmc-3d/} (see below), which does not provide the coordinate.

 The opening angle is measured by the modeling of existing molecular line emission maps, for example, of $^{12}$CO rotational transitions \citep[e.g.,][]{Arce2006}.  
 The emission distribution toward some sources suggests that  the opening angle increases with the protostellar evolutionary stage and spreads out from $\sim$10~deg to 100~deg for Class 0 and I sources \citep{Arce2006}. 
 However, if the UV-heated outflow cavity walls produce the FIR mid-$J$ CO lines, they should emerge from inner dense regions \citep[$n \ge$ 10$^6$ cm$^{-3}$; Paper I; ][] {Visser2012}. 
 These regions are within a few arcseconds and are smaller than (or comparable to) the beam sizes of mm/sub-mm wave radio telescopes even towards the nearby star forming regions. 
 An additional method to determine the opening angle is to fit the spectral energy distribution using dust continuum models \citep[e.g.,][]{Furlan2008}, which are model-dependent. 
 The opening angles derived by the latter method are generally smaller than (or similar to) those determined by the former method. 
 For example, an opening angle of 30$^\circ$ is derived for TMC1 via both methods, while the opening angle of L1551-IRS5 is 10$^\circ$ and 100$^\circ$ by the SED modeling and the CO map, respectively \citep{Furlan2008,Arce2006}.  
 Therefore, it is hard to define ``an" opening angle for a source. 
 As a result, we assume the opening angle of 30$^\circ$ for all sources, which does not change the 1D density profile significantly and fits the FIR mid-$J$ CO lines reasonably well, compared to other values. In addition, this opening angle produces one of the highest CO fluxes for a given UV luminosity, and thus we can test the contribution of the UV heated outflow cavity wall to the mid-$J$ CO emission. The effect of the opening angle will be discussed in Section~\ref{sec:dis_para}.

\subsection{PDR model}\label{sec:pdrmodel}
 We have developed a self-consistent PDR model (Paper I). 
 Our PDR model consists of four parts: the calculation of dust temperature, radiative transfer of UV photons, chemistry, and gas energetics.
 The dust temperature $T_{\rm dust}$ is calculated with the dust continuum radiative code, RADMC-3D, adopting the dust opacity for the average Milky Way dust in dense molecular clouds with $R_{\rm V}$=5.5 and C/H~=~42~ppm in PAHs \citep{Draine2003} for a given density distribution and a given bolometric luminosity, $L_{\rm bol}$.

The FUV radiative transfer is calculated in order to determine the unattenuated FUV strength $G_0$ and average visual extinction $\left<A_{\rm V}\right>$ following the method of \citet{van Zadelhoff2003} and \citet{Bruderer2009a}.  
 The FUV radiative transfer is calculated for only one representative wavelength with photon energy of 9.8~eV, in the middle of the 6 - 13.6~eV FUV band.  We then measure the FUV strength $G_0$ in units of the Habing field (ISRF; $1.6 \times 10^{-3}\, {\rm erg\, s^{-1}\, cm^{-2}}$).
 We adopt  the same dust properties used for the calculation of the dust temperature \citep{Draine2003}. 
   
 In the low-mass classical T Tauri stars, accretion shocks onto the protostar are theorized to produce the observed FUV radiation \citep{Calvet1998,Ingleby2011}, while for the intermediate mass Herbig Ae/Be stars, the central star itself can be also a FUV radiation source \citep[e.g.,][]{Meeus2012}.  
 Bow shocks or small scale-shocks inside the cavity or along the cavity wall can also produce additional local UV photons \citep{Neufeld1989,Lefloch2005}. \citet{van Kempen2009a} argued that the outflow cavity wall could be illuminated by the FUV radiation of $\sim$600 ISRF from the shocks in HH 46. In addition, some sources, such as Elias 29, GSS30-IRS1 \citep{Liseau1999}, and RCrA~IRS5 \citep{Lindberg2014} are externally illuminated by nearby bright stars. 
 
 In our tests, we assume that the only FUV source is  accretion onto the protostar.
 The FUV spectrum affects the photoelectric heating rate of PAHs and small grains \citep{Spaans1994} as well as photodissociation (and photoionization) of species \citep{van Dishoeck2006}. 
 However, because we cannot observe the FUV spectrum directly from the central protostar, we assume that it is similar to that of a black body radiation of $\sim$15,000~K (BB1.5), which represents the FUV continuum of TW Hya \citep{Herczeg2002,Yang2012} and fitted FIR mid-$J$ CO fluxes of HH46 better than a black body radiation of 10,000 K (BB1.0, Paper I).
 
 FUV observations toward classical T Tauri stars find the UV luminosity integrated from 1250~\AA\, to 1750~\AA\, ($L_{\rm uv}^{\rm Int}$) is related with the accretion luminosity $L_{\rm acc}$ as ${\rm log}_{10} { L}_{\rm UV}^{\rm Int} =   0.836\times {\rm log}_{10} { L}_{\rm acc} \,-\,1.67$ with an accuracy of 0.38 dex  \citep{Yang2012}. 
 As the FUV luminosity integrated from 912~\AA\, to 2050~\AA\, is about 2 times  $L_{\rm uv}^{\rm Int}$ for TW Hya and AU Mic \citep{Herczeg2002,Yang2012} and the accretion luminosity dominates the bolometric luminosity during the Class 0 and I stages, we adopt a reference FUV luminosity $L_{\rm UV}^{\rm Y}$,
    \begin{equation}
  \label{eq:luv}
  {\rm log}_{10} {L}_{\rm UV}^{\rm Y} =   0.836\times {\rm log}_{10} {L}_{\rm bol} \,-\,1.37.
 \end{equation}
In this paper, $L_{\rm UV}^{\rm Y}$ is used as the unit of $L_{\rm UV}$.  

When the spectrum in the FUV range is similar to BB1.5, UV photons can be radiated from BB1.5 or from the bremsstrahlung free-free emission with the temperature of $\sim$30,000 K \citep{Nomura2005}. If all $L_{\rm bol}$ is emitted by either of these mechanisms, the blackbody radiation and the free-free emission radiate 28~$\%$ and $\sim$10~$\%$ of $L_{\rm bol}$ in the FUV range, respectively. In our model, it is thus difficult to $L_{\rm UV}$ larger than 0.28 $L_{\rm bol}$, which is similar to the observed 1$\sigma$ scatter ($5~L_{\rm UV}^{\rm Y}$) at the lowest $L_{\rm bol}$.  Therefore, we assume $5~L_{\rm UV}^{\rm Y}$ as the upper limit of $L_{\rm UV}$ in our models.

 In our model, the gas-phase chemical reaction network is based on UMIST2006 database \citep{Woodall2007} modified by \citet{Bruderer2009b}. 
 For photoreaction rates, we have adjusted the attenuation factor, $\gamma$, following the method of \citet{R&ouml;llig2013} and calculated the unattenuated photoreaction rate with the photodissociation and photoionization cross sections provided by \citet{van Dishoeck2006}.  We follow the model of \hh \, formation on interstellar dust grains via physisorption and chemisorption from \citet{cazaux2002,cazaux2004,cazaux2010erratum} with the sticking coefficient of \citet{Hollenbach1979}. The neutral gas can deplete onto dust grains and evaporate by thermal and non-thermal (photon and cosmic ray) events. We also consider electron attachment to grain and cation-grain charge transfer.  The cosmic-ray ionization rate of H$_2$ is set to be $5 \times 10^{-17}$~s$^{-1}$ \citep{Dalgarno2006}. We let the chemistry evolve for 10$^5$ years.

 We consider important heating and cooling processes described in \citet{R&ouml;llig2007}.
 We adjust the photoelectric heating rates of PAHs and small grains \citep{Weingartner2001} with the correction factor given by \citet{Spaans1994}. We also reduce the \hh\, vibrational heating and cooling rate excited by the FUV photons because only UV photons in the range of 912 - 1100 \AA \, can pump H$_2$. We also calculate the H$_2$ formation heating, gas-grain cooling/heating, and atomic and molecular line cooling (for details see Paper I). 

 The chemistry and gas energetics are calculated iteratively. 
 It is very time-consuming to calculate the chemistry with the full chemical network. 
 In order to reduce the time, a subset of the full chemical network has been adopted for gas energetics. As a check we compared the 1D PDR models, with a small network with chemical species described in Table 1 of \citet{Woitke2009}, to identical models with the full  chemical network. We find that 
 gas temperatures for the two chemical networks are consistent over the range of density and FUV strength  relevant to this work. 
 The CO abundances near the surface ($A_{\rm V} < 1$), however, differ from each other by an order of magnitude. 
 Therefore, the iterative calculation of the gas energetics and chemistry use the small network, then the chemistry with the full network is calculated 
with the gas temperature determined using the small network.
 
\subsection{Line radiative transfer}
We have developed a new line radiative transfer code in general grid (RIG). 
For details, refer to Paper I. 
The most important strength of RIG is the ability to optimize the grid coordinates to a given model. RIG works in any coordinate systems, including Cartesian, cylindrical, spherical, and ($r$, $\delta$) coordinates.  
As described above, the ($r$,~$\delta$) coordinates are optimal to model  the envelope with outflow cavity walls, and thus, the grid cell number of 300 in these coordinates (30 in $r$ and 10 in $\delta$) provides adequate spatial resolution. The best-fit models with a  larger number of grid cells (100 in $r$ and 30 in $\delta$), which is comparable to (in $r$) or higher than (in $\delta$) the spatial resolutions of the model by \citet{Visser2012}, show  similar results to the models with the 300 grid cells.
 
 Collisional rate coefficients for CO are adopted from Leiden Atomic and Molecular Database\footnote{http://home.strw.leidenuniv.nl/$\sim$moldata/} \citep{Sch&ouml;ier2005} updated by \citet{Yang2010} and \citet{Neufeld2012}.
 Following \citet{Visser2012}, we fix the non-thermal Doppler width as 0.8~km~s$^{-1}$ and velocity distribution as $v(r)~=~2$~km~ s$^{-1}$~$\sqrt{r_{in} / r}$ with the inner boundary radius $r_{in} $.  
 Because the CO ladders in the PACS wavelength range are generally optically thin \citep{Manoj2013}, the velocity field does not  significantly affect the result.

In order to compare to observations, we have synthesized maps of CO spectra, viewed at face-on with 0.1$''$ spatial resolution, using a ray-tracing method; these maps are then used to predict the number of CO molecules emitting in the $J$ level with Eq.~\ref{eq:NJ} at a given pixel. Most of the mid-$J$ CO line produced by PDRs emits within a depth of $\sim$10 AU ($0''.1$ at 100 pc) from the surface of the outflow cavity walls, which can be represented in our model due to the optimized $\delta$ grid (see Section~\ref{sec:result2}). We tested the resolution of $0''.05$ and found that the difference in simulated fluxes between $0''.1$ and $0''.05$ resolutions is less than 2\% in the mid-$J$ transitions. 
 An edge-on view reduces the mid-$J$ CO fluxes by up to 25\% as a result of extinction from the dusty envelope. Most of synthesized mid-$J$ CO emission arises within the $9''.4 \times 9''.4$ central spaxel of PACS, (which is a few 1000~AU at the distance of our sources).
 
%%%%%%%%%%%%%%%%%%%%%%%
%
%   RESULT
%
%%%%%%%%%%%%%%%%%%%%%%%%%
\section{ RESULT }\label{sec:results}

\subsection{Best-fit models}
 
 Rotational diagrams from our best-fit models are plotted as black circles in Figures~\ref{fig:rd1}~-~\ref{fig:rd4}. 
 Rotational temperatures  derived from the best-fit models $T_{\rm W}(\mathrm{MODEL})$ are listed in Table~\ref{tb:result} and inside each panel of Figures~\ref{fig:rd1}--\ref{fig:rd4}. 
 Most of our best fit models reproduce the observed mid-$J$ CO emission for the sources with $|T_{\rm W}(\mathrm{MODEL}) -T_{\rm W}^{\mathrm{C}}(\mathrm{OBS})| < 3 \sigma_{\rm W}(\mathrm{OBS})$ except for Ced110-IRS4, VLA 1623-243, and  L1551-IRS5.
 Ced110-IRS4 and L1551-IRS5 have $T_{\rm W}(\mathrm{OBS}) > 400$~K, and thus, these sources might be  heated mainly by shocks, because the UV heated cavity wall cannot produce a rotational temperature above 400~K in our models. 
 Although VLA 1623-243 has $T_{\rm W}(\mathrm{OBS})$ of 347~K, shocks could also be the main contributor to the mid-$J$ CO emission because this source is known to have prominent outflow emission \citep[e.g.,][]{Bjerkeli2012}.

\begin{figure*}
\includegraphics[width=0.50 \textwidth]{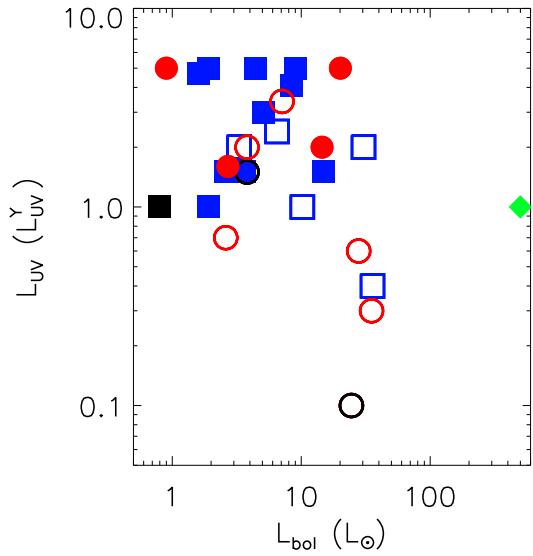}
\caption{The bolometric luminosity ($L_{\rm bol}$) and the best fit UV luminosity $L_{\rm UV}$ (in unit of $L_{\rm UV}^{\rm Y}$) of the sources. The symbol is the same as in Figure~\ref{fig:ep_source}. Red, blue, and green indicate the class 0, I, and intermediate mass sources, respectively. For some sources (Ced110-IRS4, VLA 1623-243, and L1551-IRS5), a PDR model cannot reproduce the observed $T_{\rm rot}$. The sources are represented as black symbols.
}\label{fig:ep_source_luv}
\end{figure*}

 Figure~\ref{fig:ep_source_luv} shows the best fit $L_{\rm UV}$ (in unit of $L_{\rm UV}^{\rm Y}$) versus $L_{\rm bol}$ of sources.
 From this analysis we find that the extended sources have a higher $L_{\rm UV}$ than the compact sources. 
 Sources with the upper limit of $L_{\rm UV}$ (NGC1333 IRAS 4B, Ser SMM4, TMC1, and Elias 29) are all associated with extended emission and our predictions underproduce the mid-$J$ CO fluxes. The UV heated outflow cavity wall generally radiates the compact emission (see below), thus, the extended emission is more likely generated by shocks. 

\begin{figure*}
\includegraphics[width=0.30 \textwidth]{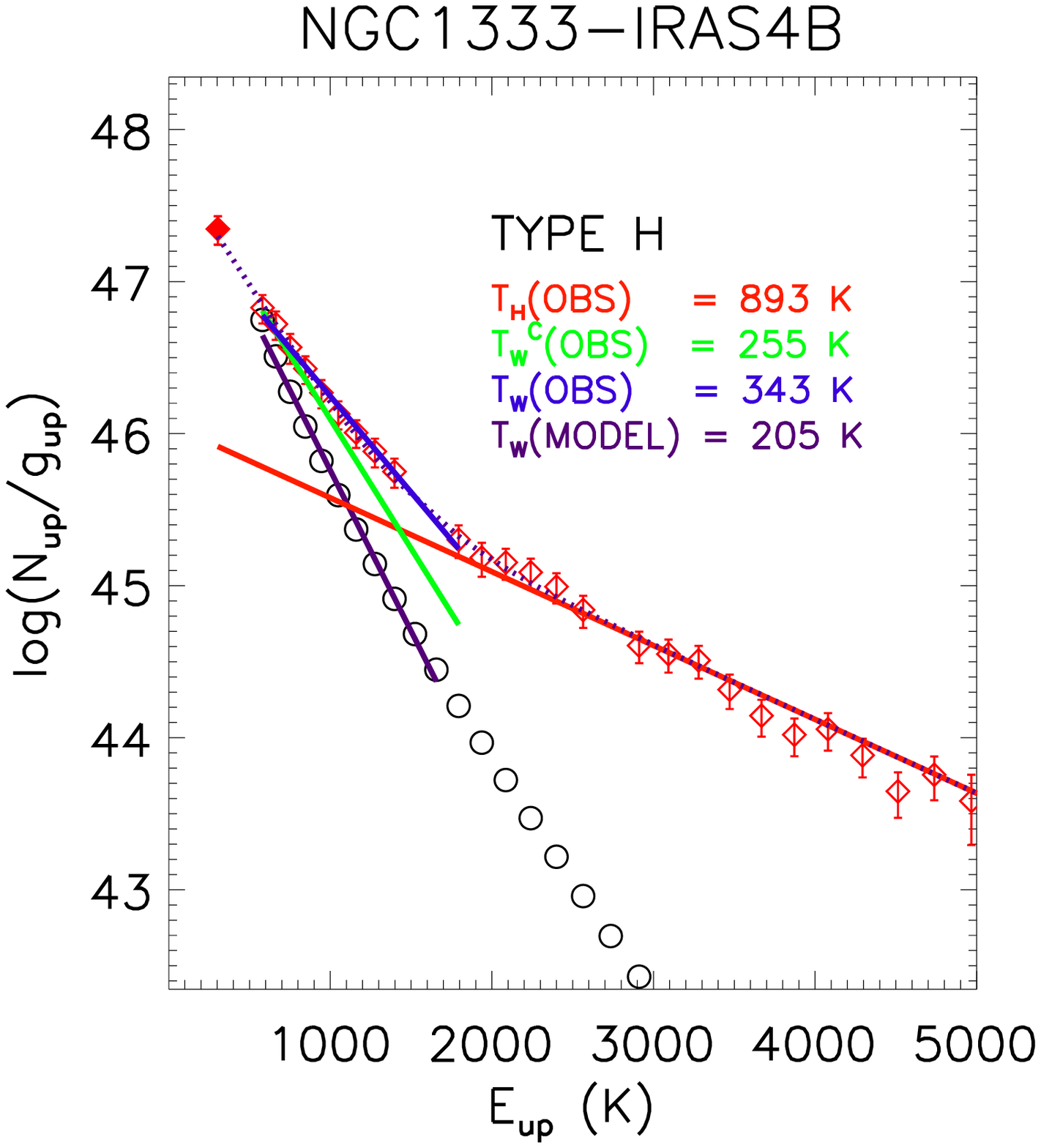}
\includegraphics[width=0.30 \textwidth]{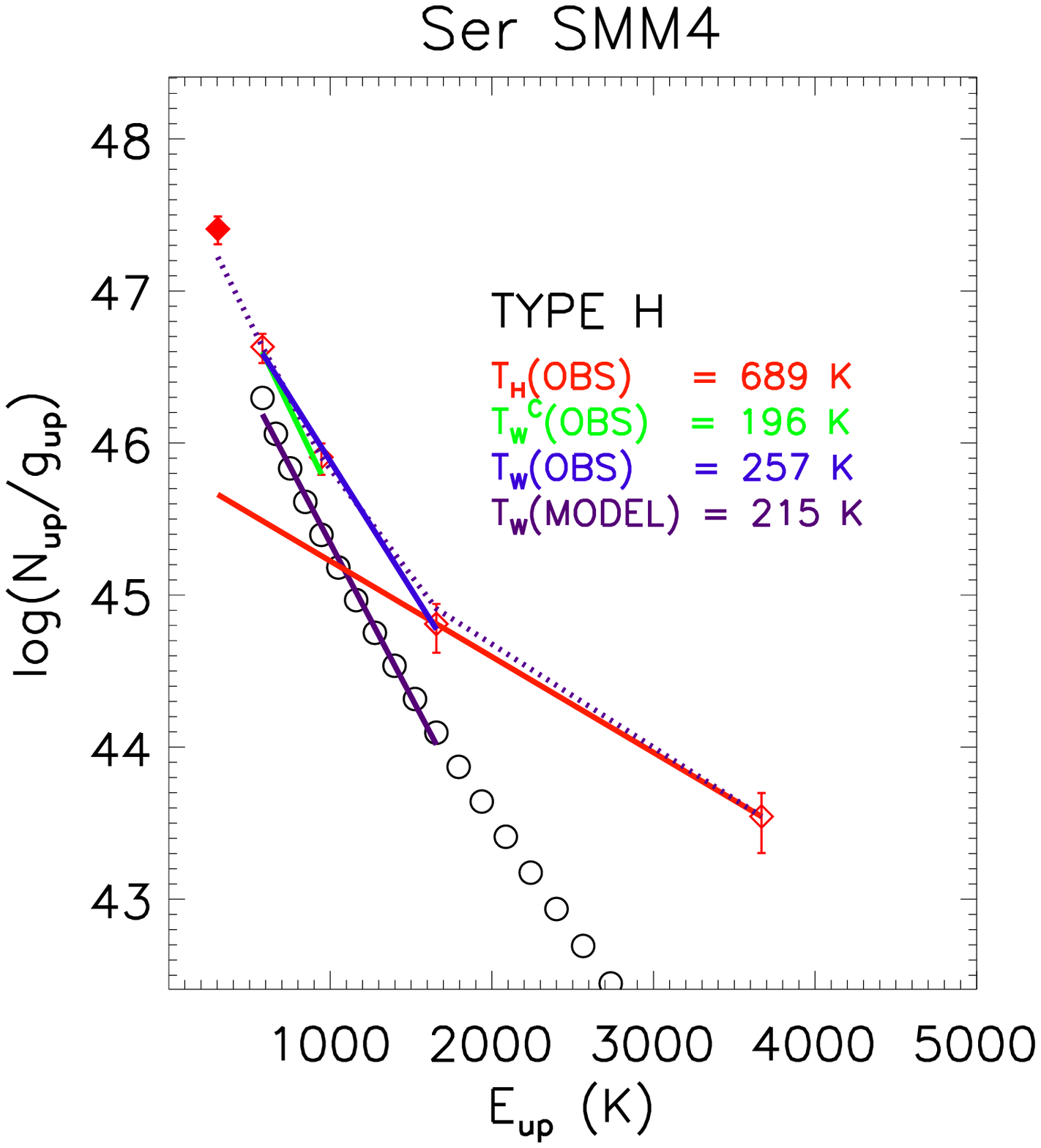}

\includegraphics[width=0.30 \textwidth]{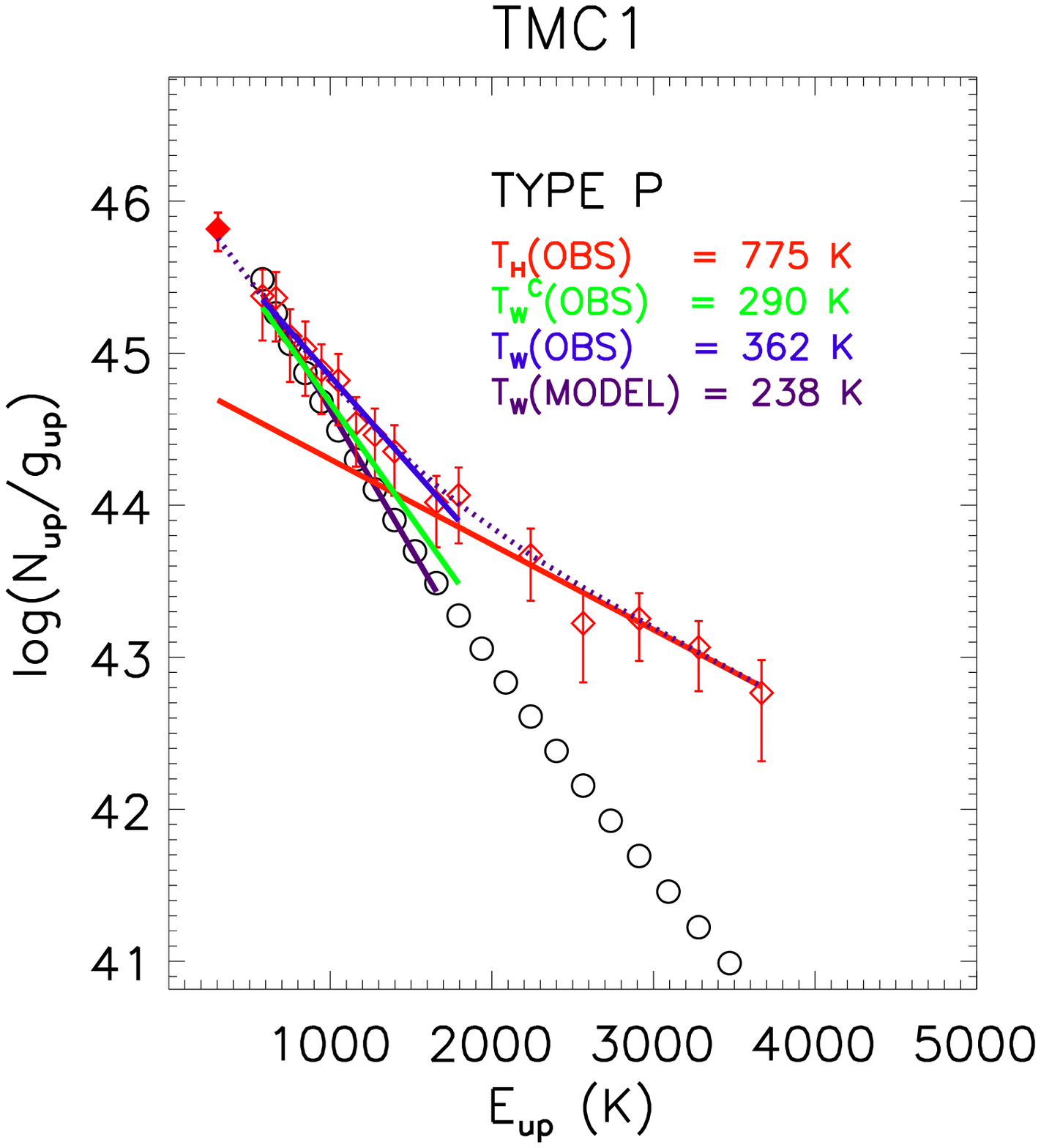}
\includegraphics[width=0.30 \textwidth]{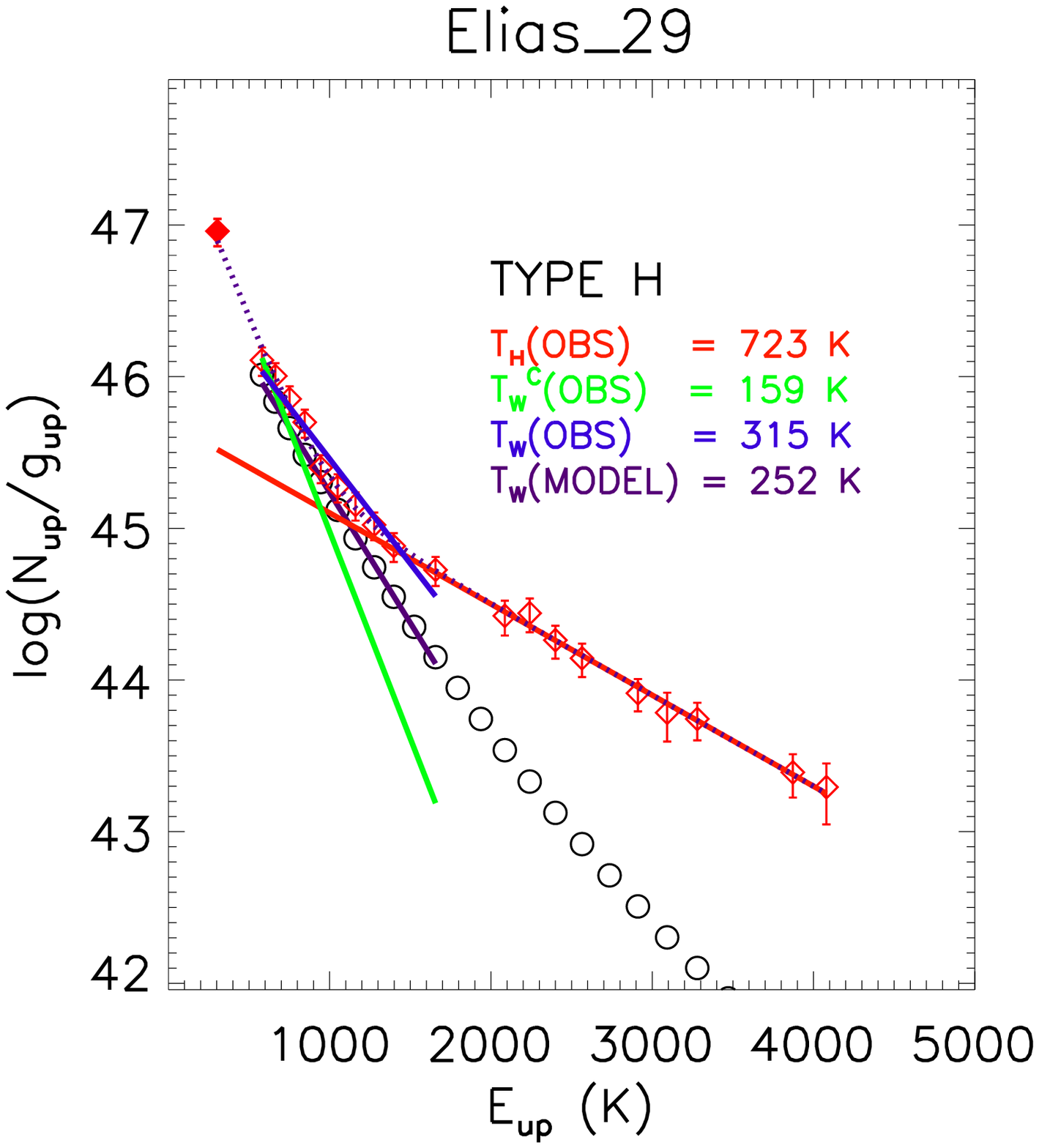}
\caption{The same as Figure \ref{fig:rd1}, but the $10^4$~K black body radiation field was used for NGC1333-IRAS4B, Ser SMM4, TMC1, and Elias 29. The emission of the HIFI $^{12}$CO $J$=~10--9 \citep{San Jos&eacute;-Garc&iacute;a2013, Y&inodot;ld&inodot;z2013} is plotted as a filled red diamond, which is a similar to the flux extrapolated from the mid-$J$ lines.}\label{fig:rd_bb1.0}
\end{figure*}

 In addition, the sources where we estimate an  upper limit to $L_{\rm UV}$, except TMC1, have a strong ``broad" velocity component in the HIFI $^{12}$CO $J$=10--9  spectrum \citep{San Jos&eacute;-Garc&iacute;a2013,Y&inodot;ld&inodot;z2013}. Furthermore, the HIFI $^{12}$CO $J$=10--9 flux is similar to the flux extrapolated from the mid-$J$ lines (see Figure~\ref{fig:rd_bb1.0}). TMC1 also shows a strong outflow detected in the PACS [O I] and [C II] lines \citep{Karska2013,Lee2014a}.
 Therefore, in this object the mid-$J$ CO emission could readily be produced by shocks as opposed to UV radiation.

 However, for a given UV luminosity, BB1.0 generates a lower rotational temperature, but higher CO fluxes than BB1.5 (Paper I). 
 When a UV luminosity of 5 $L_{\rm UV}^{\rm Y}$ with BB1.0 is used, we find a better fit for both rotation temperature and CO fluxes  for NGC1333~IRAS~4B, TMC1, and Elias~29, as shown in Figure~\ref{fig:rd_bb1.0}.  However, the CO $J$=~14--13 flux in   Ser SMM4 is still lower than the observed one by a factor of two. More than half of the observed CO $J=$~14--13 flux is radiated from the blue extended (outflow) emission in Ser SMM4 \citep{Karska2013,Dionatos2013}, thus, PDRs could reproduce only the compact emission near the protostar detected in the central spaxel, if any.

 The  intermediate mass embedded class 0 protostar NGC7129-FIRS2 has $L_{\rm  bol}$=~500~$L_{\odot}$  and the stellar mass $M_{*} = 5~{\rm M}_\odot$, which is derived by assuming that the source is at the birthline \citep{Eiroa1998,Fuente2005}. The theoretical model of a protostar with an accretion rate of $\dot{M} = 10^{-5}~{\rm M}_\odot {\rm yr}^{-1}$ shows that $L_{\rm acc}$ dominates the contribution to $L_{\rm bol}$ up to about 4 ${\rm M}_\odot$ \citep{Palla1991,Palla1992}. For the protostar with $M_{*} = 5~{\rm M}_\odot$ at the birthline, the emission from the surface of the protostar with $T_{\rm eff} \sim 10^4$~K is the primary contributor to~$L_{\rm bol}$ \citep{Palla1991,Palla1992,Palla1993}, and $L_{\rm UV}$ accounts for 7.5~$\%$ of $L_{\rm bol}$, which is an order of magnitude higher than the best fit $L_{\rm UV}$ (0.5~$L_{\rm UV}^{\rm Y}$) of the model with BB1.0. Thus, the best fit $L_{\rm UV}$ may allude that most UV photons radiated from the protostar might be blocked by the dense material located in the vicinity of the central protostar.

 When the UV spectrum of BB1.5 is adopted, the outflow cavity wall heated by 1~$L_{\rm UV}^{\rm Y}$ can reproduce the observed mid-$J$ CO emission. This might imply that the accretion luminosity dominates the contribution to $L_{\rm bol}$ for NGC 7129-FIRS2 ($\dot{M} > 10^{-5}~{\rm M}_\odot~{\rm yr}^{-1}$ ) and the relation between $L_{\rm acc}$ and $L_{\rm UV}$ derived from the T-Tauri stars could be expanded to even intermediate mass protostars. More detailed study, and likely higher resolution observations, are needed to distinguish which interpretation is adequate for the PACS observation.

 According to our models, the UV-heated outflow cavity walls could reproduce the observed ``\textit{compact}" mid-$J$ CO emission with or without the hot shock-heated components in the case when the observed rotational temperature is below 400~K. 
 We, however, note that our model provides only a possible explanation for the mid-$J$ CO emission, and a quantitative contribution of the UV-heated outflow cavity wall should be calculated by simultaneous modeling of PDR and shocks.
 
\subsection{Physical and chemical structure of the UV-heated outflow cavity wall}\label{sec:result2}

\begin{figure*}
\begin{center}
\includegraphics[width=0.7 \textwidth]{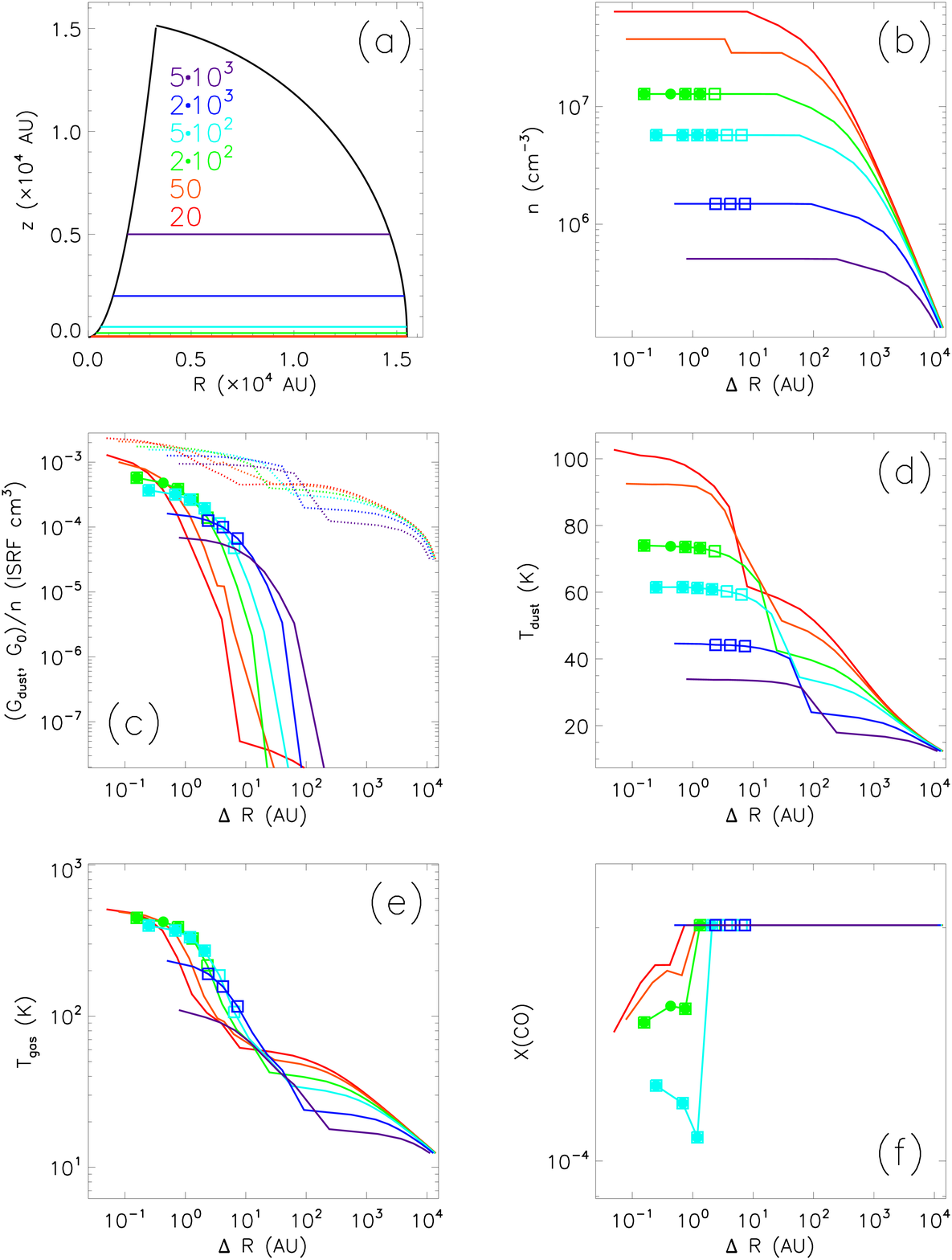}
\caption{The results of the best-fit model for Ser SMM1. 
Density (b), ratio of FUV strength to density (c), dust temperature (d), gas temperature (e), and CO abundance (f) distributions are plotted along given horizontal cuts in the envelope of Ser SMM1. Each color line indicates the physical values for a given z-height (in the unit of AU), which is represented with the same color in the panel (a). $\Delta R$ is the horizontal distance from the outflow cavity wall surface. The filled circles, open squares, and filled squares on top of the lines indicate the grid cells where emissions of $J$=24--23, 14--13, and both lines are radiated, respectively. In the panel (c), solid and dotted lines indicate the ratio of dust attenuated and unattenuated (incident) FUV strength to density, respectively. (Figures 11.1–-11.28 for the other sources are available in the online supplementary material.)}\label{fig:Ser_SMM1_model}
\end{center}
\end{figure*}

Figure~\ref{fig:Ser_SMM1_model} shows the physical and chemical properties of the best-fit model for Ser SMM1. 
The properties are plotted along the horizontal distance, $\Delta R$, from the surface of the outflow cavity wall for given z-heights, which  are marked with horizontal color lines and labels
 in top left panel of Figure~\ref{fig:Ser_SMM1_model}. Other panels show density $n$ (b), the ratio of unattenuated (attenuated) FUV strength to density $G_0/n$ ($G_{\rm dust}/n$) (c), dust temperature $T_{\rm dust}$ (d), gas temperature $T_{\rm gas}$ (e), and CO abundance X(CO) (f).
  The regions emitting the majority of flux for CO $J$=~24--23 (filled circles), $J$=~14--13 (open squares), and both transitions (filled squares) are plotted over the layers in the panels. 
  
 The ratio of the FUV strength to density ($G_{\rm 0}/n$) can be used to parameterize the dense gas PDR ($n \geq 10^6 \mathrm{cm}^{-3}$) because the physical and chemical properties are similar for a given $G_{\rm 0}/n$ \citep[][Paper I]{Kaufman1999}. 
 More directly, photoelectric heating of PAHs and small grains ($\propto G_0\,n$) and gas-grain collisional cooling ($\propto n^2$) determine $T_{\rm gas}$ (see \citealt{Visser2012}). 
 CO is destroyed by photodissociation ($\propto G_0\,n$) and forms by two-body reactions ($\propto n^2$; dissociative recombination and charge transfer) (see Paper I).  
 A higher $G_0/n$ thus gives a higher $T_{\rm gas}$, but lower X(CO) near the surface.

 In low-mass star forming regions, the power-law index in the density profile is lower than two, and  $G_0/n$ increases toward the center. 
 Therefore, as the z-height is lowered (i.e., colors from purple to red in the panel (a) of  Figure~\ref{fig:Ser_SMM1_model}), $T_{\rm gas}$  increases while X(CO) decreases near the surface. 
 However, X(CO) near the surface increases from z=500~AU downward (cyan line in the panel (f) of Figure~\ref{fig:Ser_SMM1_model}) to equatorial plane (red line). 
 Near log $G_0/n \sim -3$, UV photons photodesorb H$_2$O ice into the gas phase  preventing all oxygen from frozen onto dust grains, and a high $T_{\rm gas}$ ($>$ 300 K)  makes the formation rate of CO high enough to keep X(CO) high in the inner dense regions (see Figure~\ref{fig:Ser_SMM1_model}), where the FIR mid-$J$ CO lines form (Paper I).   
 \textit{Both H$_2$O photodesorption and the fast CO formation at $>$300~K seem important for the physical and chemical conditions in the embedded phase} (see Paper I for the detail discussion).

 We find that most of mid-$J$ CO fluxes in the best-fit models are produced within specific conditions. 
 The CO $J$=24--23 line forms in the central spaxel with log~$G_0 / n \geq$~$-4.5$ and $\mathrm{log}~G_0 \ge 3$ for all our sources (see Figure~\ref{fig:Ser_SMM1_model}). 
 These regions have a density of log~$n$~(cm$^{-3}$)$\gtrsim$~6 and depth from the outflow surface wall of $\Delta R \leq$ a few AU (average visual extinction of 0.1 $\leq \left<A_{\rm V}\right> ({\rm mag})\leq$ 1), where $T_{\rm gas} \geq$ 300 K and log~X(CO) $\geq$~-5.  
 The CO $J$=14--13 line emits from  the same gas where CO $J$=24--23 forms, but also in the gas with $T_{\rm gas} \simeq$ 100 K, which is located at a higher $\left<A_{\rm V}\right>$ and a larger distance from the protostar. 
 
 Although some sources show that the CO $J$=14--13 emission is radiated from outside  the central spaxels (see the blue squares in Figure~\ref{fig:Ser_SMM1_model}), most of the CO emission is radiated from near the protostar in our models. Therefore, the extended mid-$J$ CO emission cannot be reproduced by the UV heated gas, and is likely associated with shock-heated gas. In this work, we used the fluxes extracted over the whole PACS spaxels, and thus, for the extended sources, a higher best fit $L_{\rm UV}$ is required to reproduce the total flux. 

 %
 %  DISCUSSIONS
 %
 %
\section{Discussion}\label{sec:discussions}
\subsection{Effect of physical parameters}\label{sec:dis_para}
 In order to test the effect of physical parameters, we use the model of L1157, which is a compact source located near the median position of the density and the bolometric luminosity plot (Figure \ref{fig:ep_source_type}). 
 We explore the effect of UV luminosity, opening angle, and power index in the density distribution. 
 We set the standard UV luminosity of protostar as 2.4~$L_{\rm UV}^{\rm Y}$ which is the best-fit value for L1157. 
 
\begin{figure*}
\includegraphics[width=0.8 \textwidth]{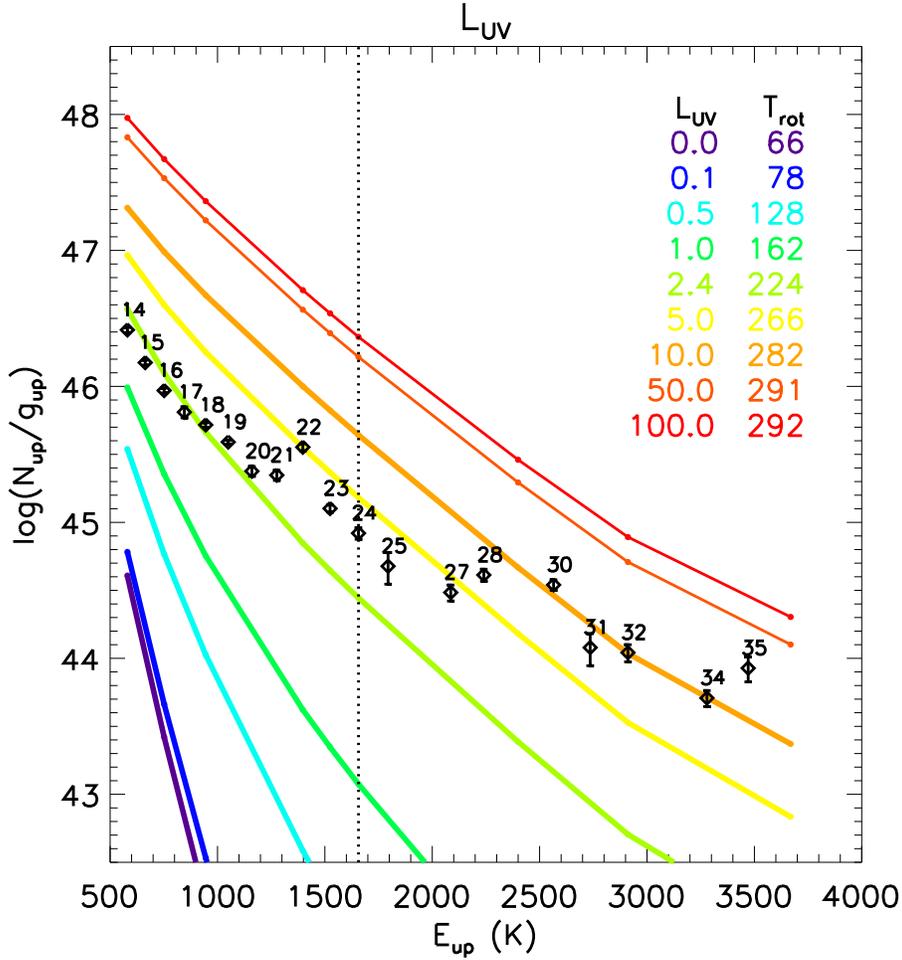}
\caption{The effect of UV luminosity in the model of  L1157. The $L_{\rm UV}$ of the standard model (yellow line)  is 2.4 $L_{\rm UV}^{\rm Y}$ (= 0.82~$L_\odot$). Each color represents the UV luminosity scaled to $L_{\rm UV}^{\rm Y}$ (see Eq.~\ref{eq:luv}). 
Color lines indicate the  rotational diagrams of models with different UV luminosities, and the observed data are plotted with open diamonds. The rotational temperatures, $T_{\rm rot}$ shown in the right top of the panel are the values fitted to the mid-$J$ CO lines of 550~K $\leq E_{\rm up}$ $\leq$ 1700~K. A vertical dashed line indicates the highest ($J$=~24) levels in the mid-$J$ CO lines, which are relevant to this work. $L_{\rm UV}> 10 L_{\rm UV}^{\rm Y}$ (thin lines) are tested to cover a high dynamic range of $L_{\rm UV}/n_{\rm1000 AU}$ although they are unrealistic for L1157 (see text).}\label{fig:ep_uv}
\end{figure*}

 \textit{The effect of $L_{\rm UV}$.} Figure~\ref{fig:ep_uv} shows the effect of UV luminosity in the range of 0.0 $\le L_{\rm UV}$/$L_{\rm UV}^{\rm Y} \le$ 100. 
 $ L_{\rm UV}$/$L_{\rm UV}^{\rm Y} > 10$ is unrealistic because $L_{\rm UV}$ cannot exceed  0.28~$L_{\rm bol}$ equivalent to $ L_{\rm UV}$/$L_{\rm UV}^{\rm Y}=~9$ for BB1.5 (see Section~\ref{sec:pdrmodel}), but we test two higher ratios of 50 and 100 to cover a high dynamic range of $L_{\rm UV}/n_{\rm 1000 AU}$. % in Figure~\ref{fig:ep_source3}.}   
 A higher  protostellar UV luminosity produces a larger number of the CO molecules in a given mid-$J$ level.  Thus, $T_{\rm rot}$ increases with $L_{\rm UV}$ up to  $L_{\rm UV} <10$~$L_{\rm UV}^{\rm Y}$, then it is nearly constant with $T_{\rm rot}\sim$300~K.
 
\begin{figure*}
\includegraphics[width=1.0 \textwidth]{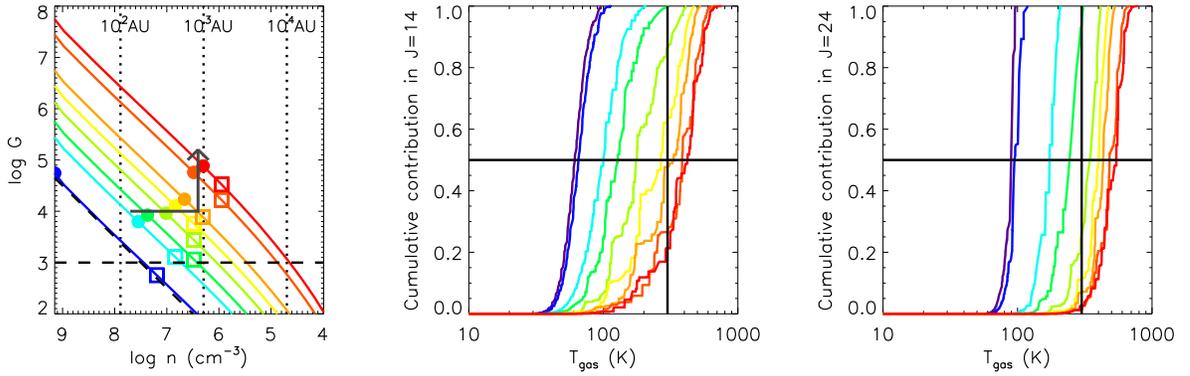}
\caption{$G_0$ at the outflow cavity wall surface for given $L_{\rm UV}$'s (left) and the Cumulative contribution of different gas temperatures to the fluxes of $J$=~14--13 (middle) and $J$=~24--23 (right). The color lines are the same as in Figure~\ref{fig:ep_uv}. The filled circles and open squares in the left panel represent the conditions where most emissions of $J$=~24--23 and 14--13 are radiated, respectively. The filled circles move along the grey solid arrow as $L_{\rm UV}$ increases. Vertical dotted lines indicate the distance from the protostar. The black dashed lines represent $\mathrm{log}~G_0/n=-4.5$ and $\mathrm{log}~G_0=~3$. In the middle and right panels, vertical and horizontal lines indicate $T_{\rm gas}$=~300~K and the cumulative contribution of 0.5, respectively.}\label{fig:ep_uv_pop}
\end{figure*}

 The left panel of Figure~\ref{fig:ep_uv_pop} shows $G_0$ estimated for the surface of the outflow cavity as a function of $L_{\rm UV}$. The density power law index is 1.6 and $G_0$ follows the inverse square law of distance from the protostar, thus $G_0/n$ increases toward the central source.
As mentioned in Section~\ref{sec:result2}, a dense region ($\mathrm{log}~n\ge 6$) exposed to the UV radiation with ${\rm log}~G_0 \ge 3$ and $\mathrm{log}~G_0/n \ge$~-4.5 can produce the warm CO gas with $T_{\rm rot}\sim 300$~K traced by the CO mid-$J$ transitions. 
As $L_{\rm UV}$ increases, the region satisfying this condition expands. In addition, the filled circles in the left panel of Figure~\ref{fig:ep_uv_pop}, where the majority of CO $J$=24--23 emission radiates, moves outward.  This is because the lowest  density satisfying the condition decreases with $L_{\rm UV}$, and there is more mass in the envelope at lower densities. Furthermore, following them as $L_{\rm UV}$ increases, the filled circles move horizontally along $\mathrm{log}~G_0\sim4$ down to log~$n\sim 6.5$. When $L_{\rm UV}$ is higher than 5 $L_{\rm UV}^{\rm Y}$ (yellow line), the largest CO $J$=24--23 emission is radiated around the 1000~AU.
The critical density of $J$=~24--23 is log~$n$=~7, and thus, CO in lower densities is not thermalized, resulting in decreases $J$=~24--23 emission in the less dense gas beyond 1000~AU.

 The UV heated cavity walls consist of a lower temperature gas component with $T_{\rm gas}\sim 100$~K as well as the warm gas that we are interested in.
Therefore, the CO $J$=~14--13 transition also traces the cool gas (Paper I).
When $L_{\rm UV}$ is low, the majority of  CO $J$=~14--13 line emission arises from this cool gas, but the contribution of the warm gas to the CO $J$=~14--13 emission increases with $L_{\rm UV}$, as shown in the right panel of Figure~\ref{fig:ep_uv_pop}. As a result, $T_{\rm rot}$ increases with $L_{\rm UV}$.

\begin{figure*}
\includegraphics[width=1.0 \textwidth]{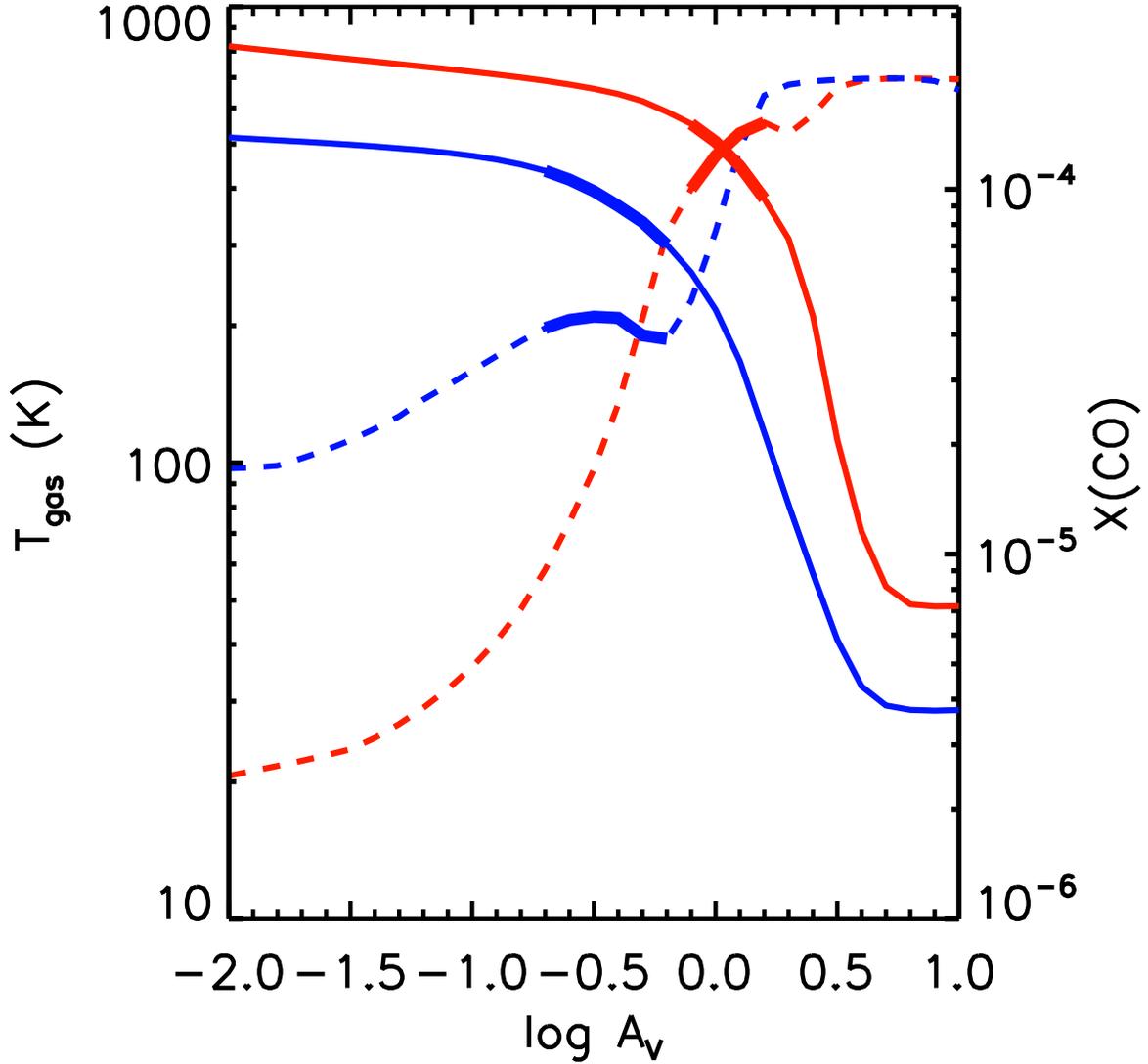}
\caption{Gas temperature (solid) and CO abundance (dotted) of 1D model with log~$n$~(cm$^{-3}$)=~6.5 for a given UV strength of log~$G_0$=~4 (blue) and 5 (red). 
Thick lines indicate the region emitting 70\% of CO $J$=~24--23  (see Paper I).} \label{fig:ep_1d_model}
\end{figure*}

 However, once $L_{\rm UV}$ is high enough to populate the CO mid-$J$ levels consistent with $T_{\rm rot}\sim$300~K, then it remains at this rotational temperature even with above this fiducial $L_{\rm UV}$.   Exploring this more deeply,
the lowest and highest mid-$J$ CO transitions ($J$=~14--13 and  $J$=~24--23) are radiated within a narrow temperature range of 200--400~K and 300--600~K, respectively, as shown in Figure~\ref{fig:ep_uv_pop}. This is an intrinsic property of a PDR; UV photons heat the gas, but also destroy  CO.  
For example, in the 1D PDR model (Paper I), much of the $J$=~24--23 flux is emitted within gas with $A_{\rm V}\sim$0.3 when log~$n$=~6.5 and log~$G_0$=~4 (blue lines in Figure~\ref{fig:ep_1d_model}). If this gas is exposed to a higher UV flux of log~$G_0$=~5 (red lines in Figure~\ref{fig:ep_1d_model}), then the gas temperature increases, but the CO abundance decreases at $A_{\rm V}\sim0.3$. As a result, the center of emission arises from  deeper layers ($A_{\rm V}\sim1$).  Here the gas temperature is similar to that of the gas with $A_{\rm V}\sim0.3$ under log~$G_0$=~4.   Thus there is some similarity in the excitation conditions even within a changing external radiation field.
  
\begin{figure*}
\includegraphics[width=1.0 \textwidth]{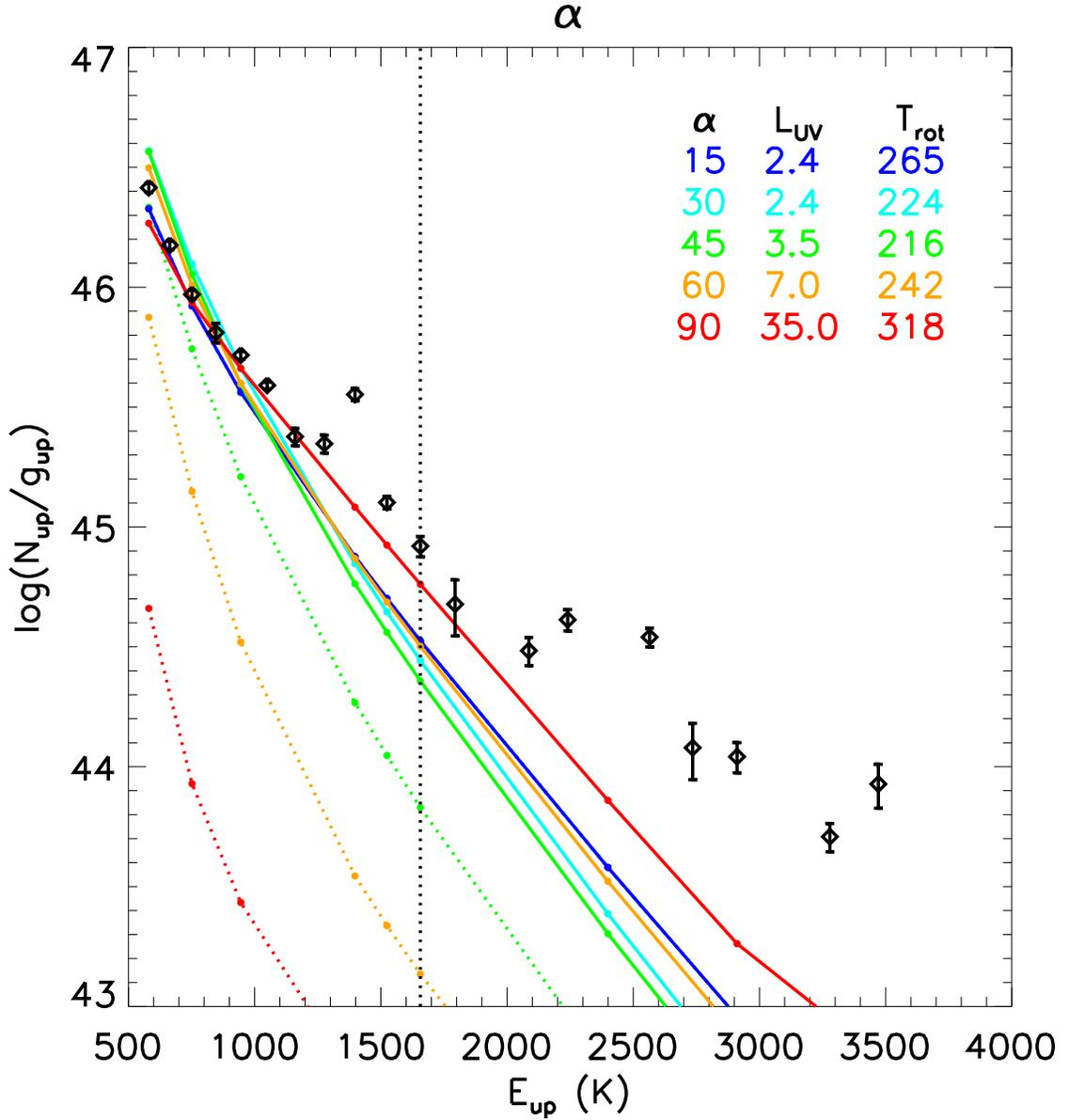}
\caption{The effect of opening angle $\alpha$ in the model of  L1157. Each color line indicates the rotation diagram from the model with a different opening angle $\alpha$. Dotted lines show the dependence of $\alpha$ at a given UV luminosity ($L_{\rm UV} = 2.4 L_{\rm UV}^Y$). However, solid lines present the best-fitted $L_{\rm UV}$ (on upper right) at a given $\alpha$. }\label{fig:ep_alpha}
\end{figure*}

 \textit{The effect of opening angle.}  As the opening angle increases, more UV photons escape and do not interact with the envelope. 
 As a result, at a given UV luminosity, the FUV strength along the wall of a larger outflow cavity is decreased, resulting in reduced CO emission and $T_{\rm rot}$ in the PACS range  (see dotted lines in Figure \ref{fig:ep_alpha}).  
 As seen in Figure~\ref{fig:ep_alpha}, an increase of $L_{\rm UV}$ is required in order to fit the observed fluxes for a wider opening angle. Therefore, our models with an opening angle of 30$^\circ$ require the minimum $L_{\rm UV}$. 
Thus, the best-fit model with a low $L_{\rm UV}$, assuming fixed opening angle of 30$^\circ$, can have an improved solution with a larger opening angle and a higher $L_{\rm UV}$.  For L1551-IRS5, a model with  the upper limit of $L_{\rm UV}= 5$~ $L_{\rm UV}^{\rm Y}$ and a large opening angle of 100~$^\circ$, which has been derived from the CO map \citep{Arce2006}, provides an improved fit  to the data.
 However, models with different opening angles in Figure~\ref{fig:ep_alpha}, will change dust continuum image and spectral energy distribution.   Thus, the density profile should be adjusted when assuming a different opening angle, which is not accounted for in our models. 
 
\begin{figure*}
\includegraphics[width=1.0 \textwidth]{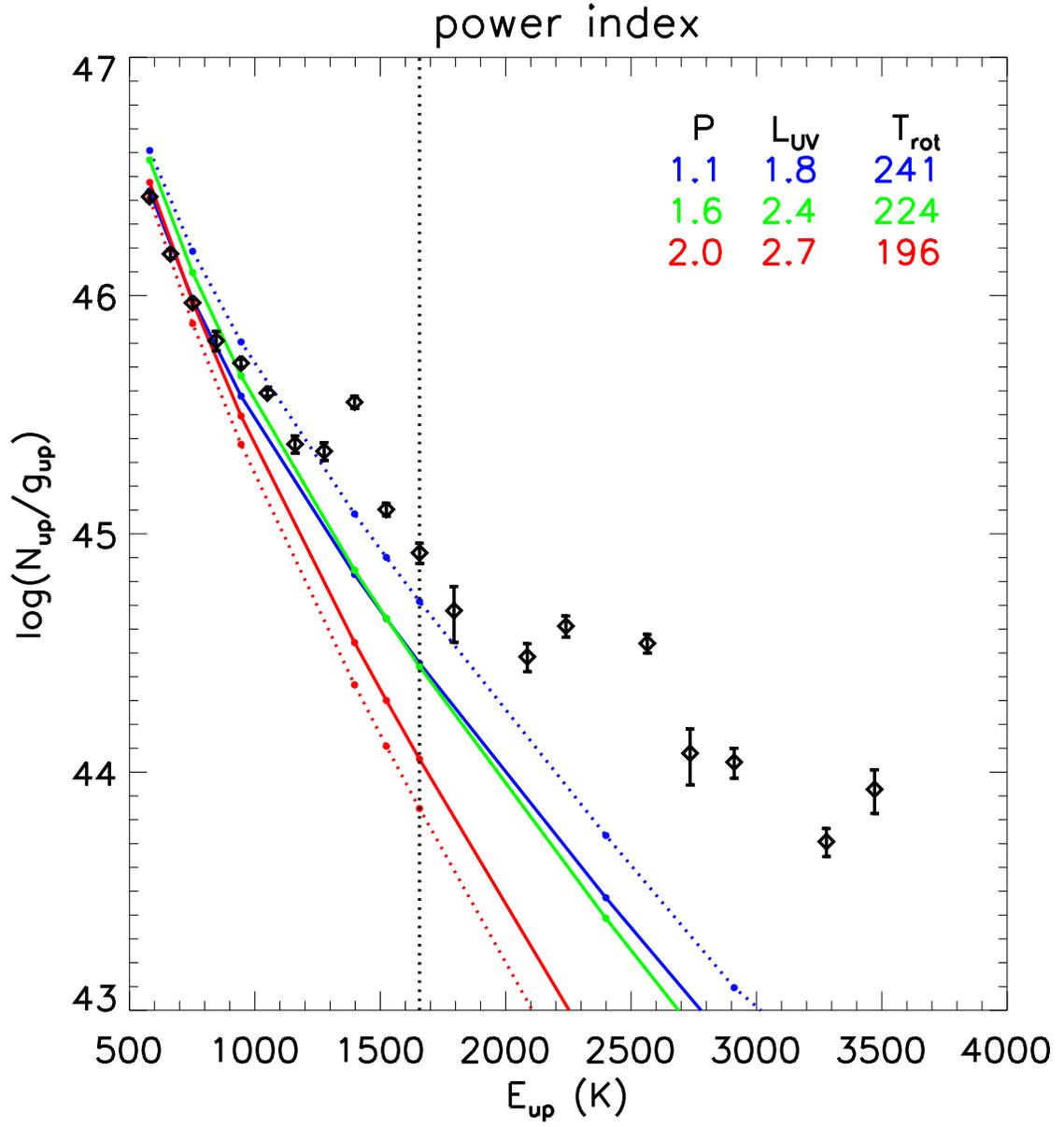}
\caption{The same as Figure \ref{fig:ep_alpha} except for the power index in density for L1157.  }\label{fig:ep_power}
\end{figure*}

 \textit{The effect of the density profile.}  A variation of the power index changes the amount and concentration of dense gas, as well as $G_0/n$, which could change the overall emission profile. In the case of L1157,  the power index in the density profile of L1157 has a minor impact on the mid-$J$ line fluxes as shown in Figure \ref{fig:ep_power}.
 The majority of mid-$J$ CO emission arises from gas near 1000~AU (see the left panel of Figure~\ref{fig:ep_uv_pop}) and at this radius changing the power index does not alter the emission appreciably. 
  However, in the case where mid-$J$ emission is generated in gas far at greater distances, then the power index significantly affects the result. 
  For example, TMC1 has a density at 1000~AU similar to that of TMC1A, and a lower $L_{\rm bol}$ than TMC1A only by a factor of 3. 
  The power indexes of TMC1 and TMC1A are 1.1 and 1.6, respectively. 
  The best-fit $L_{\rm UV}$ (in the unit of $L_\odot$) for TMC1 is  larger than that for TMC1A at least by an order of magnitude, because the lower power index of TMC1 reduces the size of  the dense region relevant to the population of mid-$J$ CO levels. 
  Therefore, a higher $L_{\rm UV}$ is needed to heat the reduced volume and produce the observed mid-$J$ CO emission.

\subsection{PDRs as a candidate mechanism to produce mid-$J$ CO emission}\label{sec:ut}  
\begin{figure*}
\includegraphics[width=0.95 \textwidth]{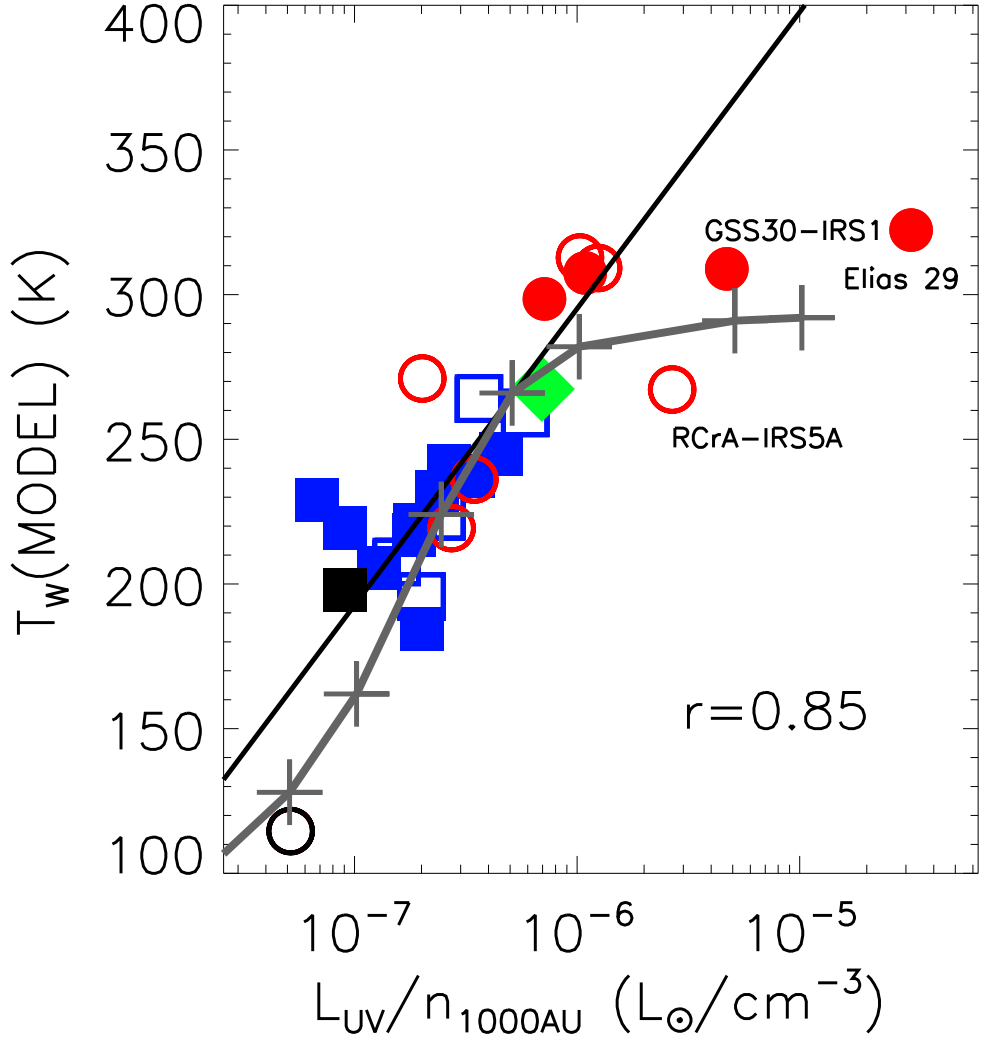}
\caption{The ratio of the best-fit UV luminosity to density at 1000~AU ($L_{\rm UV}/n_{1000 \rm AU}$) and  the rotational temperature of the warm component in the best fit model $T_{\rm W}({\rm MODEL})$. $T_{\rm W}({\rm MODEL})$ increases and has a strong correlation with $L_{\rm UV}/n_{1000 \rm AU}$ (r= 0.72 in the confidence level of 5 sigma) up to $L_{\rm UV}/n_{1000 \rm AU}\sim10^{-6}$ then nearly constant with $\sim300$~K. The grey line and symbols indicate the result of Figure~\ref{fig:ep_uv}.  The symbols are the same as Figure~\ref{fig:ep_source_luv}.
}\label{fig:ep_source3}
\end{figure*}

\citet{Manoj2013} argued that the PDRs cannot reproduce the constant $T_{\rm W}({\rm OBS})$ because $T_{\rm gas}$ is roughly proportional to $L_{\rm UV}/n_{\rm 1000 AU}$. 
However, in our modeling, $T_{\rm rot}$ is nearly constant as $\sim 300$~K when $\mathrm{log}~L_{\rm UV}/n_{\rm 1000AU} \ge -6$ as shown in Figure~~\ref{fig:ep_source3}. In addition, the variation of $T_{\rm rot}$ is within $\sim$100~K at $\mathrm{log}~L_{\rm UV}/n_{\rm 1000AU}< -6$.
Therefore, $T_{\rm rot}$ can be considered nearly constant with a scatter of 100~K when the mid-$J$ CO emission radiated from the UV heated outflow cavity walls. 

As shown in Figure~\ref{fig:Ser_SMM1_model}, most sources radiate the mid-$J$ CO emission within $\sim$1000~AU. Therefore, it is impossible that for an internal PDR to account for any extended CO emission. About a half of our sources have compact CO emission (see Table~\ref{tb:result}). In addition, even extended sources also show that a significant mid-$J$ CO flux (or emission peak) is detected near the protostar \citep{Karska2013,Green2013}. Thus, it is possible that some portion of the mid-$J$ CO emission produced by a PDR, which appears to be the case for lower-$J$ $^{13}$CO lines \citep{Y&inodot;ld&inodot;z2012}.

 $^{12}$CO $J$=10--9 HIFI data \citep{San Jos&eacute;-Garc&iacute;a2013, Y&inodot;ld&inodot;z2013} is the highest transition where published data exist that is spectroscopically resolved, except for Ser~SMM1 \citep{Kristensen2013}. 
According to \citet{Y&inodot;ld&inodot;z2013}, the contribution of the narrow velocity component to the total $J$=~10--9 flux seems to have no correlation with the evolutionary stage and varies from zero to 100~$\%$ with a median of 42~$\%$. 
In addition, the contribution of the broad velocity component increases as the rotational level ($J\le$~10) increases \citep{van Kempen2009b, Y&inodot;ld&inodot;z2012, Y&inodot;ld&inodot;z2013}. As mentioned by \citet{Manoj2013}, it is likely that a broad (shocked) component has a larger  contribution to the mid-$J$ CO emission above that seen for $J$=~10--9. 

 However, a UV-irradiated shock might be a contributor to the FIR line emission toward some embedded protostars \citep{Goicoechea2012,Kristensen2013,Lee2014a}. A portion of the outflowing gas exists between the protostar and quiescent outflow cavity walls and, depending on the presence of dust in the cavity, this gas will be irradiated by the accreting star, or perhaps by highly excited H or H$_2$ locally in the gas itself.   Thus, although it seems likely that the mid-$J$ emission arises in an outflow,  it is possible, or even probable, that the UV photons are a key additional contributor, beyond the shock itself, to  heating outflowing gas. 

 CO ladder emission must carry a tremendous amount of information on the physical state and evolution of protostars.  In gas that is not thermally coupled to dust grains (i.e. in PDR or shock heated gas) CO emissions are the main gas coolant and thus we are tracing the energy output from star formation.  This work and \citet{Visser2012} have shown that an internal PDR can contribute to mid-$J$ CO emissions, even in the case without a contribution from shock heating.   However, the broad line width detected for resolved CO lines ($J \ge 10$) implies that this gas is not quiescent, but rather must be entrained in an outflow and subject to shock. Evidence for the presence of UV emission is found in the detection of ions such as CH$^+$ and OH$^+$ \citep{Kristensen2013}, which require UV radiation in close proximity to the protostar to be created.  Furthermore, the inferred abundance of water vapor in shocked gas \citep[e.g.,][]{Santangelo2014} is below that predicted by shock models requiring some mechanism to reduce the anticipated H$_2$O creation in the shock.  A likely candidate to reduce the water abundance is UV photodissociation generated by the protostar or perhaps produced in the shock itself.   Thus, we know that both shocks and UV photons are present in this environment and detailed modeling of both, encompassing all these constraints, will be required to extract the key gas physics that is operating in the gas near stars at their birth.

\section{Summary}\label{sec:summary}

We have modeled a UV-heated outflow cavity wall as mechanism to produce the mid-$J$ CO line emission detected  by \textit{Herschel}/PACS towards protostars. We obtain the following results:\\
$\bullet$ The UV-heated outflow cavity walls can reproduce the observed compact FIR mid-$J$ CO emission, except for the sources with the rotational temperature above 400~K, alone or when combined with a hot shocked component.\\
$\bullet$ The mid-$J$ ($14\leq J \leq 24$) CO emission can be radiated from the surface (0.1~$\leq$~A$_{\rm V}$~$\leq$~1) of  a dense ($n~\geq$~10$^6$\vol) outflow cavity wall  with $\mathrm{log}~G_{\rm 0}/n \geq-4.5$ and $\mathrm{log}~G_{\rm 0} \ge 3$, where X(CO) $\geq$ 10$^{-5}$ and  $T_{\rm gas} \geq$300~K. \\
$\bullet$ Under the above conditions, the H$_2$O photodesorption and  the CO formation rates are high enough to keep CO in the warm gas, resulting in the mid-$J$ CO emission. 

 Our results could support the result of \citet{Visser2012} and the possibility that the PDR contributes at some level to mid-$J$ CO emission for sources with the UV luminosity derived from the T-Tauri stars.
 
 \acknowledgments
This research was supported by the Basic
Science Research Program through the National Research Foundation
of Korea (NRF) funded by the Ministry of Education of
the Korean government (grant No. NRF-2012R1A1A2044689).

\bibliographystyle{aa}
\bibliography{biblio}

\clearpage

\begin{deluxetable}{lcrrrrcc}
\tablewidth{0pt}
\tablecaption{Source parameters \label{tb:source} }
\tablehead{	\colhead{source} & 
			\colhead{$D$}  	& 
			\colhead{$L_{\rm bol}$} & 
			\colhead{$T_{\rm bol}$} & 
			\colhead{$p$\tablenotemark{a}} & 
			\colhead{$r_{\rm in}$\tablenotemark{b}} & 
			\colhead{$r_{\rm out}$\tablenotemark{b,c}} & 
			\colhead{n$_{\rm 1000 AU}$\tablenotemark{a,c}}\\
			
        				&
        		\colhead{pc} & 
        		\colhead{$L_{\odot}$} & 
        		\colhead{K}   &
        		            & 
        		\colhead{AU} & 
        		\colhead{AU} & 
        		\colhead{cm$^{-3}$}       }          
\startdata
            L1448-MM &  232 &  8.4 &  47 &  1.5 & 20.7 & 1.9(4) & 3.9(6)\\
      NGC1333-IRAS2A &  235 & 35.7 &  50 &1.7 & 35.9 & 1.8(4) & 1.7(6)\\
      NGC1333-IRAS4A &  235 &  9.1 &  33 & 1.8 & 33.5 & 3.4(4) & 6.7(6) \\
      NGC1333-IRAS4B &  235 &  4.4 &  28 & 1.4 & 33.5 & 2.7(4) & 5.7(6) \\
               L1527 &  140 &  1.9 &  44 & 0.9 &  5.4 & 6.5(3) & 8.1(5)  \\
         Ced110-IRS4 &  125 &  0.8 &  56 &  1.4 &  4.1 & 5.7(3) & 3.9(5) \\
               BHR71&  200 & 14.8 &  44 & 1.7 & 24.8 & 1.2(4) & 1.8(6) \\
          IRAS 15398 &  130 &  1.6 &  52 &1.4 &  6.2 & 6.2(3) & 1.6(6) \\
        VLA 1623-243 &  125 &  2.6 &  35 & 1.4 &  4.3 & 1.0(4) & 7.7(5) \\
                L483 &  200 & 10.2 &  49 &  0.9 & 12.5 & 1.3(4) & 5.1(5) \\
            Ser SMM1 &  230 &  1.9 &  26 &1.3 &  31.0 & 1.6(4) & 4.1(6) \\
            Ser SMM4 &  230 &  1.9 &  26 & 1.0 &  6.8 & 1.1(4) & 5.4(6) \\
            Ser SMM3 & 230 &  5.1 &  38 &0.8 &  8.9 & 1.1(4) & 1.1(6) \\
                L723 &  300 &  3.6 &  39 &1.2 &  8.4 & 2.4(4) & 8.0(5)\\
                B33  &  250 &  3.3 &  36 &1.4 & 9.8 & 1.2(4) & 1.5(6) \\
               L1157 &  325 &  6.5 &  39 &1.6 & 14.3 & 3.1(4) & 2.0(6) \\
\hline
               L1489 &  140 &  3.8 & 200 &1.5 &  8.4 & 6.7(3) & 1.9(5) \\
          L1551-IRS5 &  140 & 24.5 & 105 &1.8 & 28.9 & 2.6(4) & 1.2(6) \\
                TMR1 &  140 &  3.8 & 133 & 1.6 &  8.8 & 7.9(3) & 2.1(5) \\
               TMC1A &  140 &  2.7 & 118 & 1.6 &  7.7 & 6.9(3) & 2.2(5) \\
              TMC1   &  140 &  0.9 & 101 &1.1 &  3.7 & 6.7(3) & 1.8(5) \\
                HH46 &  450 & 27.9 & 104 & 1.6 & 28.5 & 2.3(4) & 1.2(6) \\
              DK Cha&  178 & 35.1 & 591 & 1.6 & 12.0 & 9.6(3) & 9.2(5) \\ 
          GSS30-IRS1 &  125 & 14.5 & 138 & 1.6 & 16.2 & 1.6(4) & 1.7(5) \\
            Elias 29 &  125 & 20.1 & 386 &1.6 & 16.2 & 1.6(4) & 8.3(4) \\
               RNO91 & 125 &  2.6 & 340 &1.2 &  6.6 & 5.9(3) & 3.3(5) \\ 
          RCrA-IRS5A & 130 &  7.1 & 126 & 0.8 & 10.1 & 1.0(4) & 2.8(5) \\
\hline
        NGC7129-FIRS2\tablenotemark{d} &1260 & 500 &-- & 1.4 & 100.0 & 1.8(4) & 1.0(7)
\enddata
\tablecomments{Sources above the horizontal line are Class 0, sources below are Class I. Physical parameters ($p$, $r_{\rm in}$, $r_{\rm out}$, and $n_{1000 \rm AU}$) are adopted from \citet{J&oslash;rgensen2002} and \citet{Kristensen2012}.}
\tablenotetext{a}{The power law index and the molecular hydrogen number density at 1000 AU for the density structure, $n~(r)~=~n_{1000\rm AU}\times \left(\frac{r}{1000 {\rm AU}}\right)^{-p}$}
\tablenotetext{b}{Inner ($r_{in}$) and outer ($r_{out}$) boundary radii.}
\tablenotetext{c}{a(b) = a $\times$ 10$^b$}
\tablenotetext{d}{Intermediate mass source \citep{Crimier2010,Fich2010}}
\end{deluxetable}
\clearpage

\begin{deluxetable}{lcccccccc}
\rotate
\tablewidth{0pt}
\tablecaption{Model results\label{tb:result}}
\tablecolumns{7}
\tablehead{	\colhead{Source} &
			\colhead{Extent}\tablenotemark{a} &
			\colhead{$T_{\rm H}\mathrm{(OBS)}$ \tablenotemark{b}} &
			\colhead{$T_{\rm W}^{\mathrm{C}}\mathrm{(OBS)}$ \tablenotemark{c}} &
			\colhead{$T_{\rm W}\mathrm{(OBS)}$ \tablenotemark{d}} &
            \colhead{$T_{\rm W}\mathrm{(MODEL)}$\tablenotemark{e}} &
			\colhead{$L_{\rm UV}$ \tablenotemark{f}} & 
            \colhead{source type\tablenotemark{g}} &
            \colhead{Ref.\tablenotemark{h}} 
 }
\startdata
      L1448-MM  & E &  771 & 224 & 324 $\pm$ 22 & 240 &  4.1 & H & 2,3 \\
NGC1333-IRAS2A  & C &  --  & --  & 518 $\pm$170 & 195 &  0.4 & S & 1 \\
NGC1333-IRAS4A  & E &  479 & 184 & 298 $\pm$ 19 & 184 &  5.0  & H & 1 \\
NGC1333-IRAS4B  & E &  893 & 255 & 343 $\pm$ 20 & 205 &  5.0\tablenotemark{i}& H & 4 \\
         L1527  & E &  634 & 248 & 299 $\pm$ 45 & 219 &  1.0 & P & 1,2,11 \\
   Ced110-IRS4  & E &  799 & 417 & 491 $\pm$ 61 & 197 &  1.0 & P & 1 \\
         BHR71  & E &  659 & 225 & 376 $\pm$ 27 & 237 &  1.5 & H & 1,2 \\
    IRAS~15398  & E &  525 & 246 & 281 $\pm$ 18 & 220 &  4.7 & P & 1 \\
  VLA 1623-243  & E &   -- &  -- & 347 $\pm$ 13 & 216 &  1.5 & S & 2 \\
          L483  & C &  719 & 289 & 355 $\pm$ 26 & 259 &  1.0 & P & 1 \\
      Ser SMM1  & C &  656 & 273 & 351 $\pm$ 31 & 264 &  2.0 & P & 5 \\
      Ser SMM4  & E &  689 & 196 & 257 $\pm$ 18 & 228 &  5.0\tablenotemark{i} & H & 1,6 \\
      Ser SMM3  & E &  653 & 195 & 291 $\pm$ 18 & 244 &  3.0 & H & 1,6 \\
          L723  & E &  581 & 219 & 350 $\pm$ 26 & 231 &  1.5 & H & 1 \\
          B335  & C &  612 & 224 & 310 $\pm$ 18 & 207 &  2.0 & H & 2 \\          
         L1157  & C &  801 & 266 & 360 $\pm$ 23 & 223 &  2.4 & H & 2 \\
\hline
         L1489  & C & 739  & 282 & 373 $\pm$ 30 & 312 &  1.5 & H & 1,2,11 \\
    L1551-IRS5  & C &  --  & --  & 436 $\pm$ 32 & 114 &  0.1 & S & 2,11 \\
          TMR1  & C &  745 & 227 & 328 $\pm$ 28 & 309 &  2.0 & H & 1,2,11 \\
         TMC1A  & E &  --  & --  & 445 $\pm$ 47 & 298 &  1.6 & S & 1,2,11 \\
          TMC1  & E &  775 & 290 & 362 $\pm$ 29 & 307 &  5.0\tablenotemark{i}& P & 1,2,11 \\
          HH46  & C &  652 & 265 & 307 $\pm$ 22 & 236 &  0.6 & P & 1,9 \\
        DK Cha  & C & 1056 & 262 & 386 $\pm$ 26 & 219 &  0.3 & H & 2,10 \\
    GSS30-IRS1  & E &  802 & 212 & 335 $\pm$ 19 & 308 &  2.0 & H & 2 \\
    Elias 29    & E &  723 & 159 & 315 $\pm$ 19 & 322 &  5.0\tablenotemark{i} & H & 2 \\
         RNO91  & C &  738 & 275 & 252 $\pm$ 25 & 270 &  0.7 & P & 1 \\
    RCrA-IRS5A  & C & 893  & 241 & 293 $\pm$ 17 & 267 &  3.4 & H & 2,7 \\  
\hline
 NGC7129-FIRS2  &-- &  797 & 250 & 336  $\pm$ 24 & 267  & 1.0 & H  & 8 
\enddata

\tablenotetext{a}{Spatial extent of CO lines. When most of mid-$J$ CO fluxes are detected in the central pixel, we define it as a compact source (C), and the other case, as an extended source (E). For sources in ref. 1, we use the extent of the CO $J$=14--13 emission. For sources in ref. 2, the sources with a smaller extent than a point spread function at CO~$J$=16--15 is considered as compact. }

\tablenotetext{b}{Rotational temperature of the ``Hot" component from the observed fluxes  above CO $J$=24--23.}

\tablenotetext{c}{Rotational temperature of the ``Warm" component from the observed fluxes  between CO $J$=14--13 and $J$=24--23 in condition of removing the contribution of the ``Hot" component to the fluxes in these transitions.}

\tablenotetext{d}{Rotational temperature of the ``Warm" component from the total observed fluxes  between CO $J$=14--13 and $J$=24--23} 

\tablenotetext{e}{Rotational temperature from the modeled fluxes  between CO $J$=14--13 and $J$=24--23.}

\tablenotetext{f}{The best-fit UV luminosity in unit of $L_{\rm UV}^Y$. (see Eq. \ref{eq:luv})}

\tablenotetext{g}{Type of Rotational Diagram for the observed data (see text).}

\tablenotetext{h}{1 : \citet{Karska2013}, 2: \citet{Green2013}, 3: \citet{Lee2013}, 4: \citet{Herczeg2012}, 5: \citet{Goicoechea2012}. 6: \citet{Dionatos2013}, 7: \citet{Lindberg2014}, 8: \citet{Fich2010}, 9: \citet{van Kempen2010a} 10:\citet{van Kempen2010b}, 11: \citet{Lee2014a}
}
\tablenotetext{i}{The upper limit of $L_{\rm UV}$ in our model. The observed mid-$J$ CO fluxes of these four sources are not fitted even with this upper limit of $L_{\rm UV}$.}

\end{deluxetable}

\clearpage

\clearpage

\end{document}